\documentclass[%
 reprint,
 nofootinbib,
superscriptaddress,
 amsmath,amssymb,
 aps,
 prc,
 floatfix,
]{revtex4-2} 

\usepackage{graphicx}
\usepackage{dcolumn}
\usepackage{bm}
\usepackage[caption=false]{subfig}
\usepackage{float}
\usepackage[stable]{footmisc}
\usepackage[english]{babel}
\usepackage[autostyle, english = american]{csquotes}
\MakeOuterQuote{"}
\usepackage[table]{xcolor} 

\usepackage{hyperref}
\hypersetup{hidelinks}

\begin{document}

\preprint{APS/123-QED}

\title{Longitudinal dynamics of large and small systems from a 3D Bayesian calibration of RHIC top-energy collision data}




\author{A.~Mankolli}
\affiliation{Department of Physics and Astronomy, Vanderbilt University, Nashville, TN 37235.}

\author{C.~Shen}
\affiliation{Department of Physics and Astronomy, Wayne State University, Detroit, MI 48201.}
\affiliation{RIKEN BNL Research Center, Brookhaven National Laboratory, Upton, NY 11973.}

\author{M.~Luzum}
\affiliation{Instituto  de  F\'{i}sica,  Universidade  de  S\~{a}o  Paulo,  C.P.  66318,  05315-970  S\~{a}o  Paulo,  SP,  Brazil. }

\author{J.-F.~Paquet}
\affiliation{Department of Physics and Astronomy, Vanderbilt University, Nashville, TN 37235.}

\author{M.~Singh}
\affiliation{Department of Physics and Astronomy, Vanderbilt University, Nashville, TN 37235.}

\author{J.~Velkovska}
\affiliation{Department of Physics and Astronomy, Vanderbilt University, Nashville, TN 37235.}


\author{S.~A.~Bass}
\affiliation{Department of Physics, Duke University, Durham, NC 27708.}

\author{C.~Gale}
\affiliation{Department of Physics, McGill University, Montr\'{e}al, QC H3A 2T8, Canada.}

\author{G.~A.~C.~da Silva}
\affiliation{Instituto  de  F\'{i}sica,  Universidade  de  S\~{a}o  Paulo,  C.P.  66318,  05315-970  S\~{a}o  Paulo,  SP,  Brazil. }

\author{L.~Du}
\affiliation{Department of Physics, University of California, Berkeley, CA 94720.}
\affiliation{Nuclear Science Division, Lawrence Berkeley National Laboratory, Berkeley, CA 94720.}
\affiliation{Department of Physics, McGill University, Montr\'{e}al, QC H3A 2T8, Canada.}

\author{L.~Kasper}
\affiliation{Department of Physics and Astronomy, Vanderbilt University, Nashville, TN 37235.}

\author{G.~S.~Rocha}
\affiliation{Instituto de F\'{i}sica, Universidade Federal Fluminense, Niter\'{o}i, Rio de Janeiro, 24210-346, Brazil}
\affiliation{Department of Physics and Astronomy, Vanderbilt University, Nashville, TN 37235.}

\author{D.~Soeder}
\affiliation{Department of Physics, Duke University, Durham, NC 27708.}

\author{S.~Tuo}
\affiliation{Department of Physics and Astronomy, Vanderbilt University, Nashville, TN 37235.}

\author{G.~Vujanovic}
\affiliation{Department of Physics, University of Regina, Regina, SK S4S 0A2, Canada.}

\author{X.~Wu}
\affiliation{Department of Physics, McGill University, Montr\'{e}al, QC H3A 2T8, Canada.}

\author{W.~Zhao}
\affiliation{Department of Physics, University of California, Berkeley, CA 94720.}
\affiliation{Nuclear Science Division, Lawrence Berkeley National Laboratory, Berkeley, CA 94720.}


\author{M.~Chartier}
\affiliation{Oliver Lodge Laboratory, University of Liverpool, Liverpool, United Kingdom.}

\author{Y.~Chen}
\affiliation{Department of Physics and Astronomy, Vanderbilt University, Nashville, TN 37235.}

\author{R.~Datta}
\affiliation{Department of Physics and Astronomy, Wayne State University, Detroit, MI 48201.}

\author{R.~Dolan}
\affiliation{Department of Physics and Astronomy, Wayne State University, Detroit, MI 48201.}

\author{R.~Ehlers}
\affiliation{Department of Physics, University of California, Berkeley, CA 94720.}
\affiliation{Nuclear Science Division, Lawrence Berkeley National Laboratory, Berkeley, CA 94720.}

\author{H.~Elfner}
\affiliation{GSI Helmholtzzentrum f\"{u}r Schwerionenforschung, 64291 Darmstadt, Germany.}
\affiliation{Institute for Theoretical Physics, Goethe University, 60438 Frankfurt am Main, Germany.}
\affiliation{Frankfurt Institute for Advanced Studies, 60438 Frankfurt am Main, Germany.}

\author{R.~J.~Fries}
\affiliation{Cyclotron Institute, Texas A\&M University, College Station, TX 77843.}
\affiliation{Department of Physics and Astronomy, Texas A\&M University, College Station, TX 77843.}

\author{D.~A.~Hangal}
\affiliation{Lawrence Livermore National Laboratory, Livermore, CA 94550.}
\affiliation{Department of Physics and Astronomy, Wayne State University, Detroit, MI 48201.}


\author{B.~V.~Jacak}
\affiliation{Department of Physics, University of California, Berkeley, CA 94720.}
\affiliation{Nuclear Science Division, Lawrence Berkeley National Laboratory, Berkeley, CA 94720.}

\author{P.~M.~Jacobs}
\affiliation{Department of Physics, University of California, Berkeley, CA 94720.}
\affiliation{Nuclear Science Division, Lawrence Berkeley National Laboratory, Berkeley, CA 94720.}

\author{S.~Jeon}
\affiliation{Department of Physics, McGill University, Montr\'{e}al, QC H3A 2T8, Canada.}

\author{Y.~Ji}
\affiliation{Department of Statistical Science, Duke University, Durham, NC 27708.}

\author{F.~Jonas}
\affiliation{Department of Physics, University of California, Berkeley, CA 94720.}
\affiliation{Nuclear Science Division, Lawrence Berkeley National Laboratory, Berkeley, CA 94720.}

\author{M.~Kordell~II}
\affiliation{Cyclotron Institute, Texas A\&M University, College Station, TX 77843.}
\affiliation{Department of Physics and Astronomy, Texas A\&M University, College Station, TX 77843.}

\author{A.~Kumar}
\affiliation{Department of Physics, University of Regina, Regina, SK S4S 0A2, Canada.}

\author{R.~Kunnawalkam-Elayavalli}
\affiliation{Department of Physics and Astronomy, Vanderbilt University, Nashville, TN 37235.}

\author{J.~Latessa}
\affiliation{Department of Computer Science, Wayne State University, Detroit, MI 48202.}

\author{Y.-J.~Lee}
\affiliation{Laboratory for Nuclear Science, Massachusetts Institute of Technology, Cambridge, MA 02139.}
\affiliation{Department of Physics, Massachusetts Institute of Technology, Cambridge, MA 02139.}



\author{A.~Majumder}
\affiliation{Department of Physics and Astronomy, Wayne State University, Detroit, MI 48201.}

\author{S.~Mak}
\affiliation{Department of Statistical Science, Duke University, Durham, NC 27708.}


\author{C.~Martin}
\affiliation{Department of Physics and Astronomy, University of Tennessee, Knoxville, TN 37996.}

\author{H.~Mehryar}
\affiliation{Department of Computer Science, Wayne State University, Detroit, MI 48202.}

\author{T.~Mengel}
\affiliation{Department of Physics and Astronomy, University of Tennessee, Knoxville, TN 37996.}

\author{C.~Nattrass}
\affiliation{Department of Physics and Astronomy, University of Tennessee, Knoxville, TN 37996.}

\author{J.~Norman}
\affiliation{Oliver Lodge Laboratory, University of Liverpool, Liverpool, United Kingdom.}

\author{M.~Ockleton}
\affiliation{Oliver Lodge Laboratory, University of Liverpool, Liverpool, United Kingdom.}

\author{C.~Parker}
\affiliation{Cyclotron Institute, Texas A\&M University, College Station, TX 77843.}
\affiliation{Department of Physics and Astronomy, Texas A\&M University, College Station, TX 77843.}


\author{J.~H.~Putschke}
\affiliation{Department of Physics and Astronomy, Wayne State University, Detroit, MI 48201.}

\author{H.~Roch}
\affiliation{University of Jyväskylä, Department of Physics, P.O. Box 35, FI-40014 University of Jyväskylä, Finland.}
\affiliation{Helsinki Institute of Physics, P.O. Box 64, FI-00014 University of Helsinki, Finland.}

\author{G.~Roland}
\affiliation{Laboratory for Nuclear Science, Massachusetts Institute of Technology, Cambridge, MA 02139.}
\affiliation{Department of Physics, Massachusetts Institute of Technology, Cambridge, MA 02139.}

\author{B.~Schenke}
\affiliation{Physics Department, Brookhaven National Laboratory, Upton, NY 11973.}

\author{L.~Schwiebert}
\affiliation{Department of Computer Science, Wayne State University, Detroit, MI 48202.}

\author{A.~Sengupta}
\affiliation{Cyclotron Institute, Texas A\&M University, College Station, TX 77843.}
\affiliation{Department of Physics and Astronomy, Texas A\&M University, College Station, TX 77843.}



\author{C.~Sirimanna}
\affiliation{Department of Physics, Duke University, Durham, NC 27708.}

\author{R.~A.~Soltz}
\affiliation{Lawrence Livermore National Laboratory, Livermore, CA 94550.}
\affiliation{Department of Physics and Astronomy, Wayne State University, Detroit, MI 48201.}

\author{I.~Soudi}
\affiliation{Fakult\"at f\"ur Physik, Universit\"at Bielefeld, D-33615 Bielefeld, Germany}

\author{Y.~Tachibana}
\affiliation{Akita International University, Yuwa, Akita-city 010-1292, Japan.}


\author{X.-N.~Wang}
\affiliation{Key Laboratory of Quark and Lepton Physics (MOE) and Institute of Particle Physics, Central China Normal University, Wuhan 430079, China.}
\affiliation{Department of Physics, University of California, Berkeley, CA 94720.}
\affiliation{Nuclear Science Division, Lawrence Berkeley National Laboratory, Berkeley, CA 94720.}

\author{J.~Zhang}
\affiliation{Department of Physics and Astronomy, Vanderbilt University, Nashville, TN 37235.}

\collaboration{The JETSCAPE Collaboration}

\begin{abstract}
A comprehensive Bayesian analysis of the 3D dynamics of high-energy nuclear collisions is presented. We perform a systematic model-to-data comparison using simulations of large and small collision systems, and a broad range of
measurements from the PHENIX, STAR, PHOBOS, and BRAHMS collaborations spanning nearly two decades of RHIC operations. In particular, we
perform fully 3D multi-stage simulations including rapidity-dependent energy
deposition with global energy conservation using the 3D Glauber model, along with relativistic viscous hydrodynamics with MUSIC. 
We calibrate the model on rapidity- and $p_T$-differential observables and
analyze the respective constraints on initial state and transport properties they
provide. We emphasize the additional constraints provided by rapidity-dependent measurements, the differences in large and small system calibrations, and
the tension exhibited by particular observables. We use our calibrated model
to make predictions of observables in p-Au and
$^3$He-Au collisions. Furthermore, we facilitate direct comparison of experimental measurements by highlighting the dependence of flow measurements on the rapidity of the regions of interest and reference, as well as the importance of the centrality selection. In particular, we examine the apparent differences between the STAR and PHENIX $v_2$ and $v_3$ measurements in small systems. 
\end{abstract}

\maketitle


\newpage

\definecolor{GHighlight}{rgb}{0.1,0.9,0.3}



\section{\label{sec:introduction}Introduction }

The strong nuclear force described by Quantum Chromodynamics (QCD) can, under extreme conditions of energy density, give rise to a deconfined state of quarks and gluons called the quark-gluon plasma (QGP). The study of QGP has been the principal goal of the relativistic nuclear collision programs at the Large Hadron Collider (LHC) at CERN and the Relativistic Heavy Ion Collider (RHIC) at Brookhaven National Laboratory. Measurements spanning over two decades at these colliders have confirmed the hydrodynamic nature of the QGP expansion~\cite{BRAHMS:2004adc, PHOBOS:2004zne, STAR:2005gfr, PHENIX:2004vcz, Muller:2012zq}. Phenomenological studies using portions of this wealth of data in a Bayesian inference framework have made significant progress in constraining the transport properties of strongly coupled QCD matter~\cite{Bernhard:2016tnd, Bernhard2019, Ke:2016jrd, Auvinen:2017fjw, JETSCAPE:2020mzn, JETSCAPE:2020shq, Nijs:2020roc, Nijs:2020ors, Nijs:2022rme, Soeder:2023vdn, Virta:2024avu, Parkkila:2021tqq, Yang:2022ixy,
Heffernan:2023gye, Heffernan:2023utr,  Jahan:2024wpj, Shen:2023awv, Shen:2023pgb, Gotz:2025wnv, JETSCAPE:2023nuf, Heffernan:2023kpm, Everett:2021ruv, Parkkila:2021thx, Paquet:2023rfd}, which remain difficult to constrain with lattice QCD calculations. 

The experimental programs have in addition provided evidence for the possible emergence of QGP in smaller collision systems~\cite{CMS:2010ifv, ATLAS:2015hzw, CMS:2012qk, ATLAS:2012cix, ALICE:2012eyl, PHENIX:2013ktj, Nagle:2018nvi, Heinz:2019dbd}. 
In particular, studies at RHIC have measured hydrodynamic flow signals in a variety of small systems and their strong dependence on the initial state of the collisions~\cite{PHENIX:2018lia, STAR:2023wmd}. Furthermore, the orientation of the event planes, and the hydrodynamic response to the initial geometry have been understood to have particularly strong longitudinal dependence in smaller collisions~\cite{STAR:2023wmd, Nie:2020trj, Jia:2024xvl}.

The aim of this study is to exploit the large number of rapidity-dependent measurements at the RHIC top energy in tandem with a (3+1)D model of nuclear collisions to better constrain the properties of the QGP formed in heavy-ion collisions, to gain insight about the nature of the longitudinal dynamics in these collisions, and to explore any tension between large and small systems within a hydrodynamic description. This information can be extracted only thanks to our use of a fully (3+1)D model that not only describes the finite rapidity region explicitly and makes possible direct comparisons with forward and backward measurements, but also allows us to incorporate the finite rapidity effects embedded in mid-rapidity measurements.

In Section~\ref{sec:model} we describe in detail the theoretical model of nuclear collisions
we employ in our study. Section~\ref{sec:bayes} covers the techniques we use to emulate our model's predictions, and gives an overview of our Bayesian analysis procedure and choices. In Section~\ref{sec:exp_data} we list and describe the experimental data we use in our calibrations and discuss our selections. In Section~\ref{sec:posteriors} we show the results of Bayesian calibrations for our default dataset as well as several topical subsets and highlight what can be learned from them; in that Section, we also discuss the tension which different observables place on our model. In Section~\ref{sec:predictions} we use our calibrated model to provide a general description of the longitudinal structure of collectivity across collision systems and to make additional predictions for measurements, including in p-Au and $^3$He-Au collisions.

\section{The (3+1)D modeling of nuclear collisions}
\label{sec:model}

Bulk observables in high-energy nuclear collisions, such as the number of produced particles and their azimuthal distributions, have long been well-described by multi-stage models that include an initial-state energy-deposition phase followed by a hydrodynamic evolution and (often) hadronic transport~\cite{Bass:2000ib,Teaney:2001av,Heinz:2013th, Gale:2013da,DerradideSouza:2015kpt} thereafter.
Typically, the models, using a variety of prescriptions, define an initial energy and momentum density distribution for a locally thermally-equilibrated hydrodynamic medium, which is then evolved for several fm/$c$ as an expanding relativistic fluid.
Then, as the systems become increasingly dilute, individual fluid cells are converted into hadrons
which scatter with each other and decay into stable states.
In the high-energy limit, longitudinal boost-invariance is a good approximation for mid-rapidity observables~\cite{Bjorken:1982qr}. It reduces computational cost in event-by-event simulations, enabling comprehensive Bayesian inference with (2+1)D models and mid-rapidity data. A number of such studies in the last decade have been shown to provide constraints on the transport coefficients of QGP~\cite{Bernhard:2016tnd, Bernhard2019, JETSCAPE:2020mzn, JETSCAPE:2020shq, Nijs:2020roc, Nijs:2020ors, Nijs:2022rme, Virta:2024avu, Parkkila:2021tqq, Yang:2022ixy, Heffernan:2023gye, Heffernan:2023utr}.
A few studies have performed Bayesian calibrations with full (3+1)D simulations and deduced additional constraints from experimental data in forward and backward rapidity regions~\cite{Auvinen:2017fjw, Ke:2016jrd, Soeder:2023vdn, Jahan:2024wpj, Shen:2023awv, Shen:2023pgb, Gotz:2025wnv}. 
Some of these studies have included models that differ substantially from ours, particularly in the prescription of the initial state~\cite{Auvinen:2017fjw, Ke:2016jrd, Soeder:2023vdn, Gotz:2025wnv}. 
Others have used the same overarching models, though with different physics assumptions and scope~\cite{Jahan:2024wpj, Shen:2023awv, Shen:2023pgb}.
In this study, we use the 3D Glauber model~\cite{Shen:2017bsr, Shen:2022oyg, Ryu:2023bmx} to describe the initial state, followed by (3+1)D viscous relativistic hydrodynamics for the plasma phase~\cite{Schenke:2010nt,Schenke:2010rr,Paquet:2015lta}, and a hadronic afterburner to model the hadronic interactions of final-state particles~\cite{Bass:1998ca, Bleicher:1999xi}.

\subsection{\label{sec:initial-state}The 3D initial state}

The initial state of heavy-ion collisions, particularly its longitudinal dependence, is still not well understood in terms of a first-principles QCD description~\cite{Chesler:2015wra,Busza:2018rrf,Schlichting:2019abc,Berges:2020fwq}. However, models that sample nucleons from density distributions and accordingly deposit energy in the transverse plane of the collision have been successful when coupled to hydrodynamics in describing the experimental data~\cite{Song:2010mg, Moreland:2014oya}. One such model is the Monte Carlo Glauber model~\cite{Miller:2007ri, Loizides:2016djv, Alver:2008aq, Loizides:2014vua}. In recent years, efforts have been made to extend the description of some such models to the energy deposited in the longitudinal direction~\cite{Shen:2017bsr, Shen:2020jwv, Shen:2022oyg, Soeder:2023vdn}. The 3D Monte Carlo Glauber model, which we use for this study, has been a result of such efforts and has been used for a number of years to successfully describe a range of heavy-ion collision data, particularly at energies probed in the RHIC Beam Energy Scan program~\cite{Shen:2017bsr, Shen:2022oyg}.

For a given impact parameter in a collision, nucleons are first sampled according to a 3D Woods-Saxon distribution (such that collisions may take place at finite $z$ positions to an extent modulated by the collision energy). Any given nucleon is modeled as a collection of three valence quark hotspots and one soft gluon hotspot, each of which may lose and deposit energy in a collision~\cite{Shen:2022oyg, JETSCAPE:2024dgu}. The longitudinal momenta related to the valence quark hotspots are sampled from the nucleons' parton distribution functions.\footnote{Nuclear parton distribution modifications are also included in this treatment (EPS09)~\cite{Eskola:2009uj}} Their spatial positions are sampled from a 3D Gaussian with width $\sqrt{B_G}$, a parameter we vary in the Bayesian analysis to constrain sub-nucleonic fluctuations,
\begin{equation}
\label{eq:quarkprob}
P(\vec{r}) \propto e^{-r^2 / (2 B_G)}.
\end{equation}

For producing the bulk medium, not all binary collisions between incoming nucleons in the heavy-ion collision participate independently. We model this many-body effect with a parameter $\alpha_\mathrm{shadowing}$ that can vary between 0 and 1 to control the probability of string production in individual nucleon-nucleon collisions.

In a nucleon-nucleon collision, one or multiple strings may be formed between pairs of hotspots. These hotspots lose energy and decelerate along the longitudinal direction with a constant string tension~\cite{Shen:2017bsr}. The lost energy is distributed in the longitudinal direction and is eventually deposited into hydrodynamic fields as source terms after a formation time $\tau_\mathrm{form}$ in the collision pair rest frame. We calibrate the string formation time in the Bayesian analysis.

\begin{figure}[tb]
\includegraphics[scale=0.445]{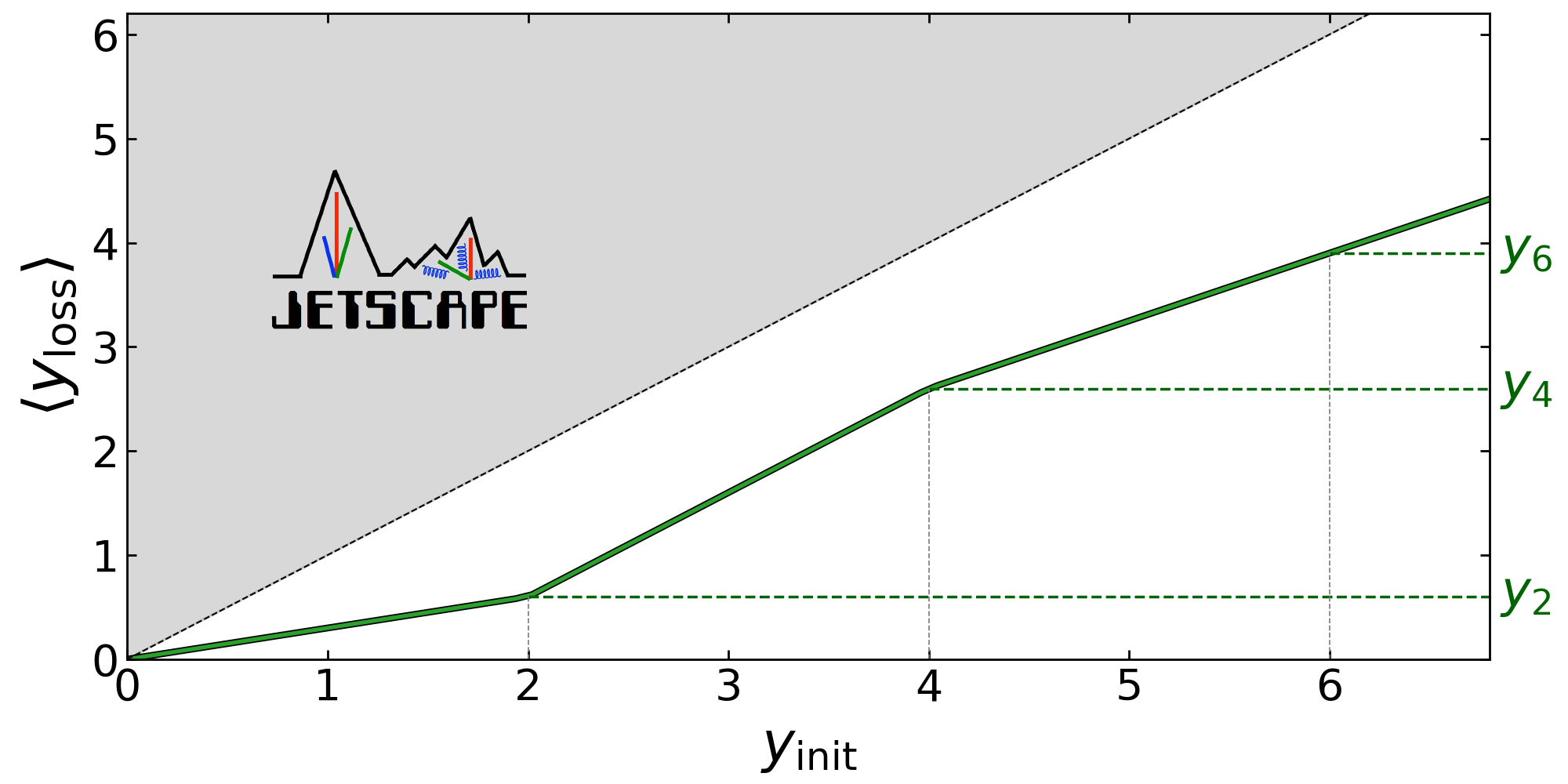}
\caption{Average rapidity loss parametrization as a function of incoming beam rapidity $y_\mathrm{init}$ given by Eq.~\eqref{eq:ylossparam}. The gray area is restricted because $y_\mathrm{loss}$ must be less than $y_\mathrm{init}$.}
\label{fig:param_yloss}
\end{figure}

The amount of energy lost by colliding nucleons is modeled according to an average rapidity loss $\langle y_\mathrm{loss} \rangle$. This is a
quantity generally modeled to describe collisions at different energies (or, equivalently, incoming beam rapidities
$y_\mathrm{init}$ = 
$y_\mathrm{beam}$ = arccosh($\sqrt{s_\mathrm{NN}}$/(2$m_N$)). In the analysis, we vary its value at three points of incoming beam rapidity, namely $y_2$ (at $y_\mathrm{init}$=2), $y_4$ (at $y_\mathrm{init}$=4), and $y_6$ (at $y_\mathrm{init}$=6), in a piece-wise parametrization that in principle allows for constraining the $\sqrt{s_\mathrm{NN}}$-dependence of the rapidity loss. Illustrated in Fig.~\ref{fig:param_yloss}, the value of the function at every point of $y_\mathrm{init}$ is thus given by:
\begin{equation}
\label{eq:ylossparam}
   \langle y_\mathrm{loss} \rangle =
   \begin{cases} 
      (y_{2})(y_\mathrm{init})/2 & 0<y_\mathrm{init}\leq2 \\
      
      y_{2} + (y_{4}-y_{2})\frac{y_\mathrm{init}-2}{2} & 2<y_\mathrm{init}\leq4 \\
      
      y_{4} + (y_{6}-y_{4})\frac{y_\mathrm{init}-4}{2} & 4< y_\mathrm{init} 
   \end{cases}
\end{equation}

Since in this study we restrict ourselves to experimental data collected at $\sqrt{s_\mathrm{NN}}$ = 200 GeV, corresponding to incoming beam rapidity $y_\mathrm{beam}$=5.36, we expect to have strong sensitivity to the latter two parameters, and limited sensitivity to $y_{2}$.\footnote{At the same time, the parametrization allows for posterior degeneracy between two parameters if the constraining region falls, as it does for our $\sqrt{s_\mathrm{NN}}$ = 200 GeV collisions, at an intermediate incoming rapidity between the two parameters. We take the step of imposing on top of this parametrization a monotonicity condition at the level of the Bayesian likelihood that forces $y_4$ $\leq$ $y_6$ (and $y_2$ $\leq$ $y_4$). This step also ensures that the $y_2$ posterior, due to the correlation with $y_4$, will not be uniform.} The event-by-event fluctuations on rapidity loss are introduced by a restricted Gaussian distribution with mean $\langle y_\mathrm{loss} \rangle$ and variance $\sigma_\mathrm{y_{loss}}$~\cite{Shen:2022oyg}.

The wounded nucleons lose part of their energy and momentum during the collision. The remainder are treated as collision remnants. 
These remnants do not have partners to form a string, yet they also deposit energy and momentum into the fluid fields according to a scaled down rapidity loss relative to that of the collision strings. The proportionality constant is given by the parameter $\alpha_\mathrm{rem}$ such that $\langle y_\mathrm{loss}^\mathrm{rem} \rangle$ = $\alpha_\mathrm{rem}$$\langle y_\mathrm{loss} \rangle$ with $\alpha_\mathrm{rem}\in [0, 1]$. If $\alpha_\mathrm{rem}$=0 the remnants deposit no energy.

The energy and momentum from the string-forming hotspots and from the wounded nucleons form a source current, $J^\nu$, whose spatial distribution is smeared according to Gaussian profiles in the transverse plane and a flux-tube-like profile in the longitudinal direction with corresponding widths in the transverse and longitudinal directions ($\sigma_x$ and $\sigma_{\eta}$, respectively). An additional parameter, $\alpha_{shift}$, modulates the longitudinal dependence of the string transverse profiles~\cite{Shen:2022oyg}.

Including energy-momentum source terms from both collision strings and remnants ensures strict energy-momentum conservation when mapping the initial state of the collision to the subsequent hydrodynamic phase, a central feature of this model~\cite{Shen:2017bsr, Shen:2020jwv, Shen:2022oyg}. All the energy and momentum present in the incoming beam are thus accounted for throughout the collision evolution.

\subsection{\label{sec:hydrodynamics}Relativistic hydrodynamics}

The contributed energy from strings and collision remnants as described above is ultimately deposited as a source term, $J^{\nu}$, to the energy-momentum tensor at a given time $\tau_{form}$.

\begin{equation}
\label{eq:tmunu}
\partial_{\mu} T^{\mu \nu} = J^\nu 
\end{equation}

\begin{figure}[tb]
\includegraphics[scale=0.45]{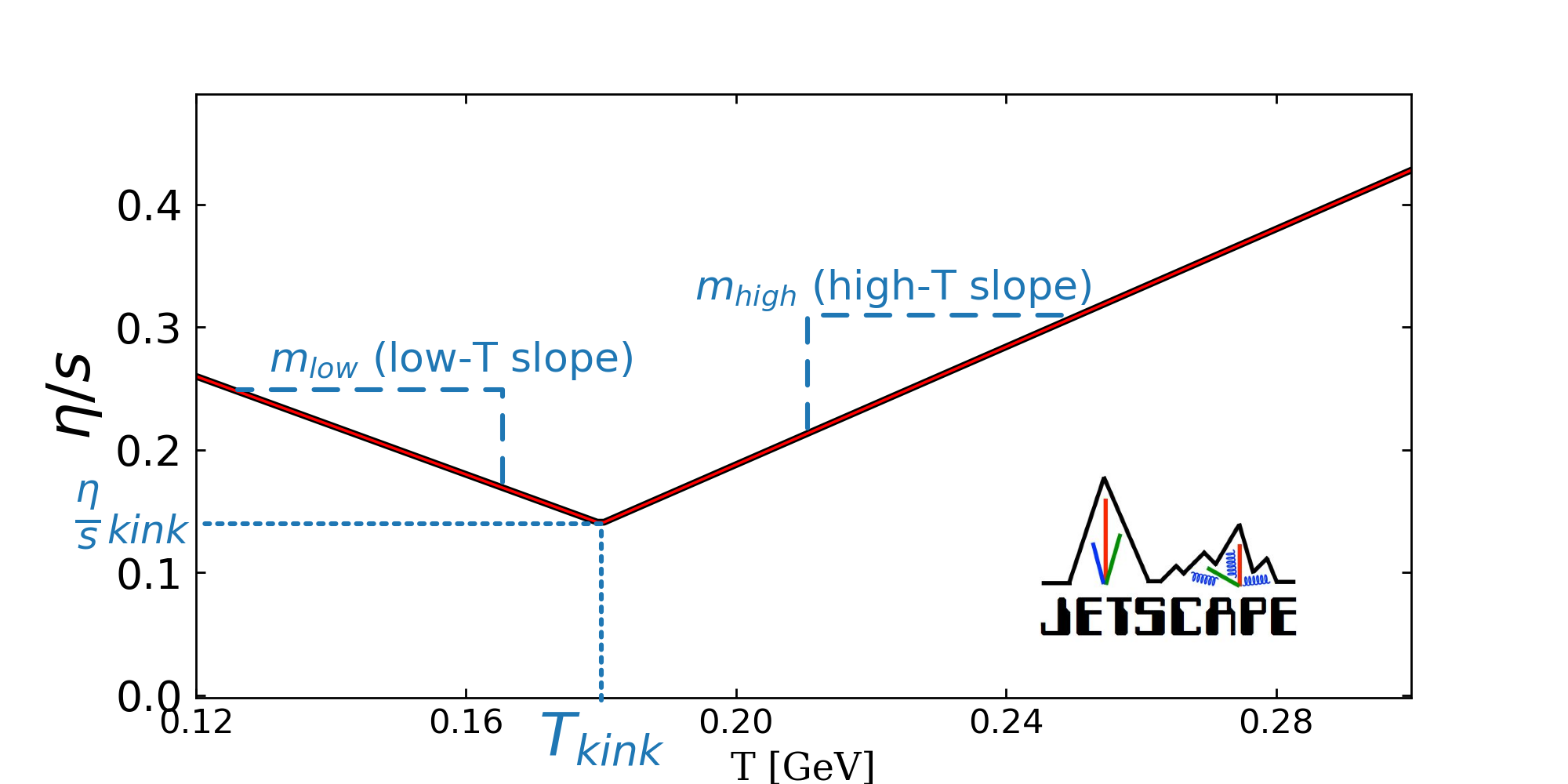}
\includegraphics[scale=0.45]{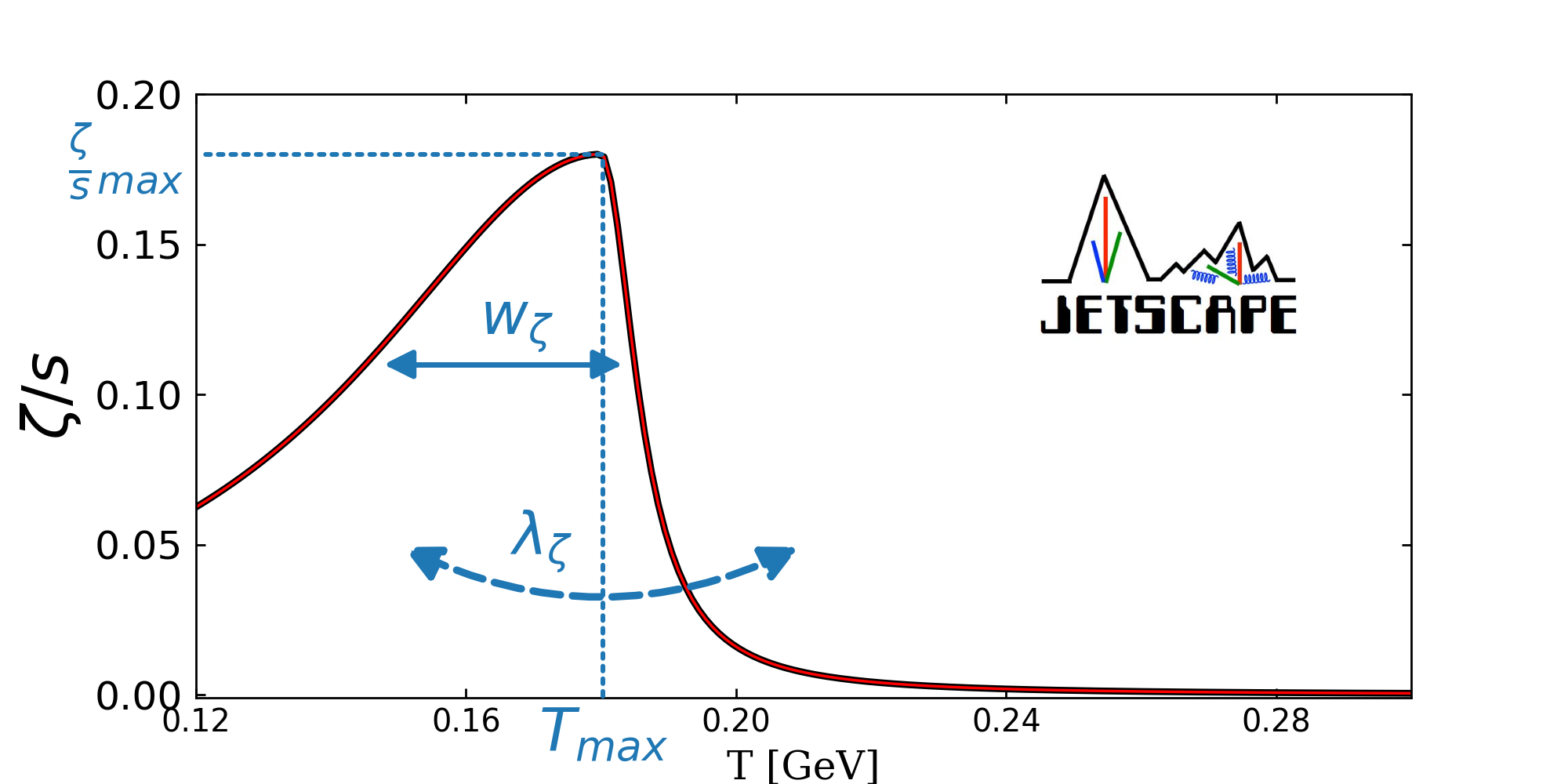}
\caption{Viscosity parametrization as a function of temperature for the specific shear viscosity given by Eq.~\eqref{eq:etaparam} (top) and specific bulk viscosity given by Eq.~\eqref{eq:zetaparam} (bottom).}
\label{fig:param_viscosity}
\end{figure}

We then evolve the system with Israel-Stewart relativistic viscous hydrodynamics in 3+1 dimensions solved using the MUSIC code~\cite{Schenke:2010nt,Schenke:2010rr,Paquet:2015lta} in the iEBE-MUSIC framework~\cite{Shen:2023awv}. The hydrodynamic properties that we aim to constrain with the data are the first-order transport coefficients $\eta/s$ and $\zeta/s$, the specific (scaled by the entropy density) shear and bulk viscosity, respectively. They are understood to have a non-trivial temperature dependence which is parametrized in this work, as was done in recent analyses~\cite{JETSCAPE:2020mzn, JETSCAPE:2020shq,Heffernan:2023utr,Heffernan:2023gye}, in a way that incorporates a few basic assumptions about the behavior near the deconfinement transition. These include the specific shear viscosity featuring a change in slope in the deconfinement region, and the specific bulk viscosity decreasing to zero at large temperature. Each transport coefficient is encoded by four parameters, illustrated in Fig.~\ref{fig:param_viscosity}. For the specific shear viscosity we have:

\begin{equation}
\label{eq:etaparam}
   \frac{\eta}{s} (T) =
   \begin{cases} 
      \frac{\eta}{s}_{kink} + (m_{low})(T_{kink}-T) & T < T_{kink} \\
      
      \frac{\eta}{s}_{kink} + (m_{high})(T-T_{kink}) & T > T_{kink} \\
    
   \end{cases}
\end{equation} where $T_{kink}$ gives the temperature of the kink and $(\eta/s)_{kink}$ its value. The parameters $m_{low}$ and $m_{high}$ give the slopes in the two temperature regions. On top of this parametrization we also enforce positivity ($\eta/s$ = max(0, $\eta/s$)) to reject unphysical viscosities. For the specific bulk viscosity we use a skewed Cauchy distribution:

\begin{equation}
\label{eq:zetaparam}
\frac{\zeta}{s} (T) = \frac{ (\zeta/s)_{max} \Lambda^2  }{\Lambda^2 + (T-T_{max})^2}
\end{equation} with \begin{equation}
\label{eq:zetaparamlambda}
\Lambda = w_{\zeta} [ 1 + \lambda_{\zeta}\mathrm{sign}(T-T_{max}) ]
\end{equation} where $T_{max}$ gives the temperature of the maximum and $(\zeta/s)_{max}$ its value. The parameter $w_{\zeta}$ encodes the width of the distribution and $\lambda_{\zeta}$ encodes the asymmetry around $T_{max}$. 

The parametrizations result in correlations between the values of the coefficients at different temperatures. These correlations are ultimately embedded in the analysis posteriors. A number of recent analyses have used different parametrizations for the viscosity coefficients, though no detailed studies on the impact of the parametrization on the posteriors have yet been done (a brief discussion of this question can be found in Ref.~\cite{Paquet:2023rfd}).

For the equation of state, we use the common prescription of lattice calculations at high temperature matched to a hadron gas calculation at low temperature~\cite{HotQCD:2014kol}. We assume $\mu_B$=0. The equation of state, particularly in the transition region, is not explicitly varied and represents one source of uncertainty not accounted for by the analysis. We choose to take the equation of state as a given and to focus on constraining the initial state and the transport coefficients of QCD (as was done in most recent analyses, including ones focusing on the finite-$\mu_B$ region).

\subsection{\label{sec:particlization}Particlization and the hadronic phase}
As the fluid cools down and crosses the deconfinement transition region, the relevant microscopic degrees of freedom become once again the confined hadrons. In order to describe hadronic interactions that take place at this stage, and to ultimately compare with experimental data, we convert the fluid to particles using the standard Cooper-Frye prescription~\cite{Cooper:1974mv} implemented in the iSS package~\cite{Shen:2014vra}. This is done on a hypersurface of constant energy density given by $e_{switch}$, the value of which is varied in the analysis. 

Furthermore, we use Grad's approximation~\cite{CPA:CPA3160020403, Israel:1976tn, Israel:1979wp, Monnai:2009ad, Denicol:2012cn} for the viscous corrections to the equilibrium momentum distribution of particles in Cooper-Frye. In the JETSCAPE 2D analysis~\cite{JETSCAPE:2020mzn}, the theoretical uncertainties related to the choice of viscous corrections were explored in detail. In this study we do not explicitly account for this uncertainty, although we use the model which the 2D analysis showed was most favored by the data. We did perform checks of the impact of particlization model on various 
observables with our calibrated model and find little sensitivity.

The hadronic phase, then, is described using UrQMD \cite{Bass:1998ca, Bleicher:1999xi}, which simulates hadronic scatterings and decays until interactions cease between particles, that is, until ``kinetic freeze-out.'' 

\section{\label{sec:bayes}Model emulation and the Bayesian comparison with data}

We work in a Bayesian statistical framework in order to perform a systematic and rigorous comparison of the model with experimental data.
Following Bayes' Theorem of conditional probability,

\begin{equation}
\label{eq:bayes}
\mathcal{P}(\theta|D) = \frac{\mathcal{P}(D |\theta) \mathcal{P}(\theta )}{\mathcal{P}(D)}
\end{equation}
where $\mathcal{P}(\theta )$ denotes the prior knowledge about the model parameters $\theta$, $\mathcal{P}(D |\theta)$ the likelihood that the data $D$ are consistent with a given set of parameters, and $\mathcal{P}(D)$ = $\int $$d\theta$ $\mathcal{P}(D |\theta) $ $\mathcal{P}(\theta ) $ the normalizing Bayes evidence. The posterior $\mathcal{P}(\theta|D)$ provides quantitative probabilistic constraints on the parameters.

The posterior is a probability distribution whose dimensionality is set by the number of parameters. It is a complicated function that must be sampled numerically. Probing accurately the posterior through sampling and marginalization is challenging, since it requires obtaining the model calculations of all calibration observables---quantities averaged over many simulated events---at a large number of parameter sets. The required number of posterior samples at different parameter sets strongly depends on the number of model parameters--about 10 million samples for 20 model parameters, in our case.

From the standpoint of direct model calculations, this task was already computationally insurmountable even for the analyses exploiting boost invariance to model the physics using (2+1)D hydrodynamic simulations. It remains very much the case for our full (3+1)D simulations. To get around this challenge, Gaussian process emulators (GPs) have been used as a fast surrogate for the model's predictions in all large-scale Bayesian analyses of heavy-ion collisions. In this analysis we use a very similar emulator to the one used in the JETSCAPE 2D Bayesian calibration. Details of the emulation are described in subsection \ref{sec:bayes:emulator} and more thoroughly laid out in \cite{JETSCAPE:2020mzn}. 

\subsection{\label{sec:bayes:priors}Priors}
We choose, as previous analyses, to use uniform distributions as priors for our parameters~\cite{JETSCAPE:2020mzn,JETSCAPE:2020shq}. This includes the parameters used to vary the transport coefficients and their temperature dependence (Eqs.~\eqref{eq:etaparam} and \eqref{eq:zetaparam}).

It is important to note that using uniform priors does not necessarily mean using weak priors, and that the prior ranges encode important prior information. For example, uniform priors of the viscosity parameters still encode a significant amount of information about their temperature dependence. Our approach in this analysis is to use wide ranges for the priors, generally consistent with the priors of previous analyses, to help understand the impact of rapidity-dependent observables. We underscore that we do not incorporate the \textit{posteriors} obtained in previous heavy-ion calibrations for common parameters in the choice of our priors due to overlapping calibration datasets and differences in modeling. The 20 model parameters and their prior ranges are listed in Table \ref{tab:table1}.

\begin{figure*}
\includegraphics[scale=0.365]{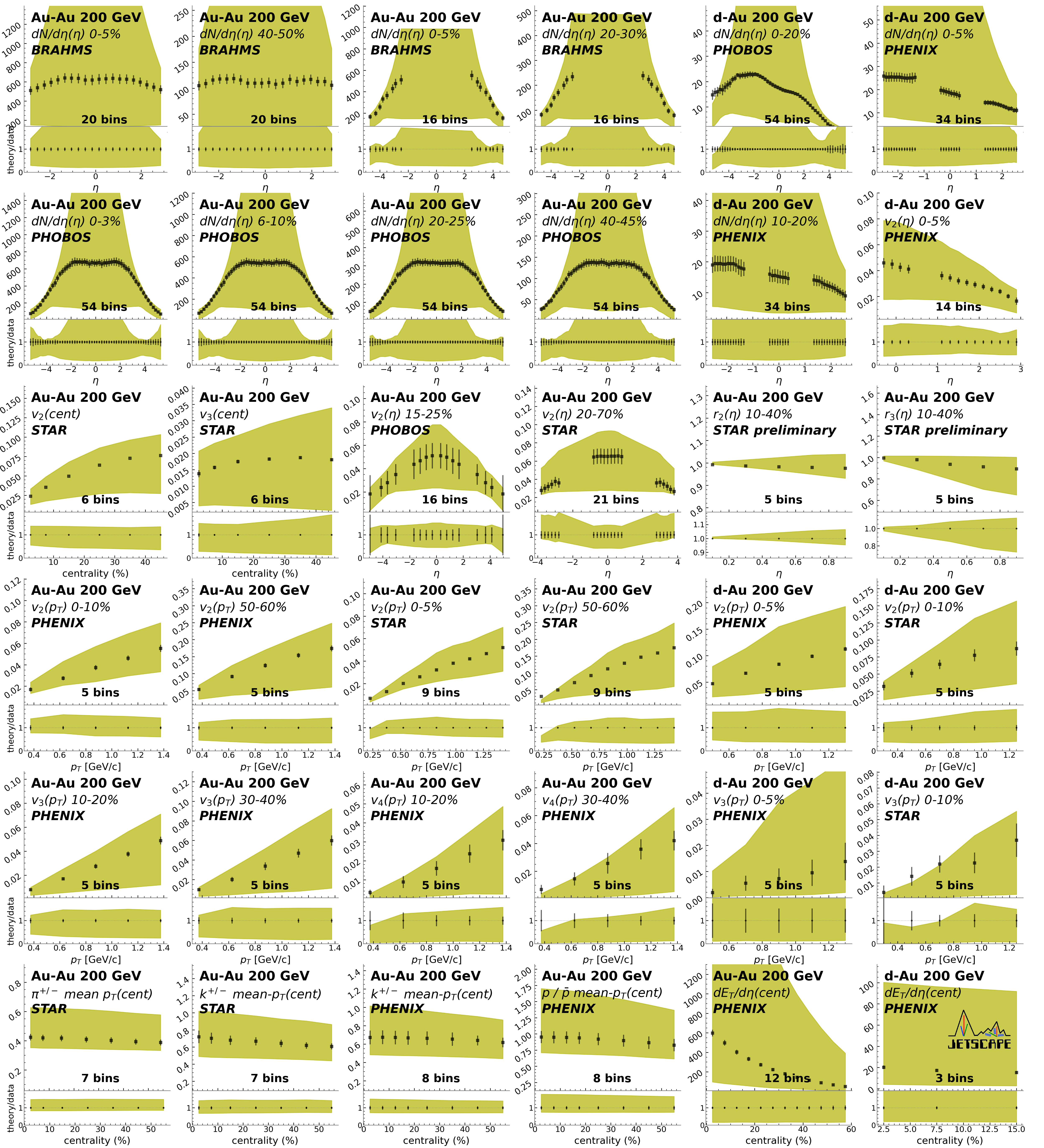}
\centering
\caption{ The observable prior for the data included in the calibration as well as some not included. The solid markers are the experimental data, while the bands show a 90\% interval of the prior calculated using the full model at the design points. Each plot is labeled with the number of data bins corresponding to each observable in a given plot.
}
\label{fig:observable_prior}
\end{figure*}

The observable prior is the representation of the assumed prior knowledge about the model parameters at the level of individual observable probability distributions. We show an approximation\footnote{Putting aside the statistical uncertainty on individual model calculations, the figure is also an approximation of the prior distribution in that it is sampled at only about 400 Latin hypercube design points.} of this in Fig.~\ref{fig:observable_prior} for a representation of all\footnote{This figure, as well as later figures in the same style showing the observable posteriors, will include a \textit{representative} subset of the full observable set considered in the analysis, omitting additional centrality classes of a number of observables. However, every observable \textit{type} is represented. For a full summary of the data points used in calibration refer to Table~\ref{tab:table2}.} the observables considered in our two calibration systems (including those observables not included in calibrations for technical reasons discussed below, but with which the model will still be compared). The plot includes a 90\% credible interval for each observable bin predicted in the Latin hypercube sample (described below) of model simulations. We see a generally wide observable prior that covers essentially all the data within experimental uncertainties. 

\begin{table}[t]
\caption{
All 20 model parameters described in Section~\ref{sec:model} and varied in the analysis along with the corresponding stage of the collision and the prior range.}
\label{tab:table1}
\begin{ruledtabular}
\begin{tabular}{ccccc}

Parameter & Collision Stage& Prior Range \\
\hline

$y_{2}$& Initial State & [0,2]
\\
$y_{4}$& Initial State  & [$y_{2}$,4]
\\
$y_{6}$& Initial State & [$y_{4}$,6]
\\
$\sigma_{y_{loss}}$& Initial State & [0,1]
\\
$\alpha_{rem}$& Initial State & [0,1]
\\
$\alpha_{shadowing}$& Initial State  & [0,1]
\\
$B_G$ [GeV$^{-2}$]& Initial State & [2,25]
\\
$\sigma_x$ [fm]& Initial State & [0.1,0.5]
\\
$\sigma_\eta$& Initial State & [0.1,0.8]
\\
$\alpha_{shift}$& Initial State  & [0,1]
\\
$\tau_{form}$ [fm]& Hydrodynamization  & [0.2,1]
\\
$(\eta/s)$ $T_{kink}$ [GeV]& Hydro & [0.13,0.3]
\\
$m_{low}$ [GeV$^{-1}$]& Hydro & [-2,1]
\\
$m_{high}$ [GeV$^{-1}$]& Hydro & [-1,2]
\\
$(\eta/s)_{kink}$& Hydro & [0.01,0.2]
\\
$(\zeta/s)_{max}$& Hydro  & [0.01,0.2]
\\
$(\zeta/s)$ $T_{max}$ [GeV]& Hydro & [0.12,0.3]
\\
$w_{\zeta}$ [GeV]& Hydro & [0.025,0.15]
\\
$\lambda_{\zeta}$& Hydro & [-0.8,0.6]
\\
$e_{switch}$ [GeV/fm$^3$]& Particlization & [0.1,0.6]
\end{tabular}
\end{ruledtabular}
\end{table}

\subsection{\label{sec:bayes:emulator}The model emulator}
The emulator acts as a fast surrogate for the model, and estimates the calculations' statistical and interpolation uncertainties (including correlated uncertainties). To ensure the most efficient emulator training, the chosen prior ranges for the parameters coincide with the bounds of the parameter space sampled to train the emulator. The training space is sampled using a space-filling, Latin hypercube algorithm, as was done in previous analyses~\cite{Heffernan:2023utr, Heffernan:2023gye}. Exceptionally, for two parameters, $y_{4}$ and $y_{6}$, the emulator training parameter ranges were  wider than the Bayesian prior ranges ([0,4] and [0,6] instead of [$y_2$,4] and [$y_4$,6], respectively). This was due to the need to enforce a physical, monotonic behavior to the rapidity loss as a function of incoming rapidity and the particular choice of our parametrization, while avoiding these correlations in the Latin hypercube sampling of the training space. 

The model is used to simulate approximately 400 parameter sets, or design points, spanning the training space for each of the two collision systems (d-Au and Au-Au). We train Gaussian Processes for each of 10 principal components (PCs) per collision system, into which the normalized observable data have been decomposed. The principal component analysis (PCA) rewrites the observables as a linear combination of ``projections'' along the new principal components. A significantly smaller number of these are used in the analysis compared to the number of original observable bins (roughly a reduction by two orders of magnitude), after truncating the linear combination past a reasonable number of terms. The reduction allows for significantly more efficient emulation and, in all cases in this analysis, the principal components that are used collectively account for well over 99\% of the variance in the observables across the prior parameter space. This is possible due to the high degree of redundancy in the information content of observables. 

The principal component analysis transformation is performed once at training time, such that the emulators are trained on the principal components. The inverse transformation is performed each time the emulators are prompted to predict the model calculations at a given parameter point, meaning that the emulator output is rotated back to the original observable space before comparing with data. The analysis choices related to PCA are described in additional detail in Appendix~\ref{PCA}.

In addition, we perform two other transformations in observable and parameter space to improve the performance of the emulator. The emulator is trained using transformed parameters of the viscosity parametrization. Instead of training on the four shear and four bulk parameters, we train on 10 values of the viscosity as a function of temperature for each respectively.\footnote{The sampling of the two viscosity coefficients is still done according to the distributions of the eight (four for each) original parameters in Eq.~\eqref{eq:etaparam} and Eq.~\eqref{eq:zetaparam}. For any given sample, before the emulator is called, the values of those parameters are converted to 20 (10 for each) values of the viscosity, spanning uniformly the temperature range according to the corresponding parametrization.} We also transform (before principal component analysis) observables which feature a relatively large dynamical range, namely the charged hadron multiplicity and the transverse energy, training on their logarithm. These procedures are further discussed in Appendix~\ref{emulator_validation}, where we also show emulator validation done with a subset of the simulated design points.

\subsection{\label{sec:bayes:mcmc}Comparison with data and parameter estimation}

In order to efficiently sample the parameter space, we use a Markov Chain Monte Carlo (MCMC) algorithm implemented in the emcee package~\cite{Foreman-Mackey:2012any}. A given number of MCMC ``walkers'' (in our analysis, 200 of them as default) are initialized by uniformly sampling the 20 model parameters. The likelihood function, shown below, and the Bayesian posterior probability are computed at each step for each walker to determine which walkers in the ensemble are updated to the proposed position and which remain unchanged. Through many iterations, the MCMC chain eventually converges to a representative sample of the posterior distribution. In principle, the representative sample approaches and becomes the exact probability distribution in the limit of infinite steps; however we arbitrarily terminate the algorithm at a practical stopping point. The MCMC phase up to convergence is called the ``burn-in.''

The optimal length of MCMC chains may be determined by directly checking for convergence in the probability distributions of parameters. In Appendix \ref{mcmc_convergence} we show the convergence checks performed for the default calibration of this analysis, for which we ultimately use 20,000 burn-in steps. We see that after roughly that many steps, the distributions for all parameters converge.

To quantify the agreement with data of any given observable prediction of the model (or, in our case, its surrogate), we define the likelihood function as the following multivariate normal distribution:

\small
\begin{equation}
\label{ml:likeli}
\mathcal{P}(D |\theta) =
\frac{1}{\sqrt{(2\pi)^n \det \Sigma}} \\
\exp\left\{-\frac{1}{2}\Delta Y^T \Sigma^{-1} \Delta Y\right\}
\end{equation}
\normalsize

where $n$ is the number of data points, $\Delta Y$ = $Y_{exp}$ - $Y_{th}$($\theta$), the difference between the experimental observables and the model prediction at a parameter point $\theta$, and $\Sigma$, the covariance matrix. It is a standard likelihood choice used in previous Bayesian analyses. It incorporates the uncertainties on the model and experiment in the form of covariance matrix $\Sigma$ = $\Sigma_{exp}$ + $\Sigma_{th}$. The covariance matrix includes along its diagonal the uncertainties on any given observable bin. It also encodes correlations in the uncertainties between different observable bins in its off-diagonal elements.

The model portion of the uncertainty in our analysis is comprised of systematic and statistical uncertainties in the model predictions, as well as emulation uncertainty. Since the emulator is trained on model calculations limited by finite statistical uncertainty, the emulator's estimate of its uncertainty includes the statistical model uncertainty. We do not separately estimate the theoretical systematic uncertainties that remain uncaptured by our calibrated model. As a consequence, the posteriors in this analysis should be interpreted as indicating the theoretical systematic uncertainty only to the extent that it is probed by the varied model parameters.

To construct the experimental covariance matrix, which is correspondingly the sum of systematic and statistical uncertainties, we have very limited input from measurements on any correlations in the uncertainty across observable bins. Therefore, we make the assumption of uncorrelated experimental uncertainties--that is, a diagonal experimental covariance matrix. Following previous analyses, we rely on the emulator to estimate the model correlations comprising the theoretical contribution to the covariance. We show an example of these emulator-estimated correlations in Appendix \ref{covariance_matrix}.

As mentioned in the previous section, before emulator training, we perform a transformation of certain observables--namely the multiplicity and the transverse energy--by taking the logarithm in order to improve the emulator performance. This yields emulator predictions in the transformed "log space," with a corresponding covariance matrix also in such space. Given the options of transforming these predictions and covariance matrices to the original observable space, or transforming the experimental data and its diagonal covariance matrix to log space, we choose the latter and ultimately compute the likelihood in log space. 

\subsection{\label{sec:bayes:predictions}Sampling the posterior and making predictions}

Having obtained the parameter posterior distribution we can sample it and construct the observable posterior, the representation of our knowledge at the level of observables after comparing with data. While we can make predictions using any one parameter sample from the posterior---indeed, for example, using the Maximum A Posteriori (MAP), i.e. most likely parameter set---the width of the observable posterior is a meaningful measure of the model uncertainty. This variation within the posterior in many cases exceeds the statistical uncertainty that can be reasonably achieved on the predictions of any given observable with the emulator for a particular parameter set. Therefore, in our predictions with the calibrated model we emphasize using many samples of the posterior to account for this uncertainty. 

We use a large number of samples of the posterior---around 1000---evaluated by the emulator, to construct a 90\% credible interval for all of the observables in the calibration systems. This gives an efficient estimate that can be readily used for any of a number of posteriors from calibrations using different datasets, for example. 

In addition, in order to obtain more precise calculations independent of the emulator uncertainty, we simulate 10,000 collisions with the full model for each of ten posterior samples from our default calibration. We do this for each of our two calibration systems, Au-Au and d-Au, as well as for p-Au and $^3$He-Au. The results in Section~\ref{sec:predictions} are all obtained using these ten samples, five of which we plot individually in the figures, as described in that Section. The values of the parameters for each of the five plotted samples are shown in Appendix~\ref{posterior_samples}.

\subsection{\label{sec:bayes:closure}Closure tests}

An essential check of any Bayesian comparison with data is to use, in place of the data, the model predictions for a given parameter set not used in training. The so-called "pseudo-data" may be assigned arbitrary "experimental" uncertainty, and varying the magnitude of this uncertainty can help gauge the sensitivity of constraints. If the uncertainty is set to be relatively small, the closure is limited by the emulation uncertainty. For the closure tests in this analysis we assign the true experimental uncertainty for the corresponding observables in order to make the most direct analogy to the real calibration. We performed closure tests, shown in Appendix \ref{closure_tests}, for a number of parameter sets from the simulated design points that were not used to train the emulator. The posterior distributions for the parameters are obtained with an identical procedure to that used for the default calibration against real data, but since the truth parameter values are a priori known, they may be compared with the posterior constraints to gauge the performance of the analysis. In Appendix \ref{closure_tests} we show a selection of marginalized posteriors for three closure points. We see generally consistent posterior credible intervals with the truth values of parameters across different closure points, offering confidence in the robustness of the analysis. We additionally see the sensitivity of particular parameters to the observables reflected in the closure test posteriors, with the rapidity loss parameters in particular yielding strongly peaked distributions.

\section{\label{sec:exp_data}The experimental data  }

\begin{table}[t]
\caption{
Listing of all the experimental data used in this work for the default calibration. For each observable we indicate the collision system, the experimental collaboration, the number of centrality bins used, and the total number of data points used to calculate the likelihood.}
\label{tab:table2}
\begin{ruledtabular}

\begin{tabular}{lrrcccc}

 observable &system & Experiment & cent. bins && data points \\ 
\hline

$dN_{ch}$/d$\eta$($\eta$)& Au-Au & PHOBOS~\cite{PHOBOS:2010eyu} & 11 && 594
\\
$v_2$($\eta$)& Au-Au & PHOBOS~\cite{PHOBOS:2004vcu} & 3 && 48
\\
$v_2$($\eta$)& Au-Au & STAR~\cite{STAR:2004jwm} & 1 && 21
\\
$v_2$($p_T$)& Au-Au & PHENIX~\cite{PHENIX:2011yyh} & 6 && 30
\\
$v_2$($p_T$)& Au-Au & STAR~\cite{STAR:2004jwm} & 7 && 56
\\
$v_2$(cent.)& Au-Au & STAR~\cite{STAR:2004jwm} & 6 && 6
\\
$v_3$(cent.)& Au-Au & STAR~\cite{STAR:2013qio} & 6 && 6
\\
$\pi^{+/-}$ $\langle p_T \rangle$(cent.) & Au-Au & STAR~\cite{STAR:2008med} & 7 && 7
\\
$K^{+/-}$ $\langle p_T \rangle$(cent.) & Au-Au & STAR~\cite{STAR:2008med} & 7 && 7
\\
$K^{+/-}$ $\langle p_T \rangle$(cent.)& Au-Au & PHENIX~\cite{PHENIX:2003iij} & 8 && 8
\\
$p/\bar{p}$ $\langle p_T \rangle$(cent.)& Au-Au & PHENIX~\cite{PHENIX:2003iij} & 8 && 8
\\
$dN_{ch}$/d$\eta$($\eta$)& d-Au & PHENIX~\cite{PHENIX:2018hho} & 3 && 102
\\
$v_2$($\eta$)& d-Au & PHENIX~\cite{PHENIX:2018hho} & 1 && 14
\\
$v_2$($p_T$)& d-Au & PHENIX~\cite{PHENIX:2018lia} & 1 && 5
\\
$v_2$($p_T$)& d-Au & STAR~\cite{STAR:2022pfn} & 1 && 5
\cr
\\
Total&  &  &  && 917 &
\end{tabular}
\end{ruledtabular}
\end{table}

\begin{table}[t]
\caption{
Listing of the experimental data \textit{not} used in calibration but nevertheless computed and compared with in this analysis. For each observable we indicate the collision system, the experimental Collaboration, the number of centrality bins used, and the total number of data points considered.}
\label{tab:expdata_not_calib}
\begin{ruledtabular}
\begin{tabular}{lrrcccc}
 observable &system & Experiment & cent. bins && data points \\
\hline

$dN_{ch}$/d$\eta$($\eta$)& Au-Au & BRAHMS~\cite{BRAHMS:2001llo} & 6 && 216
\\
$v_3$($p_T$)& Au-Au & PHENIX~\cite{PHENIX:2011yyh} & 6 && 30
\\
$v_4$($p_T$)& Au-Au & PHENIX~\cite{PHENIX:2011yyh} & 5 && 25
\\
$p/\bar{p}$ $\langle p_T \rangle$(cent.) & Au-Au & STAR~\cite{STAR:2008med} & 7 && 7
\\
$\pi^{+/-}$ $\langle p_T \rangle$(cent.)& Au-Au & PHENIX~\cite{PHENIX:2003iij} & 8 && 8
\\
d$E_T$/d$\eta$(cent.)& Au-Au & PHENIX~\cite{PHENIX:2013ehw} & 12 && 12
\\
$dN_{ch}$/d$\eta$($\eta$)& $^3$He-Au & PHENIX~\cite{PHENIX:2018hho} & 1 && 26
\\
$v_2$($\eta$)& $^3$He-Au & PHENIX~\cite{PHENIX:2018hho} & 1 && 24
\\
$v_2$($p_T$)& $^3$He-Au & STAR~\cite{STAR:2022pfn} & 1 && 7
\\
$v_2$($p_T$)& $^3$He-Au & PHENIX~\cite{PHENIX:2018lia} & 1 && 8
\\
$v_3$($p_T$)& $^3$He-Au & STAR~\cite{STAR:2022pfn} & 1 && 7
\\
$v_3$($p_T$)& $^3$He-Au & PHENIX~\cite{PHENIX:2018lia} & 1 && 8
\\
$dN_{ch}$/d$\eta$($\eta$)& d-Au & PHOBOS~\cite{PHOBOS:2004fzb} & 1 && 54
\\
d$E_T$/d$\eta$(cent.)& d-Au & PHENIX~\cite{PHENIX:2013ehw} & 3 && 3
\\
$v_3$($p_T$)& d-Au & PHENIX~\cite{PHENIX:2018lia} & 1 && 5
\\
$v_3$($p_T$)& d-Au & STAR~\cite{STAR:2022pfn} & 1 && 5
\\
$r_2$($p_T$)& d-Au & STAR~\cite{Nie:2020trj} & 1 && 5
\\
$r_3$($p_T$)& d-Au & STAR~\cite{Nie:2020trj} & 1 && 5
\\
$dN_{ch}$/d$\eta$($\eta$)& p-Au & PHENIX~\cite{PHENIX:2018hho} & 1 && 26
\\
$v_2$($\eta$)& p-Au & PHENIX~\cite{PHENIX:2018hho} & 1 && 24
\\
$v_2$($p_T$)& p-Au & STAR~\cite{STAR:2022pfn} & 1 && 7
\\
$v_2$($p_T$)& p-Au & PHENIX~\cite{PHENIX:2018lia} & 1 && 8
\\
$v_3$($p_T$)& p-Au & STAR~\cite{STAR:2022pfn} & 1 && 7
\\
$v_3$($p_T$)& p-Au & PHENIX~\cite{PHENIX:2018lia} & 1 && 8
\cr
\\
Total&  &  &  && 535 &
\end{tabular}
\end{ruledtabular}
\end{table}

A critical component of the Bayesian analysis is curation of the experimental data. Our analysis used a wide range of measurements from RHIC's BRAHMS, STAR, PHENIX and PHOBOS Collaborations. For pragmatic reasons, such as limitations of our model or of our emulators, or due to limitations in experimental data reporting (see Appendix~\ref{experimental_methods}), we did not include all curated data in our Bayesian analysis. We divided the measurements into a calibration set, listed in Table~\ref{tab:table2}, and a separate set used for additional comparisons, listed in Table~\ref{tab:expdata_not_calib}. We explain below how the classification into the two categories was made.

Another key component of the analysis is writing and validating analysis scripts to calculate each observable, matching the experimental definition. This includes reproducing as closely as possible each measurement's centrality determination, kinematic cuts, and (in the case of particle correlation observables) the reference regions; we discuss below some challenges we faced in doing so. In general, (3+1)D simulations, as compared to boost-invariant (2+1)D simulations, allow us to match experimental definitions more precisely for nearly all measurements.

In a final subsection, we compare the magnitude of the emulator uncertainty to the experimental uncertainty, using a type of visualization that we believe would have great benefits to be adopted in the community.

\subsection{Data selection}

In selecting the data, we make conservative choices with regard to the ability of our model to describe the physics. That is, we limit the measurements up to 60\% centrality in Au-Au (with the exception of the 20-70\% centrality STAR $v_2$ measurement) and up to 20\% in d-Au. We also set an upper bound of 1.5 GeV for the $p_T$-dependent observables. 

Given our assumption of vanishing net baryon density, we choose to focus on charged hadrons and do not in general include measurements of identified particle spectra or anisotropies. We make an exception to this rule to include the mean transverse momentum of identified particles from STAR and PHENIX~\cite{STAR:2008med, PHENIX:2003iij}, due to the lack of a corresponding charged hadron measurement and the known sensitivity of the transport coefficients to these observables. Specifically, we use the STAR $\pi^{+/-}$ and $K^{+/-}$ mean $p_T$, and the PHENIX $K^{+/-}$ and $p/\bar{p}$ mean $p_T$.\footnote{The STAR $p/\bar{p}$ and the PHENIX $\pi^{+/-}$ measurements did not remove the weak decay contribution, while in our training simulations we do not include this contribution. Though the effect is estimated to be within a few percent, we err on the side of accuracy by not including these two measurements in calibration.} 

For these mean $p_T$ measurements, we averaged the data values and the uncertainties between particles and anti-particles. Specifically, since the systematic uncertainty dominates in these measurements from both STAR and PHENIX, we assume the errors to be completely correlated when combining them.

On the measurements broadly considered within the model's scope, we made a further selection about whether to use them for calibration or to simply compare them with the posterior. The factors going into the selection were specific to each measurement, though we did make a general assessment of emulator accuracy across all observables.

Through careful emulator validation, we determined that our emulator's accuracy varied significantly across observables. Using a set of validation points, we performed emulator validation by comparing the emulator's prediction of its uncertainty to its actual performance in describing different observables. We used the following guiding principle: when the emulator significantly underestimated the true error in predicting an observable, the emulator was deemed insufficiently accurate to describe that observable for the purposes of parameter estimation, and the observable was excluded from the calibration. For more details on this procedure see Appendix \ref{emulator_validation}.

As a result of this emulator validation, we had to exclude many higher-order harmonic measurements, including differential $v_3$ measurements in both Au-Au and d-Au, as well as differential $v_4$ (and potentially higher orders) in Au-Au. We also did not include in calibration the transverse energy, d$E_T$/d$\eta$(cent.), in neither Au-Au nor d-Au. This was purely due to a technical reason: our calculation--configured at run-time across large-scale simulations--included only the transverse energy of charged hadrons, while the PHENIX~\cite{PHENIX:2013ehw} definition included the total hadronic transverse energy. The difference between the two definitions is expected to be around $3/2$ given that pions tend to dominate.
We choose not to include the observable in the calibration, but we do scale our calculations by the $3/2$ correction factor to compare our calibrated analysis with the PHENIX measurements.

Another observable that we compared with, but did not calibrate on, is STAR's decorrelation coefficients $r_n$~\cite{Nie:2020trj}, since the data are preliminary.

The exclusion of $^3$He-Au and p-Au measurements from calibration was due uniquely to a limited computational budget (and, thus, ultimately, emulation accuracy). Since each calibration system's design points had to be simulated separately, we opted to maximize the number of design points for a single large and a single small system, given finite computational resources. For this reason, all model predictions for $^3$He-Au and p-Au observables in this paper are done using full model simulations, and do not rely explicitly on model emulation.

Despite observables being excluded from calibration, we do compare with them on multiple occasions, sometimes using emulation\footnote{Even observables for which we did not deem the emulator precise enough for calibration are still, in most cases, sufficiently well-emulated to offer some qualitative insight in comparisons of the posterior with data.}, often using direct calculations from the model.

\subsection{Matching calculations to data}

For all the measurements included in this analysis, the centrality determination, kinematic cuts, and other specifications of the experimental measurements were matched when computed. We give additional detail below regarding the centrality selection and the calculation of momentum anisotropies, and explain the approach we used when insufficient information was available from the experimental publications.

\paragraph{Centrality determination}
As pointed out throughout the paper, the precise determination of centrality in accordance with each measurement's procedure is one of the key capabilities of a 3D analysis. For every measurement we compare with, the rapidity window matched that used in experiment. Additionally, the relevant quantity used for sorting the events measured in that region--whether the multiplicity or the transverse energy--correspondingly matched experiment as well. 

The transverse momentum (or energy) cut for centrality determination is a detail not always stated in the papers we considered. When mentioned, it is often the case that no cut was implemented. Consequently, as a generic choice across all observables we calculate, we do not include any momentum or energy cut. We verified (not shown) that the effect from reasonable cuts on transverse momentum or energy on the multiplicity- or $E_T$-based sorting of events, and ultimately on the observable, is generally small ($\sim 0.1$\%) across observables.

\paragraph{$v_n$ observables}
For all calculations of $v_n$ in this analysis, we use the scalar product method.\footnote{See Appendix \ref{same_rap} for a definition of the scalar product method used in the analysis, and for a discussion of potential prediction discrepancies with slightly different definitions.} The rapidity cuts and resulting gaps follow the experimental choices regardless of method. For all measurements we considered except one, the scalar product method is expected to be an accurate match to the measurement.
The one exception is the Au-Au PHOBOS $v_2$~\cite{PHOBOS:2004vcu} measured with the event-plane method.
The discrepancy cannot necessarily be evaluated precisely, though, informed by previous estimates~\cite{Luzum:2012da}, we attempt to account for it by increasing in the analysis the experimental uncertainty for the PHOBOS Au-Au $v_2(\eta)$ by 10\% of the mean value of the observable. For small systems with low multiplicity, the effect is expected to be small~\cite{Luzum:2012da} and we do not increase the uncertainty.

We encountered other challenges with the STAR $v_n$ measurements from Ref.~\cite{STAR:2004jwm}, from which we used multiple datasets. The analysis presented a large number of related azimuthal anisotropy measurements, cross-checks, and systematic studies performed using a variety of methods and kinematic cuts, with in-text measurement details not always unambiguously identified to a particular plot. For the $v_2(\eta)$ measurement, one source of confusion concerned the regions of interest and reference, as it was not entirely clear whether the points at mid-rapidity and those at forward/backward rapidity, which were measured using different detectors, shared a reference region. 
For our calculation we assumed that they did share a reference. 
The largest potential source of uncertainty, however, concerned the averaging over the large centrality bin of 20-70\%. Indeed, the extension to peripheral collisions up to 70\% centrality was a singular exception in our analysis. The ambiguity was in large part related to weights, or lack thereof, applied to smaller centrality bins in constructing the observable in the large bin. We were unable to obtain a conclusive prescription. Moreover, when comparing the mid-rapidity bins of this measurement to the integrated mid-rapidity $v_2$ measurement presented in that same paper~\cite{STAR:2004jwm}, we find an apparent inconsistency between them, regardless of the weighting scheme used to average smaller centrality bins. As a result, we use this dataset cautiously, and we increase in the analysis the experimental uncertainties by 12\% of the mean observable value, just enough to allow for apparent consistency with the integrated dataset. (Scaling the mid-rapidity values of the rapidity-dependent measurement by a factor of 1.12 matches them to the centrality average of the integrated, mid-rapidity measurement.) We decide in favor of its inclusion in the calibration largely on the basis of general scarcity of $v_2(\eta)$ RHIC data. We discuss the inter-relationship between these datasets, and the rest of the measurements in the analysis, in greater detail in Subsection~\ref{subsec:tension}.

The measurement of $v_2$($p_T$) in Au-Au from STAR~\cite{STAR:2004jwm} presented a unique challenge for the analysis: the lowest $p_T$ bin (0.15 to 0.3 GeV/c) was significantly outside of our observable prior. All combinations of model parameters that we investigated (our prior) predicted a $v_2$ smaller than the STAR measurement, especially for peripheral events, where the prior is multiple $\sigma$ away from the measurement. 
Given that this is the lowest bin of the measurement, we believe the tension may arise from missing physics in our model, or some issue with the measurement.
 As a result of these considerations, this bin was ultimately not included in calibration. It is still shown in all the observable posterior comparison plots in the paper and, as expected, rather poorly described by the constrained model.

\subsection{Emulator and experimental uncertainties}

\begin{figure}[tb]
\includegraphics[scale=0.6]{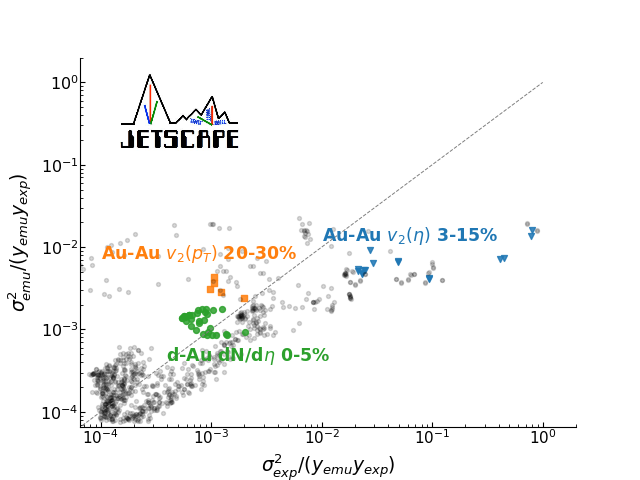}
\centering
\caption{Comparison of experimental and emulation uncertainties for all observables in the calibration. Each point represents one observable bin, with coordinates given by the scaled experimental variance (horizontal axis) and scaled emulation variance (vertical axis). The variances (diagonal elements of the covariance matrices) are scaled by the product of the mean experimental measurement $y_{exp}$ and the mean emulator prediction $y_{emu}$. The emulation uncertainties combine interpolation and statistical uncertainties, evaluated at the maximum a posteriori (MAP) parameter point. Points near the diagonal indicate balanced contributions from experimental and emulation uncertainties, while points above the diagonal are dominated by a combination of interpolation and statistical uncertainty from the emulator. The three observables shown in Figs.~\ref{covariance_ratio} and \ref{covariance_emu} are indicated in color and labeled.}
\label{covariance_all_points}
\end{figure}

Whenever available and reported separately by experiments, the statistical and systematic uncertainties were added in quadrature. Whenever this breakdown was not reported in experimental papers, on the HEPData repository, or on individual experiment websites, we made reasonable use of what was available. 

In Fig. \ref{covariance_all_points} we show a 2D comparison plot of the emulation and experimental variance, respectively, scaled by the mean prediction and measurement, for all observable bins ultimately included in the analysis (Table~\ref{tab:table2}). The multiplicities, in particular in Au-Au, and their corresponding uncertainties are in log-space and tend to populate the bottom left corner as a result. We mark in color and label three example observables that feature in Figs. \ref{covariance_ratio} and \ref{covariance_emu} in the Appendix \ref{covariance_matrix} discussion on the covariance matrix and correlations between observables. 

We see that the uncertainties on most points are rather balanced between the emulator uncertainty and the experimental uncertainty.
The location of points on the 2D plane relative to the dashed diagonal line indicates the extent to which the information content in experimental data is exploited in the analysis: data sets far above the diagonal have much larger emulation uncertainty than experimental uncertainty, and thus the analysis is not using efficiently the information they contain. Since the emulator uncertainty contains the statistical uncertainty, one cannot differentiate in Fig.~\ref{covariance_all_points} if the uncertainty is large because of statistical uncertainty (larger number of events required to reduce uncertainty), or if the emulator is struggling to interpolate the observable in the parameter space.
We point out that this situation is not unique to the present Bayesian analysis, but it has generally not been illustrated like this in the past.

\section{\label{sec:posteriors} Bayesian Posteriors  }

\subsection{\label{sec:default-calibration}The ``default'' multi-system calibration}

\begin{figure*}
\includegraphics[scale=0.365]{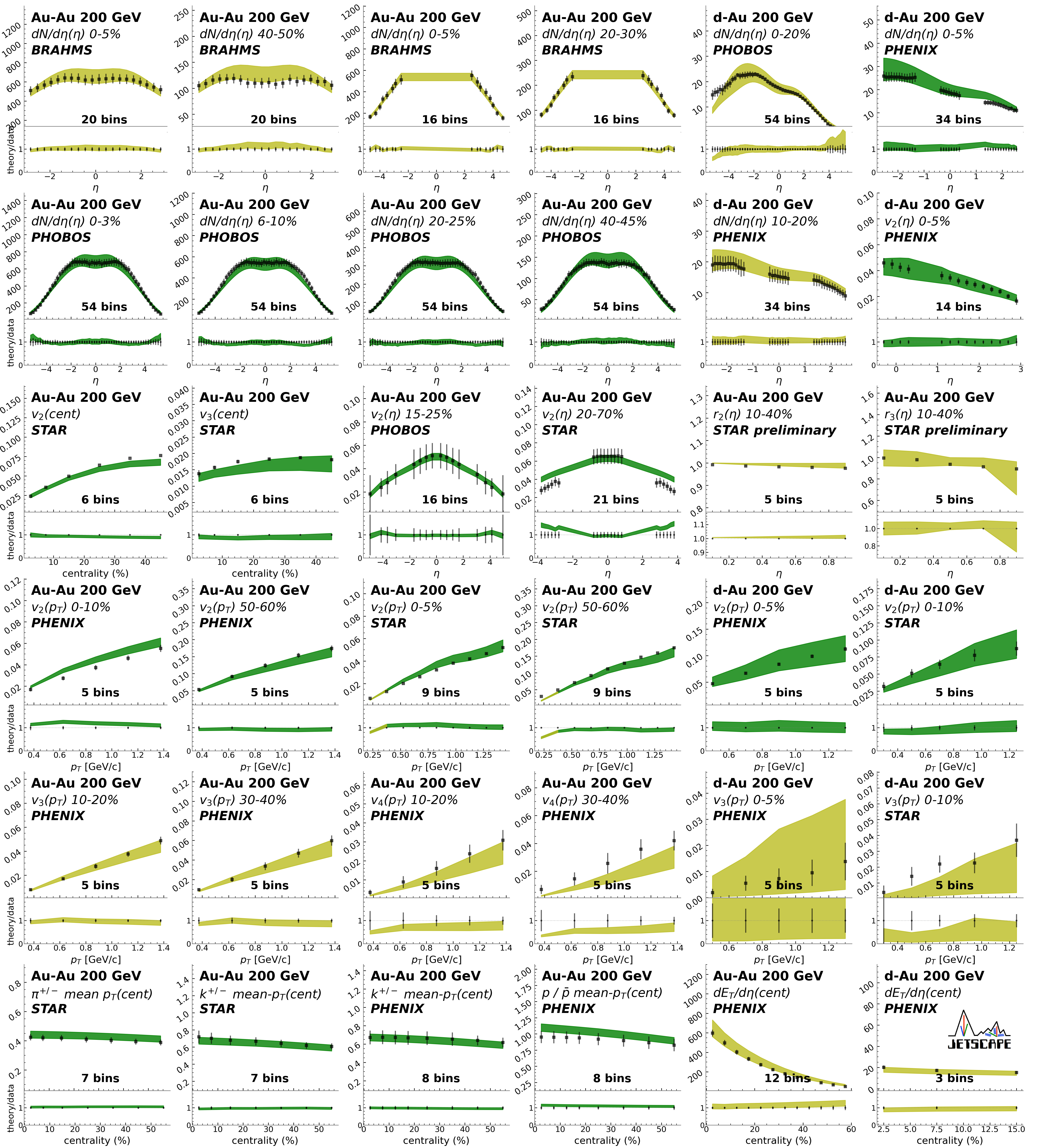}
\centering
\caption{The observable posterior for the data included in the calibration, plotted in green, as well as predictions for some not included, plotted in olive. The solid markers are the experimental data, while the bands show a 90\% interval of the posterior calculated using the emulator. Each plot is labeled with the number of data bins corresponding to each observable in a given plot.
}
\label{fig-observable-posterior}
\end{figure*}

The default calibration includes all the selected experimental data within the physics scope of the analysis and which the model is expected to reasonably describe. We include d-Au data in this calibration under the hydrodynamic assumption for small systems.\footnote{If the present calibration is used for future studies of heavy-ion physics, the Au-Au-only tune, also analyzed here, may be preferred depending on the scope and assumptions of the studies.} A detailed overview of the data and the selection criteria is provided in the previous section (Section~\ref{sec:exp_data}). We show in Fig.~\ref{fig-observable-posterior} the observable posterior, indicating with solid colored bands the 90\% credible interval computed with the emulator using one hundred thousand samples of the posterior. The bands are plotted in green for data included in calibration; for data not included in the calibration, olive bands (lighter color) are used. The experimental data and associated total uncertainties are plotted as black markers. In the same figure, in the lower section of each panel, we show the theory-to-data ratios (that is, posterior-band to experimental points), which more readily allow a visual comparison of the fit across observables. The uncertainties on the ratio plot still indicate the width of the posterior, correspondingly scaled by the data.

\begin{figure}
\includegraphics[scale=0.375]{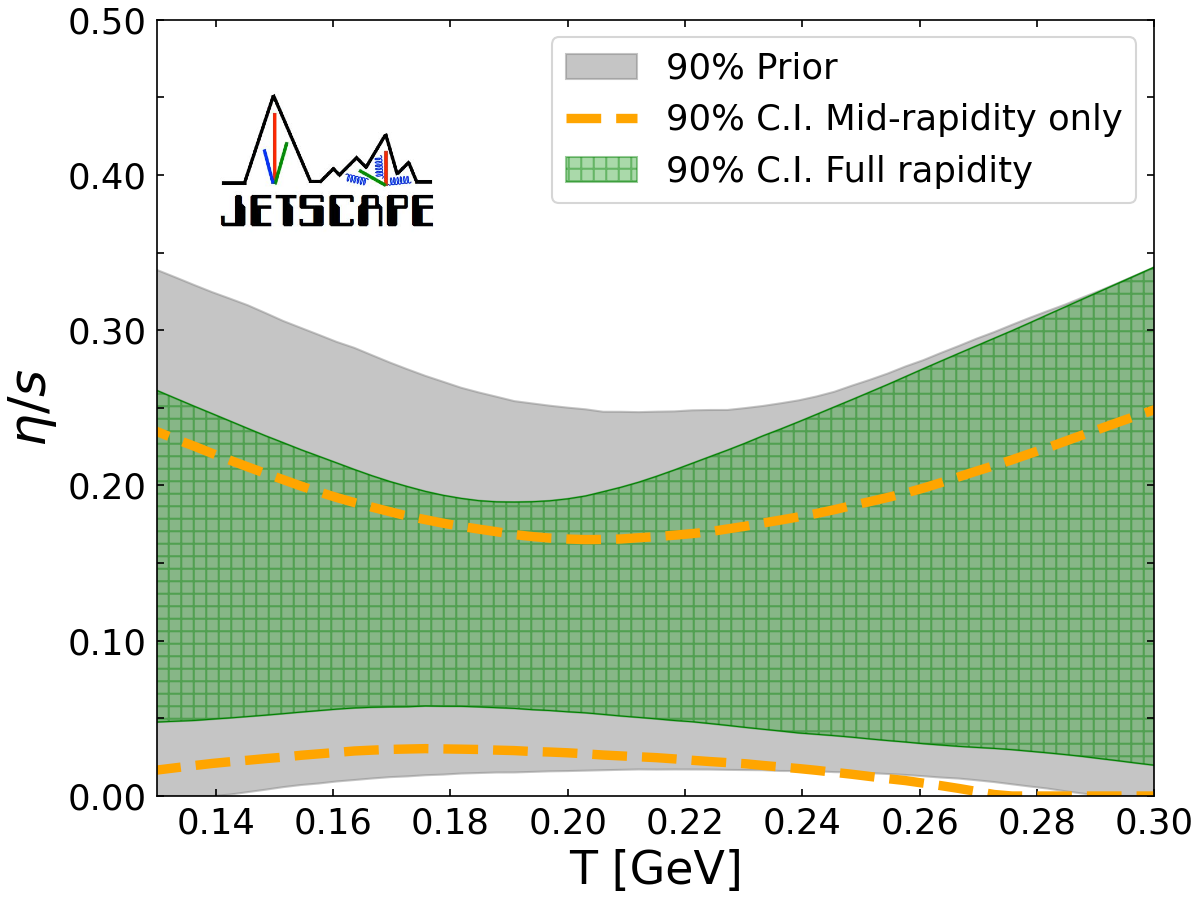}
\includegraphics[scale=0.375]{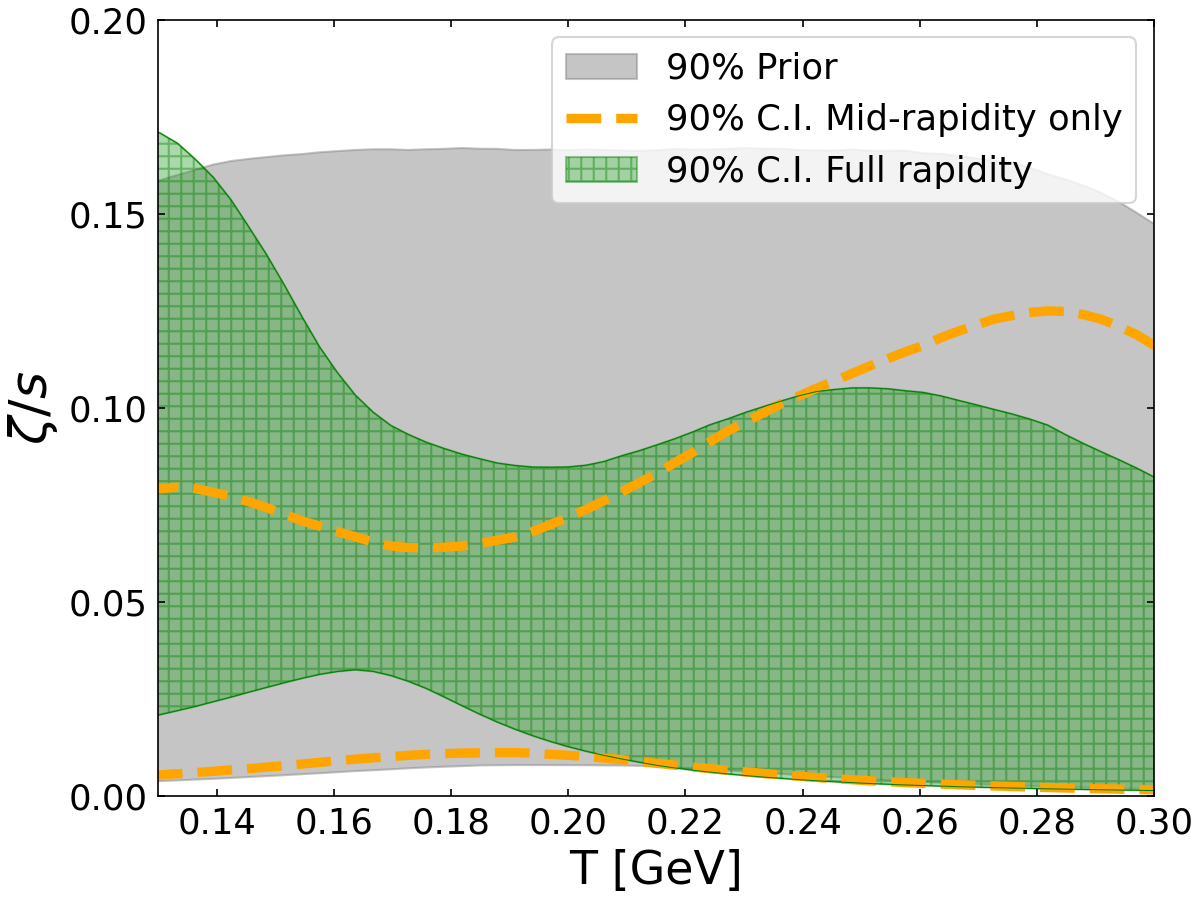}
\centering
\caption{The posterior for the specific shear (top) and bulk (bottom) viscosity as a function of temperature for the default calibration (in green) and the calibration using only the mid-rapidity data (in orange).
}
\label{fig_visc_full_midrap}
\end{figure}

\begin{figure*}
\includegraphics[scale=0.36]{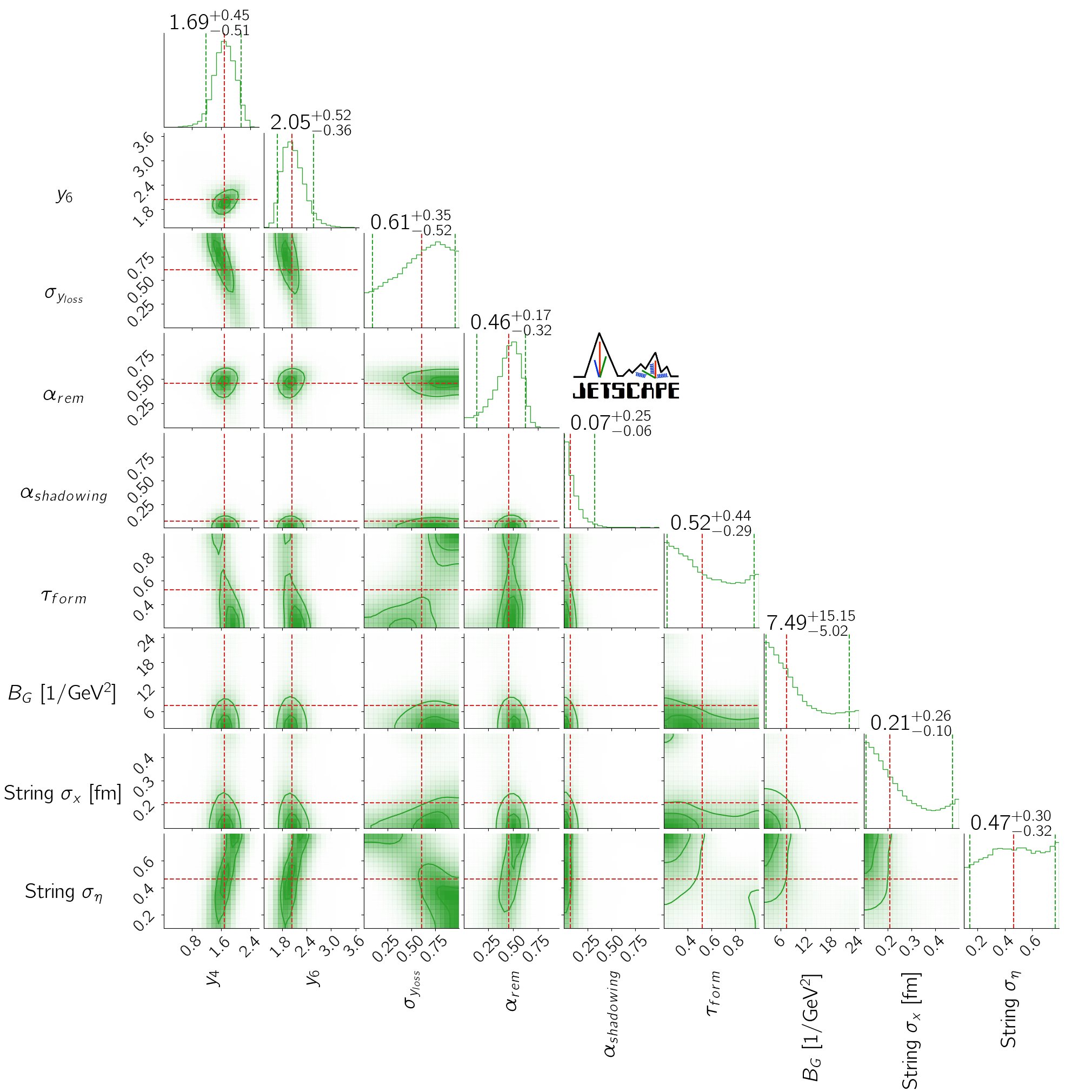}
\centering
\caption{The parameter posteriors for the default calibration for the nine parameters related to the initial state and hydrodynamization. The diagonal plots show the one-dimensional marginalized posteriors for each of the parameters. The off-diagonal plots show the two-dimensional correlations between pairs of parameters.
The dashed red line indicates the median values of each marginalized 1D distribution and the dashed green lines the corresponding 90\% bounds.
}
\label{fig_param_posterior_default}
\end{figure*}

We see a good fit to data across experiments, systems, and observables, showing the descriptive range of the model. In particular we see largely consistent fits to the multiplicity measurements both in Au-Au and d-Au and extending to large pseudo-rapidity. Notably, the width of the posterior is larger around mid-rapidity compared to finite rapidity, reflecting the corresponding difference in the emulator uncertainty in those two regions. The mean transverse momentum shows a good fit to the data for protons, kaons and pions, although slightly overestimated for the protons. The fits to the $v_2$ and $v_3$, integrated as well as $p_T$- and $\eta$-differential, are overall good, though they put the model under some tension. Specifically, the simultaneous $p_T$ and centrality dependence of the $v_2$ in Au-Au presents a challenge for the model. In addition, the rapidity dependence of the STAR $v_2$ is poorly reproduced, even though the analogous PHOBOS measurement is well-described. We explore the origins of some of this tension in more detail in Subsection \ref{subsec:tension}. 

Remarkably, the fit to a host of measurements \textit{not} included in calibration is quite good: specifically the $v_3$($p_T$) and $v_4$($p_T$) in Au-Au from PHENIX, even though no other $p_T$-differential measurements of $v_3$ and no measurements of $v_4$ at all are included in calibration. In addition, the transverse energy d$E_T$/d$\eta$ 
as a function of centrality is well described for both Au-Au and d-Au. Lastly, even though the posterior is relatively wide, the PHENIX and STAR data for $v_3$($p_T$) in d-Au are close to the model prediction. We come back to this small system data in Subsection~\ref{sec:small_systems}, where we place it in the context of other small-system measurements.

The posteriors of the viscosity parameters are shown as a function of temperature in Fig.~\ref{fig_visc_full_midrap} (in green), where we also show the posteriors for the mid-rapidity-only calibration (in orange). The mid-rapidity-only calibration features a proper subset of the default dataset, including all integrated mid-rapidity observables, as well as the mid-rapidity bins (defined in general as $|\eta|<1$, with a few exceptions extending to $|\eta|=2$ for Au-Au measurements) of observables measured as a function of $\eta$. In the default calibration we find a preference for finite shear and bulk viscosity, particularly at low temperature. The inclusion of the forward and backward regions, which correspond to a relatively larger portion of the lifetime spent at lower temperature, produces a stronger effect in that region. The preference for both finite shear and finite bulk viscosity is consistent with findings in the 2D JETSCAPE calibration which combined RHIC and LHC data~\cite{JETSCAPE:2020mzn}. However, the RHIC-only results of that calibration showed much weaker support for large viscosity than the combined result, suggesting a selection of mid-rapidity RHIC data does not show a significant preference for a viscous hydrodynamics in at least one other model. We expand on these conclusions and the effect of rapidity-dependent data in the next section.

The parameter posteriors from the calibration are shown in a corner plot in Fig.~\ref{fig_param_posterior_default} for nine of the initial state and hydrodynamization parameters.\footnote{The corner plot featuring all 20 varied parameters is included in Appendix \ref{corner_full}.} Across the diagonal we show the marginalized distributions for each of the parameters, while in the off-diagonal region we show the 2D-correlations between parameters. In each plot we indicate with dashed red line the median values of each marginalized 1D distribution, for which we also plot in dashed green lines the 90\% bounds of the posterior and, above, the numerical values of the median and those bounds. We note that most of these parameters show significant constraints relative to the flat priors, with some especially so: notably, the rapidity loss and shadowing parameters.
All posteriors from the default calibration--which again includes all of the selected data considered here from both Au-Au and d-Au at the RHIC top energy listed in Table \ref{tab:table2}--are consistent with those found in a Beam Energy Scan calibration of this model on Au-Au data (including at 200 GeV collision energy)~\cite{Jahan:2024wpj} for all comparable parameters between the two studies.

\subsubsection{\label{subsec:rapidity}The constraints from forward/backward rapidity}

\begin{figure*}
\includegraphics[scale=0.365]{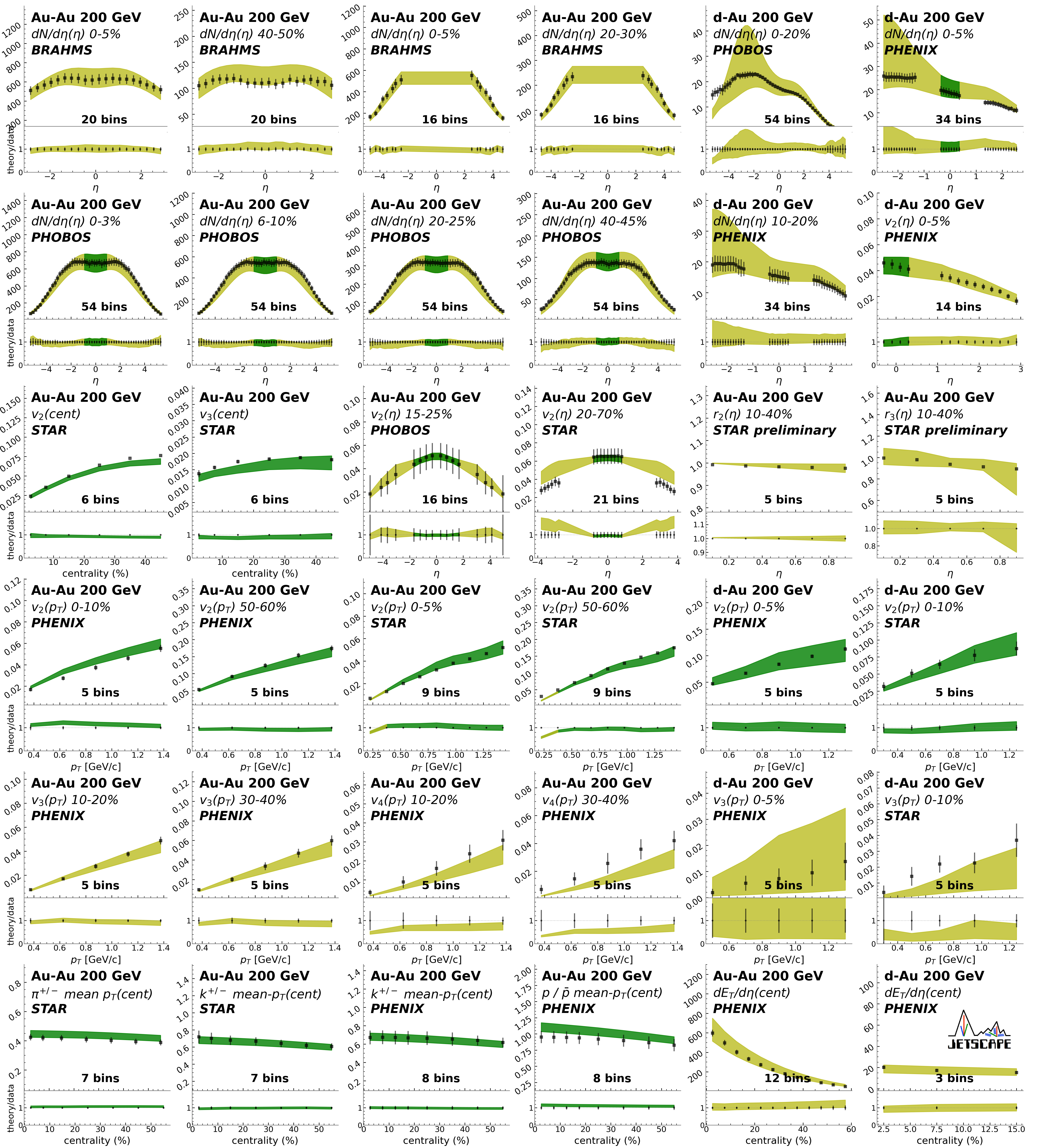}
\centering
\caption{ The observable posterior for the mid-rapidity calibration. The data included in the calibration---namely, the subset of the default calibration, including the mid-rapidity points of the rapidity-dependent measurements, and all rapidity-integrated observables---are plotted in green. The remaining data not used in calibration (including all forward/backward rapidity points) are plotted in olive. The solid markers are the experimental data, while the bands show a 90\% interval of the posterior calculated using the emulator. Each plot is labeled with the number of data bins corresponding to each observable in a given plot.
}
\label{fig_obs_post_midrap}
\end{figure*}

\begin{figure*}
\includegraphics[scale=0.36]{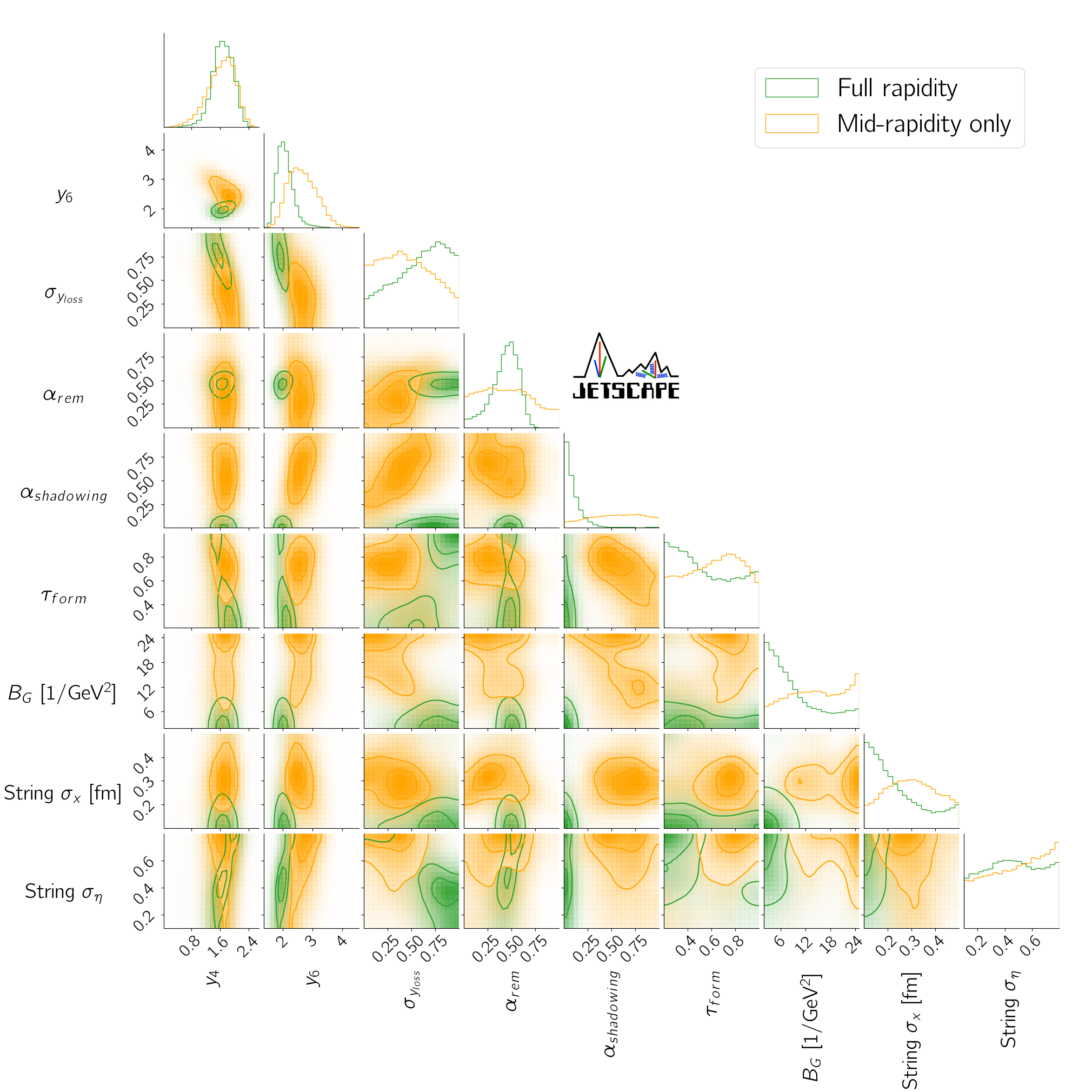}
\centering
\caption{The parameter posteriors for the default (green) and the mid-rapidity-only (orange) calibrations for the nine parameters related to the initial state and hydrodynamization. The data included in the mid-rapidity-only calibration is the subset of the default calibration including the mid-rapidity points of the rapidity-dependent measurements, and all rapidity-integrated observables. The diagonal plots show the one-dimensional marginalized posteriors for each of the parameters. The off-diagonal plots show the two-dimensional correlations between pairs of parameters.
}
\label{fig_corner_post_midrap}
\end{figure*}

A key finding from this analysis is that forcing our 3D model to fit, simultaneously, the experimental data at all rapidities yields different constraints on model parameters than if comparing only to mid-rapidity. We have already seen this in the viscosity posteriors in Fig.~\ref{fig_visc_full_midrap}. We might consider that we could readily glean conclusions from looking at the comparison between the data at all rapidities with predictions from the model constrained on only the mid-rapidity data. This is shown in Fig.~\ref{fig_obs_post_midrap}, again computed with the emulator using one hundred thousand samples of the posterior.\footnote{While we exclude in this "sub-calibration" any data points explicitly measured at forward/backward rapidities, as discussed in previous sections, a number of the rapidity-integrated measurements implicitly contain forward/backward rapidity information which we also incorporate in our calculations. In other words, our mid-rapidity calibration is not "purely" a mid-rapidity one in the sense of previous 2D calibrations which lacked access to this information.} 
If introducing finite rapidity measurements introduces a tension in the model, we might expect that the calibration using both mid-rapidity and finite rapidity data would describe more poorly the mid-rapidity points. We do not see clear visual evidence of that. That said, if examining more closely, we can still note differences. In particular, the width of the posterior, both at mid-rapidity and at finite rapidity, is significantly reduced when using the full dataset. This is especially noticeable for the multiplicity in d-Au and the $v_2$($\eta$) in Au-Au. These differences, which we might otherwise miss in "chi-by-eye" fits, are exactly the ones the Bayesian calibration allows us to extract quantitative information from, in the form of the parameter posteriors.  

As with the viscosity, we see quantitative confirmation when we examine the individual parameter posteriors for a selection of the initial state parameters. In Fig.~\ref{fig_corner_post_midrap} we show again the marginalized posteriors and 2D correlation plots of the default calibration from Fig.~\ref{fig_param_posterior_default} in green, but additionally overlay in orange the corresponding posterior information for the mid-rapidity-only calibration. While the posteriors are always consistent between the two calibrations, we see significant differences for some of the parameters, most notably for $\alpha_{rem}$, the energy loss of the collision remnants, and $\alpha_{shadowing}$, the nuclear shadowing factor. The posteriors for the rapidity loss ($y_4$ and $y_6$) are similar between the two calibrations (though narrower for the default calibration), likely due to those parameters determining the energy density profile near mid-rapidity. On the other hand, $\alpha_{rem}$ contributes more directly to energy deposition (and thus ultimately multiplicity) away from mid-rapidity, and is constrained only when including the forward/backward data. Similarly, we see a broad distribution of the $\alpha_{shadowing}$ parameter when using only mid-rapidity data. In the mid-rapidity calibration, we can also see a clear positive correlation between this parameter and the $y_6$, due to their opposing effects on the mid-rapidity multiplicity, which are shown in sensitivity plots in Appendix \ref{observable_sensitivity}. The correlation represents a degeneracy in the model that is only resolved with the inclusion of forward/backward rapidity data. We note that for the default calibration this degeneracy is no longer prominent: the 2D $y_6$-$\alpha_{shadowing}$ posterior is localized with both parameters relatively well-constrained.

\subsubsection{\label{subsec:tension}Observable tension}

\begin{figure}[bt]
\includegraphics[scale=0.5]
{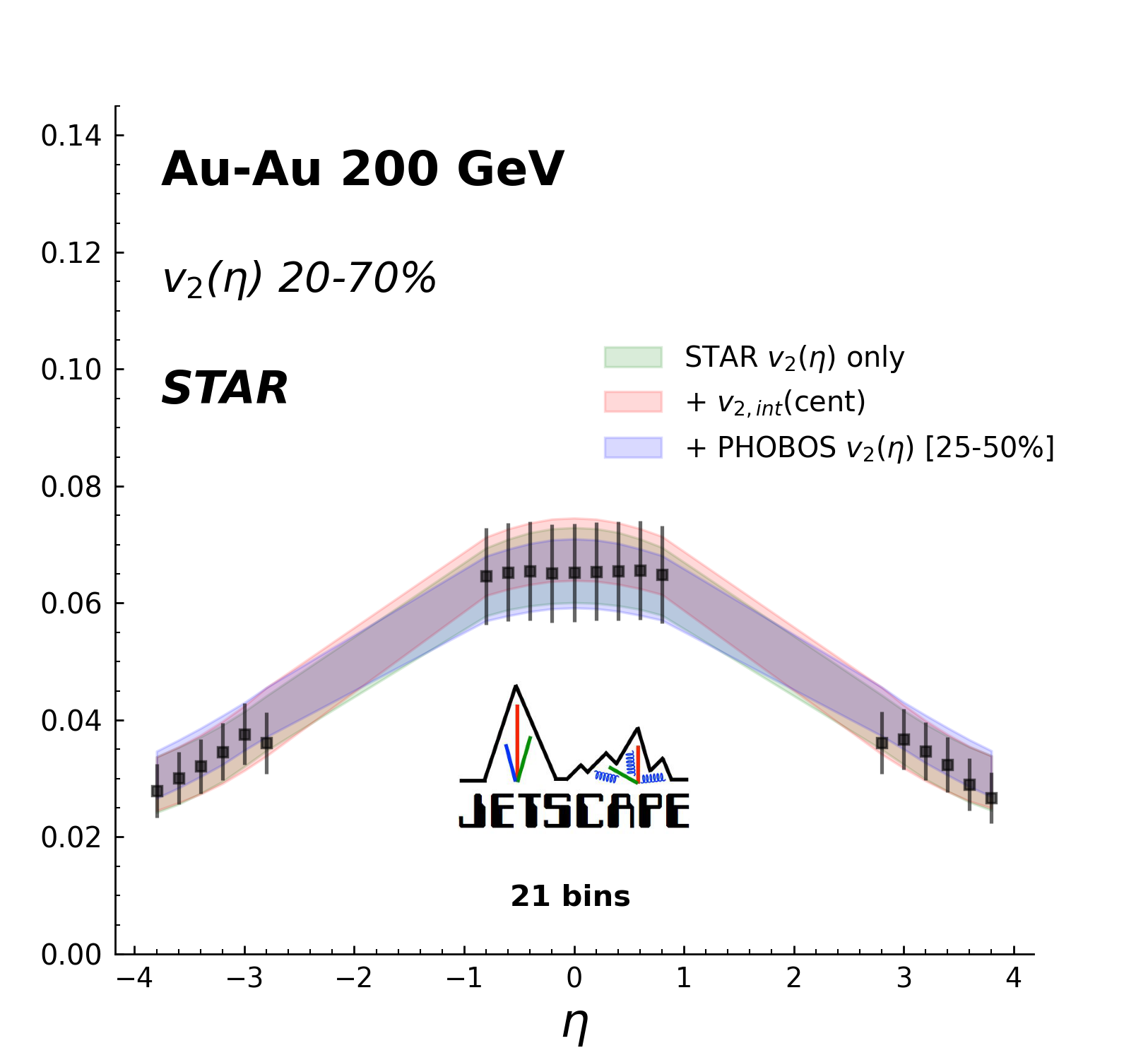}
\includegraphics[scale=0.5]{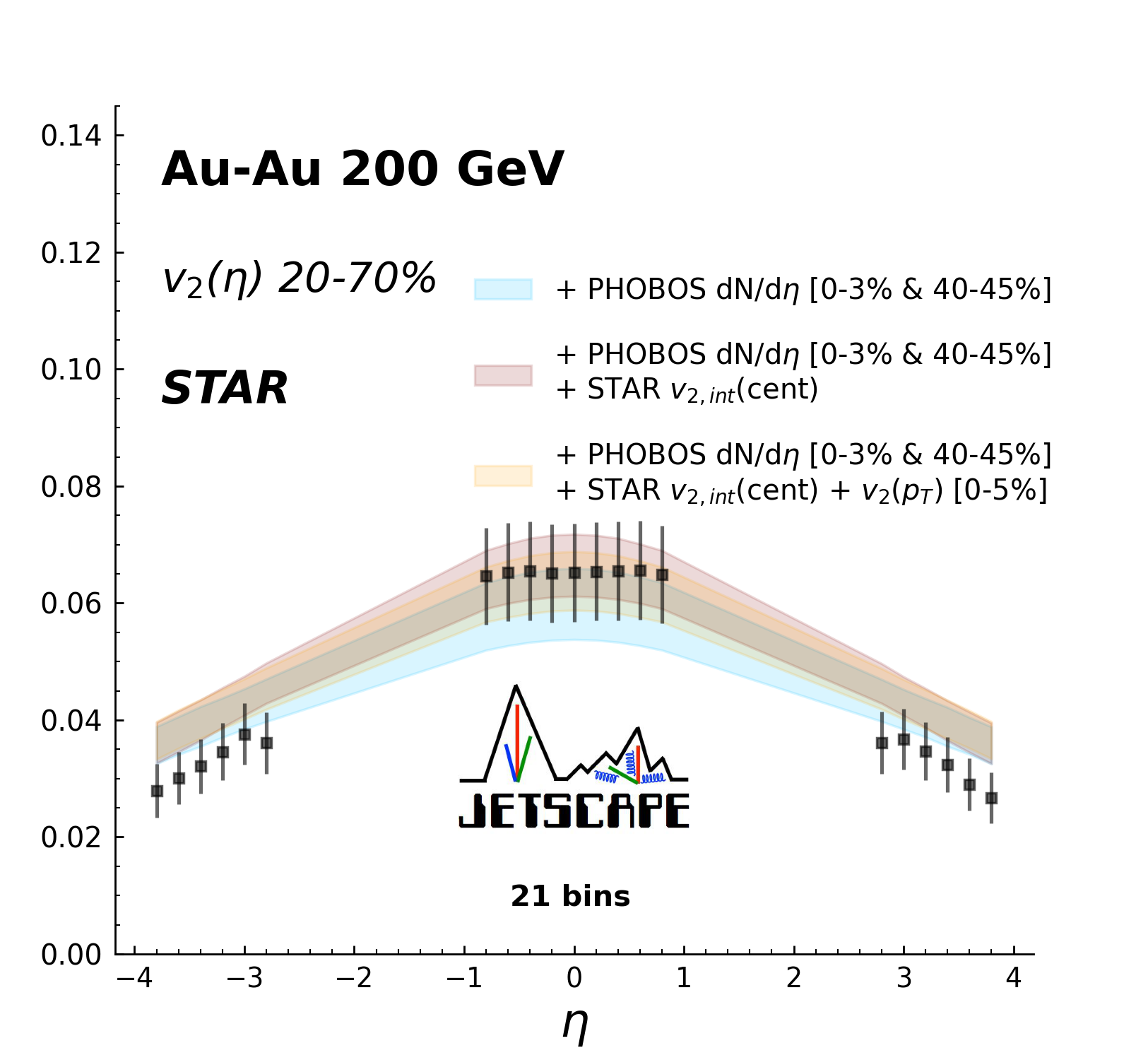}
\caption{The comparison of the STAR $v_2$($\eta$) posterior with the measurement, showing calibrations with various subsets of the data and highlighting the model tension between certain observables and the fit to the STAR measurement. The top plot shows calibrations using subsets of $v_2$ data only. The bottom plot shows calibrations that include subsets of multiplicity data. All data in these calibrations are from Au-Au measurements.
}
\label{fig:v2eta_tension}
\end{figure}

It is clear from the comparison between data and the observable posterior in Fig. \ref{fig-observable-posterior}, that the default dataset places our model under tension. In particular, the tension may be readily seen in that figure in the plots of the STAR $v_2$($\eta$) in Au-Au, and some of the lower $p_T$ bins of the $v_2$($p_T$) in Au-Au for both STAR and PHENIX measurements. For all remaining data bins used in calibration, some part of the 90\% band of the posterior is always in close proximity (and in most cases on top) of the measurements. 

\begin{figure}[bt]
\includegraphics[scale=0.5]
{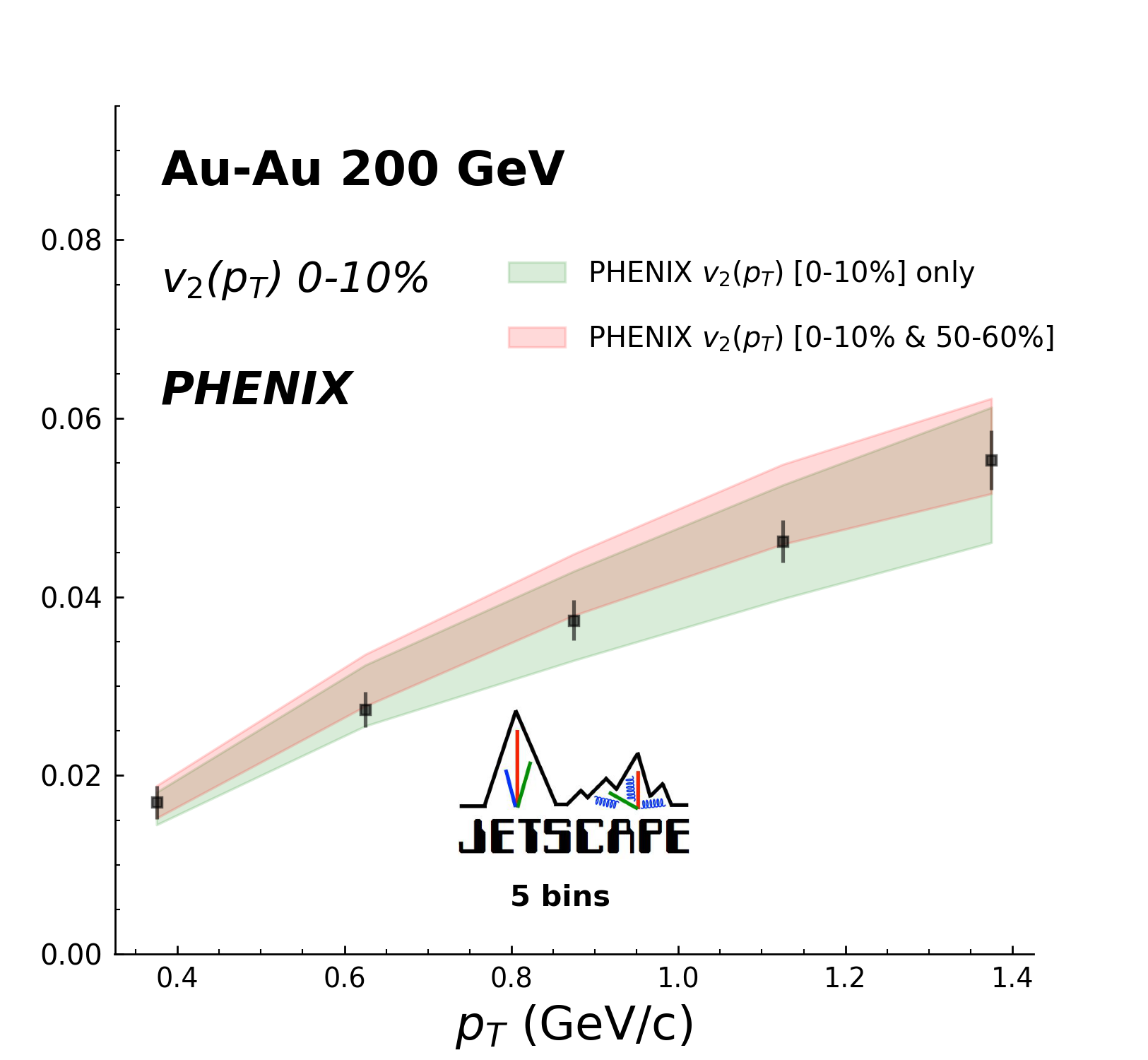}
\includegraphics[scale=0.5]{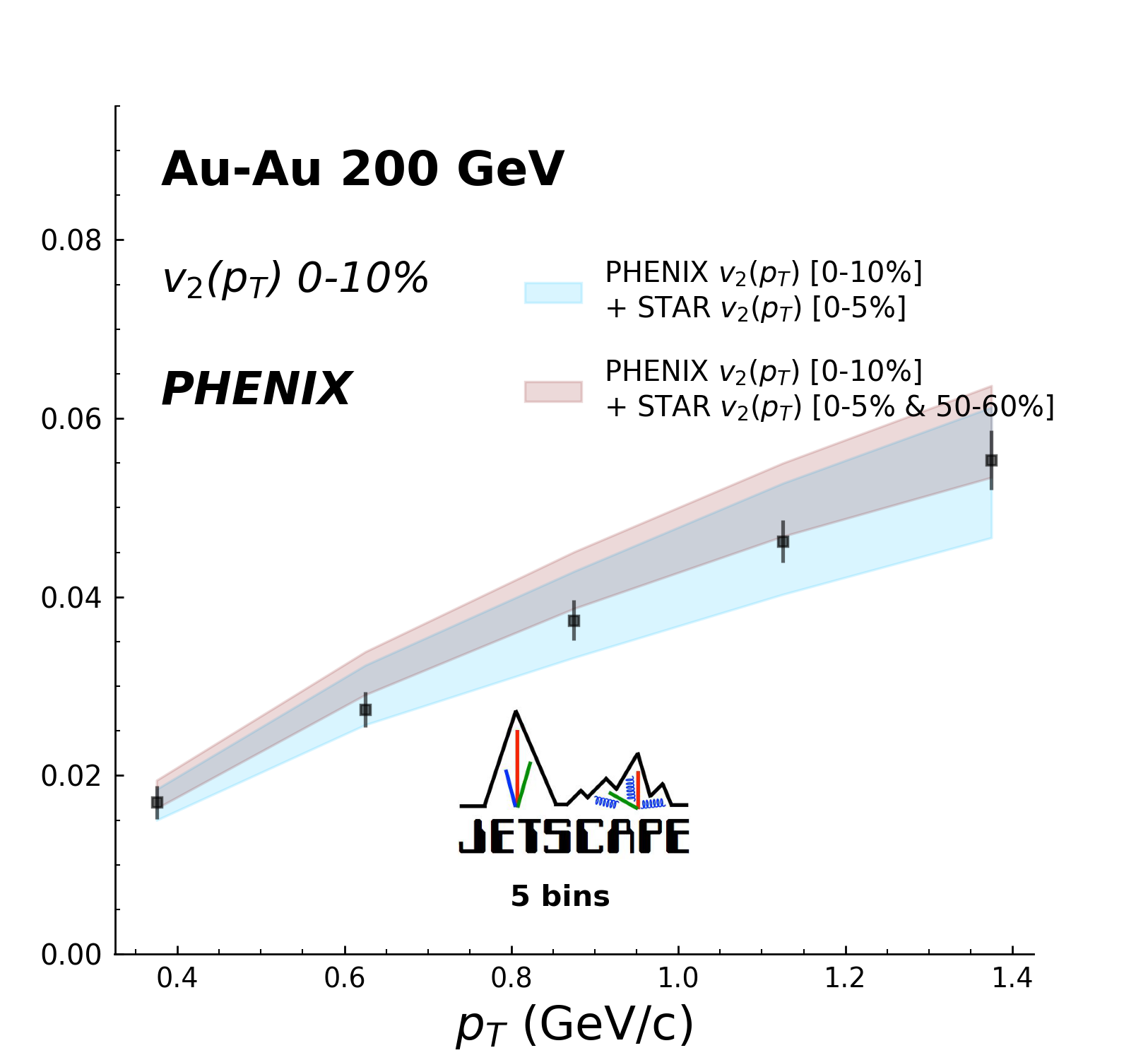}
\caption{The comparison of the PHENIX $v_2$($p_T$) posterior for central (0-10\%) Au-Au collisions with the measurement, showing calibrations with various subsets of the data and highlighting the model tension between different bins of $v_2$ data and the fit to the PHENIX measurement. The top plot shows calibrations using subsets of PHENIX $v_2$ data only. The bottom plot shows calibrations that include subsets of STAR data. All data in these calibrations are from Au-Au measurements.
}
\label{fig:v2pT_tension}
\end{figure}

In Figs. \ref{fig:v2eta_tension} and \ref{fig:v2pT_tension} we examine the origin of some of this tension by calibrating on various subsets of observables. For each of the cases mentioned above, we aim in this subsection to identify which additional observables in our default dataset are chiefly responsible for forcing the model into a preferred region of parameter space at the expense of accurately describing said dataset.

First we look at the STAR measurement of $v_2$($\eta$) in Au-Au. As we noted in Section~\ref{sec:exp_data}, this dataset presents several challenges in matching precisely to the experimental procedure, particularly from the aggregation into a large centrality bin (20-70\%). Nevertheless, as noted in that section, we carefully attempted to account for any methodological uncertainty related to it, and therefore expect our model to fit the data within uncertainties. We can see in the top plot of Fig.~\ref{fig:v2eta_tension} the 90\% interval of the posterior for this observable compared to the data for three minimally restrictive calibrations. In green, we show the posterior if using only the  STAR $v_2$($\eta$) measurement itself in constraining the model, and note that we can describe all the points with accuracy over the full range in pseudo-rapidity. In subsequent bands we show the posterior when adding (indicated by a ``+") various other observables to the calibration in addition to the one plotted. The description remains accurate, then, if we pair that measurement with the integrated $v_2$ as a function of centrality also measured by STAR. That posterior (shown in red) prefers a larger value at mid-rapidity (where the integrated measurement is taken), while a calibration using the STAR $v_2$($\eta$) and the corresponding $v_2$($\eta$) from PHOBOS (shown in purple), yields only mild differences. We conclude that the model is not placed under significant tension from the $v_2$($\eta$) measurements alone, or from the integrated $v_2$.

However, once we include multiplicity measurements extending to large pseudo-rapidity and from multiple centralities, as shown in the bottom plot of that figure, the constraints tighten and the description of the STAR $v_2$($\eta$) worsens, particularly at large $|\eta|$. Once the multiplicity is included, the description at large $|\eta|$ tends to remain poor, regardless of additional calibration observables, as shown in the last two bands. Adding the integrated $v_2$ or the $v_2$($p_T$) from STAR will modulate the mid-rapidity value, but will not improve the forward and backward region. Additional centrality bins of these observables not included in these minimal calibrations reinforce these effects. There appears, then, a tension between the model and this ensemble of data, highlighted by the difficulty to describe the correlation between $v_2$  and dN/d$\eta$ across pseudo-rapidity for all $v_2$ measurements. While the posteriors for the corresponding d-Au observables do not clearly exhibit this same tension, higher precision in measurements or emulation may bring it to the fore.

Another observable we highlight is the central bin (0-10\%) of the PHENIX $v_2$($p_T$) measurement in Au-Au. This measurement is overestimated for a few of the $p_T$ bins by the default posterior. In Fig.~\ref{fig:v2pT_tension}
we show that the model tension here is specifically between the centrality bins of $v_2$($p_T$). First, in the top plot of that figure we show in green the posterior when calibrating on only this central bin. In the same plot, we see that adding the most peripheral bin  (the 50-60\%) in that set of measurements to the calibration immediately shows a strong effect on the 0-10\% bin, leaving most of the posterior to overestimate the experimental uncertainty markers for the intermediate $p_T$ bins.

\begin{figure*}
\includegraphics[scale=0.31]{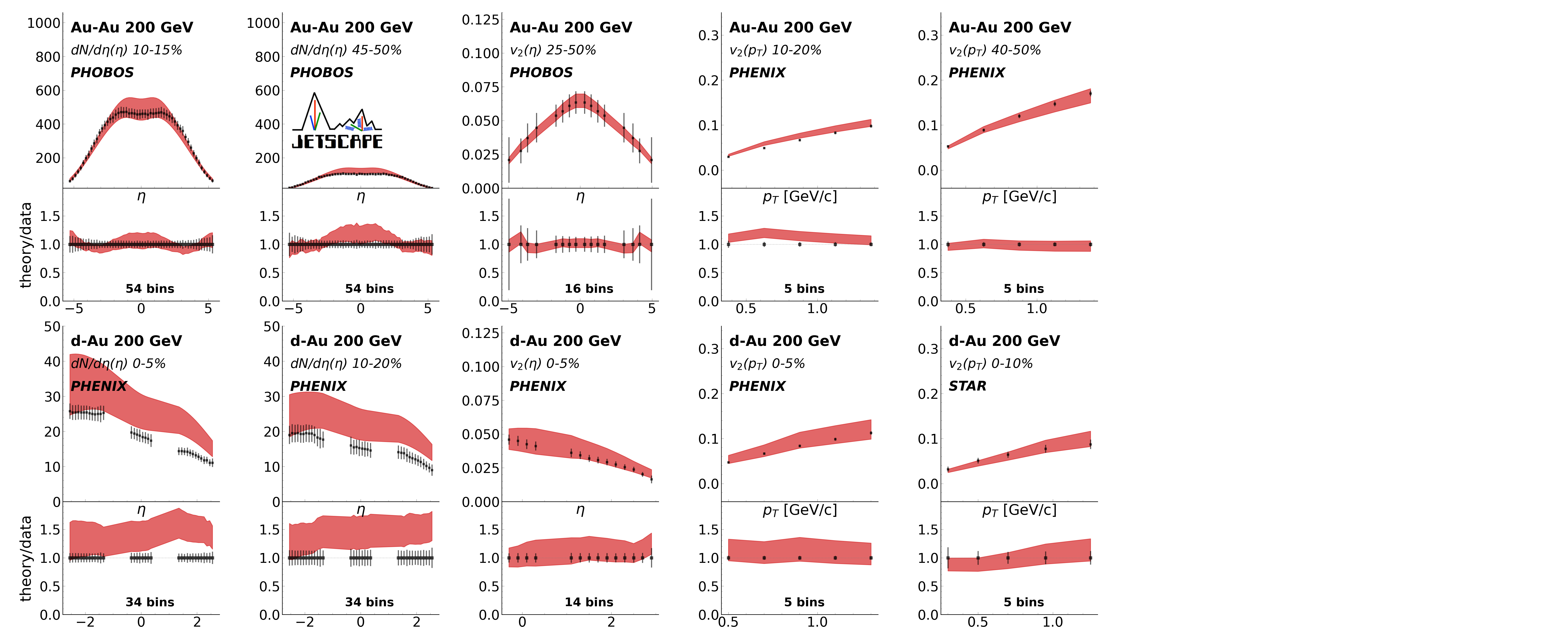}
\centering
\caption{The model posterior for select observables constrained using the Au-Au subset of the default observable set. The band represents the 90\% credible interval of the posterior, and the bottom panel in the figures shows the model posterior scaled by the values of experimental data in each bin.
}
\label{fig_AuAu_obs_post_mini}
\end{figure*}

\begin{figure*}
\includegraphics[scale=0.31]{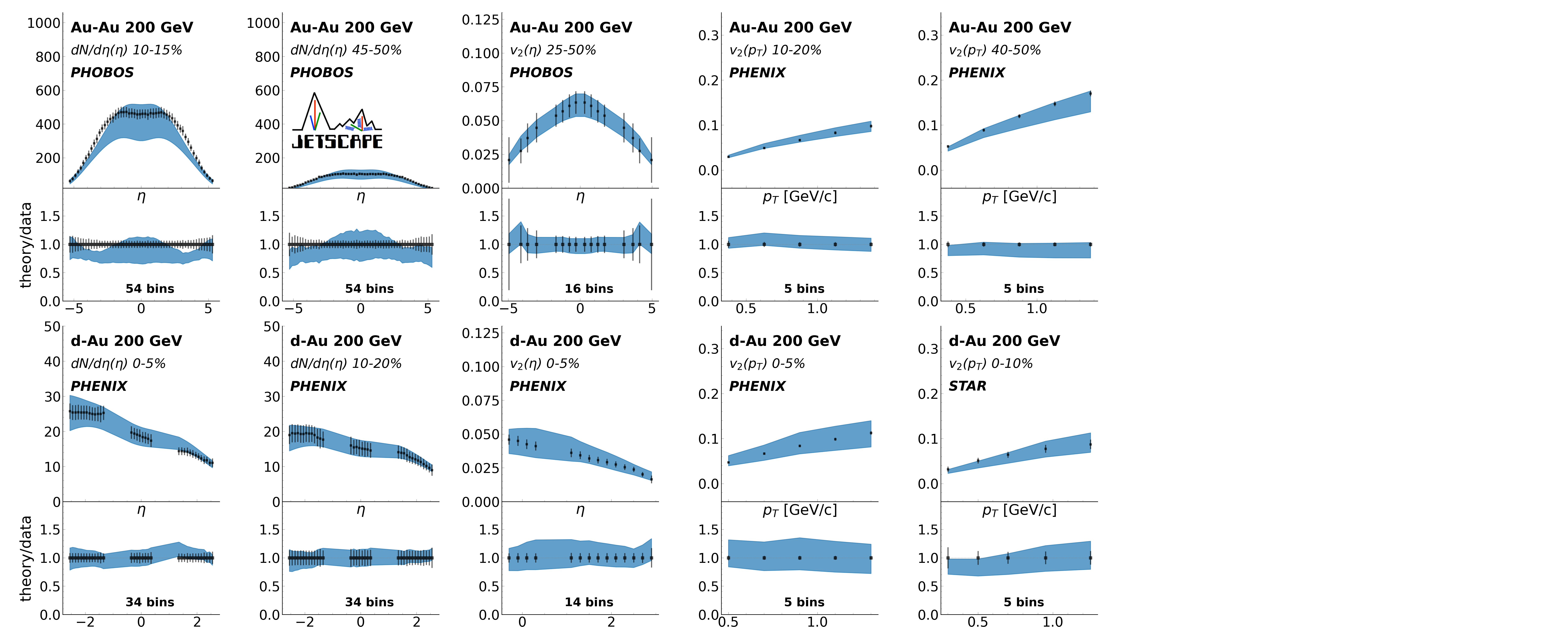}
\centering
\caption{The model posterior for select observables constrained using the d-Au subset of the default observable set. The band represents the 90\% credible interval of the posterior, and the bottom panel in the figures shows the model posterior scaled by the values of experimental data in each bin.
}
\label{fig_dAu_obs_post_mini}
\end{figure*}
Then, in the bottom plot of that figure, we show in blue the joint calibration with the most central bins of the $v_2$($p_T$) measured by STAR and PHENIX. While the measurements differ in their kinematics, we show that the model has no trouble describing the PHENIX measurement when constrained also on the STAR data. The crucial effect comes again from the peripheral bins. In the bottom plot we show the effect of the STAR peripheral bin of that set of measurements being included in calibration. In much the same way as in the top plot, this calibration shows a strong preference for larger values of the PHENIX observable, indicating once more that the model is placed under significant tension by the simultaneous $p_T$ and centrality dependence of $v_2$ in Au-Au. This behavior may be a consequence of the centrality dependence of the initial state eccentricity of our initial condition model (see e.g. Fig.~2 of~\cite{Heinz:2013th} for a discussion of this effect in 2D initial condition models). 

What we have shown in this subsection is the tendency of the model to struggle to describe certain groups of observables simultaneously. We have demonstrated how the tension apparent in the many-observable default calibration can be roughly pinned down to a few observables by means of successive calibrations with subsets of the data.

\subsection{\label{sec:system_calibrations}Comparison of independent Au-Au and d-Au calibrations}

In this subsection we show how our model performs when calibrated with data from systems of different size and how the parameter posteriors change. We answer two overarching, related questions. First, how well can the model, calibrated on data from only a small (large) system, describe various measurements from the corresponding larger (smaller) system? Second, which parameters show similar constraints, and which show the most difference? If we assume the same model---in particular, a hydrodynamic QGP---describes physics for both systems (and that the differences among the two in the initial state, size of the system, etc., are well-described), constraining on data from one system should yield a consistent (though not necessarily precise) description of data from the other. It could, of course, be the case that the differences between the systems outside of the hydrodynamic description are not fully captured by the model. The conclusions from this comparison, and indeed throughout the paper, are always drawn with that fact in mind.

We show in Figs.~\ref{fig_AuAu_obs_post_mini} and \ref{fig_dAu_obs_post_mini} (using one hundred thousand samples of the respective posteriors and the trained emulator) illustrative comparisons to a select few measured observables for each of the independent system calibrations (we show a more complete picture of the observable posterior for each calibration in Appendix \ref{observable_posteriors}). We can see from the top row of Fig.~\ref{fig_AuAu_obs_post_mini} that a model constrained on only the Au-Au data shows a level of consistency with that system's data (top row) similar to the default calibration. The predictions it makes for d-Au observables however, are in some cases off the mark, as shown in the bottom row of that figure. While the d-Au $v_2$ measurements, both as a function of $\eta$ and $p_T$, and for both PHENIX and STAR, are consistent with predictions, the multiplicity, particularly on the deuteron-going side, is significantly overestimated by the 90\% posterior band. This result highlights the necessity of calibrating the model on small system data in order to provide a consistent description of observables in such systems.

Meanwhile, in Fig.~\ref{fig_dAu_obs_post_mini} we show the posterior for select observables for a model constrained on only the d-Au data. Here, on the top row, we see that its predictions of Au-Au observables are, remarkably, generally consistent with measurements within uncertainties, including for the multiplicity. Nevertheless, they are significantly different compared to the predictions of the default or Au-Au-only calibrations in that the posteriors are much wider for most observables. For example, much of the posterior band includes mid-rapidity multiplicity predictions underestimating the data.
In both the default and the Au-Au calibrations such multiplicity values are nearly absent from the narrower posterior. Looking at the d-Au observables in the bottom row of that figure, we see that predictions for all observables are now consistent with the data. However, we note that the posterior bands for the $v_2$ observables are not significantly narrower compared to the Au-Au-only calibration (though we can still discern some slight differences). That is, whether we calibrate on Au-Au or on d-Au data exclusively, we obtain a similar description of the d-Au $v_2$, despite the significant difference in the corresponding multiplicity descriptions. While this result gives credence to the assumption of describing both systems hydrodynamically, we also note that the relatively large emulation uncertainty for d-Au observables, and in particular the d-Au $v_n$, does limit the information that can be extracted from the data.

The posteriors of twelve of the model parameters for each of the calibrations discussed in this section are shown in Fig. \ref{fig_systems_12_params} in Appendix \ref{systems_posteriors}. We summarize here that the individual system posteriors are generally consistent with each other and with the default posterior for all parameters. For some parameters, like the $y_6$ and the $\alpha_{shadowing}$, the particular system constraints reinforce each other, leading to narrower posteriors for the combined default calibration. Notably, the posterior for $\tau_{form}$ is narrower for the Au-Au-only calibration than for the combined default, a case where the particular system constraints appear to be in tension.

\begin{figure*}
\includegraphics[scale=0.4]{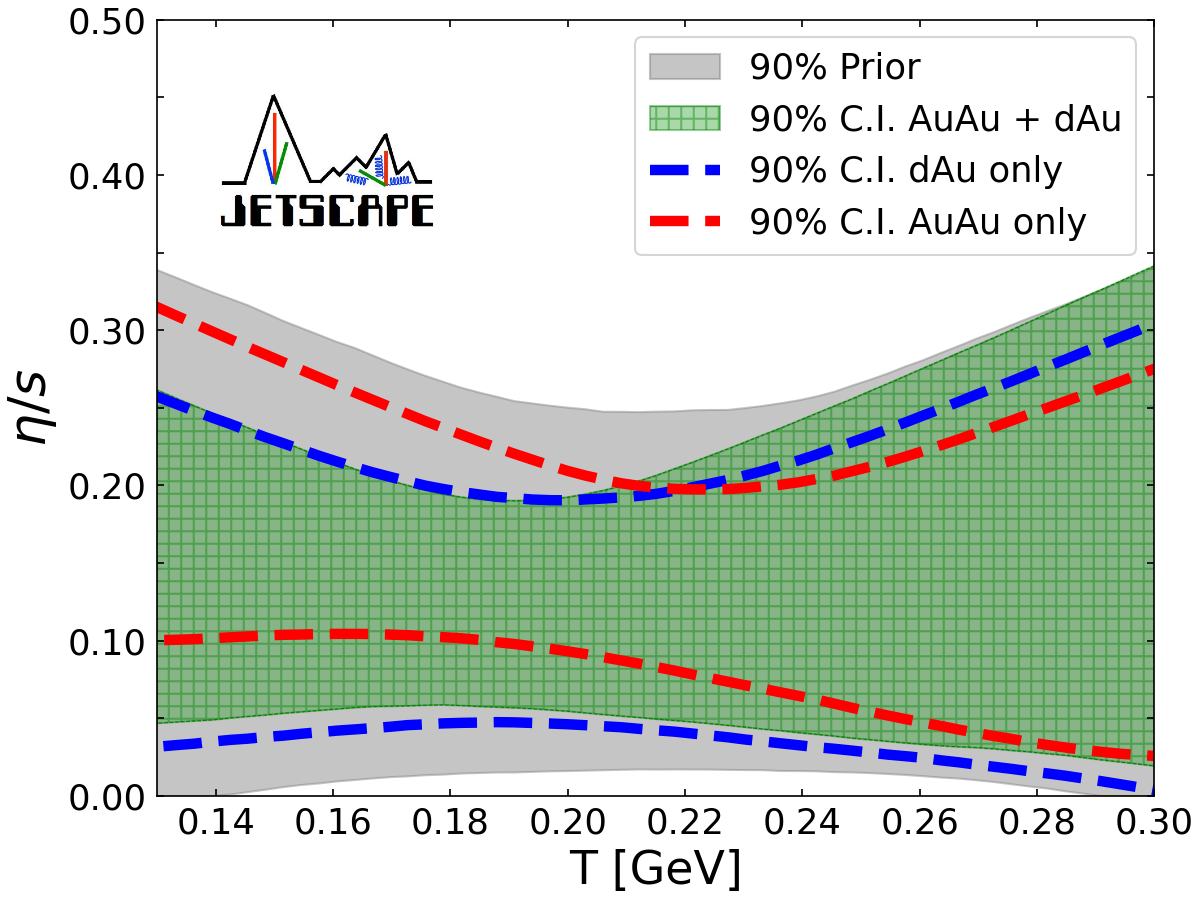}
\includegraphics[scale=0.4]{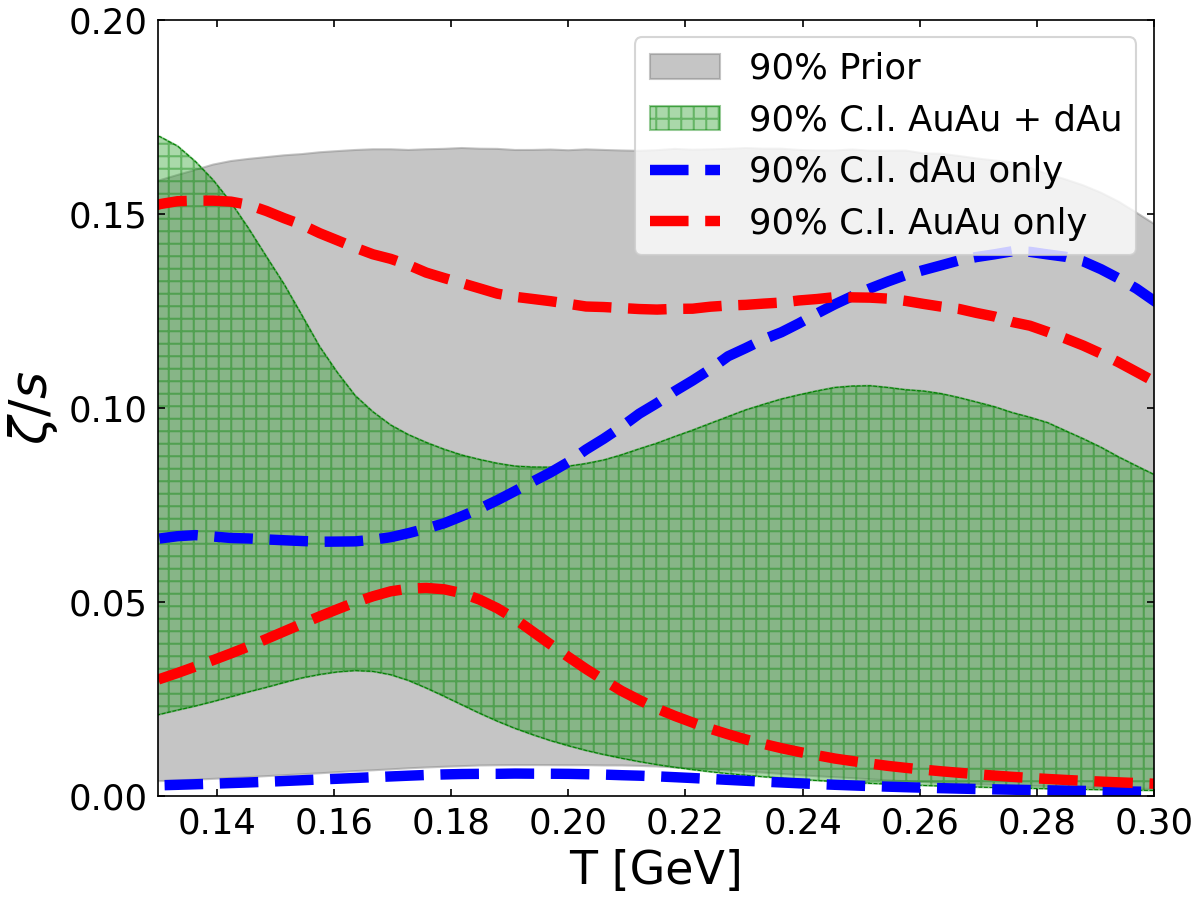}
\centering
\caption{The posterior for the specific shear (left) and bulk (right) viscosity as a function of temperature for three calibrations: the default (green), the Au-Au subset (red), and the d-Au subset (blue).
}
\label{fig_system_visc_post}
\end{figure*}

In Fig.~\ref{fig_system_visc_post} we highlight the posteriors of the viscosity parameters as a function of temperature for the individual system calibrations, as well as the default result. The Au-Au-only calibration shows a preference both for finite shear and finite bulk viscosity. This is consistent with the preference for finite shear and bulk viscosity shown in the RHIC calibration of the 2D JETSCAPE analysis, which used only Au-Au STAR data~\cite{JETSCAPE:2020mzn}. Of course, in our Au-Au calibration we include the finite rapidity data, which we showed in Subsection \ref{subsec:rapidity} to be largely responsible for this preference. The d-Au-only posterior for the shear viscosity is similar to the Au-Au, though it (and ultimately the combined posterior) prefers a slightly lower shear viscosity at low temperature. The bigger difference is in the bulk viscosity posteriors, where the two calibrations are only partly overlapping from about 190 MeV and below. The d-Au posterior is narrower than the Au-Au in this region, it prefers a low value, and does not exclude a zero bulk viscosity. Meanwhile the Au-Au posterior prefers larger values. The combined calibration posterior ultimately covers both tails and is narrowest around the point of minimal overlap of the individual system posteriors. As in previous studies of bulk viscosity with sufficiently flexible ansatz~\cite{JETSCAPE:2020mzn, JETSCAPE:2020shq, Heffernan:2023gye, Heffernan:2023utr}, constraints here tend to be stronger in the deconfinement temperature range predicted by Lattice QCD. The tightness of constraints is enhanced in the combined calibration due to some degree of tension between the individual d-Au and Au-Au results. The differences between these individual system posteriors may be due to a number of contributing factors. For one, the Au-Au calibration is constrained on mean $p_T$ data of identified hadrons, including protons, which show significant sensitivity to the bulk viscosity, and in particular a preference for larger values, as highlighted in the plots in Appendix \ref{observable_sensitivity}. In addition, the Au-Au calibration has the benefit of constraining on data from a wide range of centralities, and thus temperature profile evolutions, as well as, in general, lower uncertainties on both the experimental data and the model emulation.

\section{\label{sec:predictions} Predictions with the calibrated model}

\begin{figure}
\includegraphics[scale=0.60]{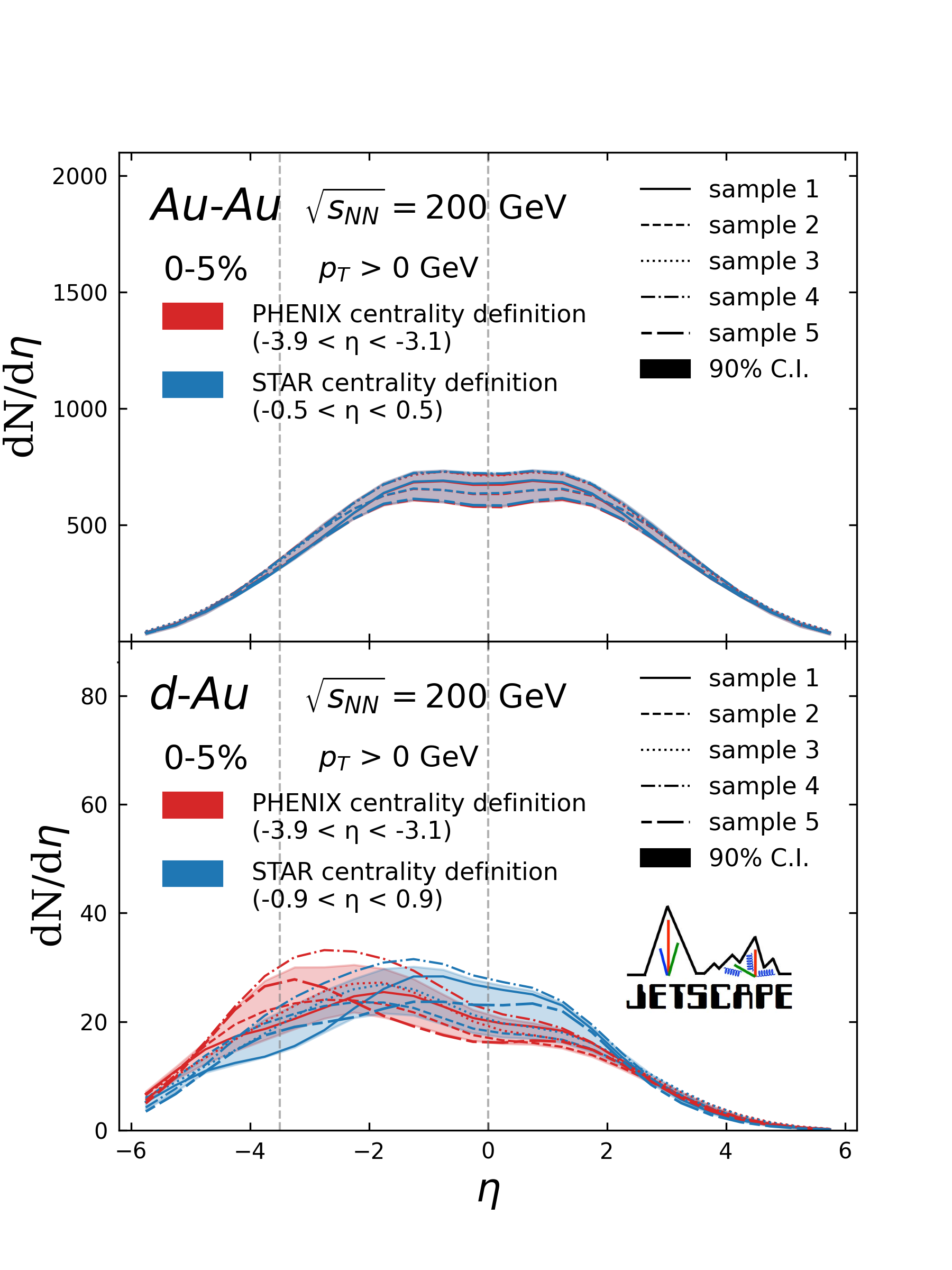}
\centering
\caption{The calibrated model calculations for dN/d$\eta$ in Au-Au (top) and d-Au (bottom) for two generic choices of centrality selection. The blue calculations use the mid-rapidity region (often used in STAR measurements) for centrality selection and the red calculations use a backward region (the Beam-Beam Counter acceptance used in many PHENIX measurements). The calculations are plotted for five individual parameter samples of the posterior simulated for ten thousand events each. In addition, a 90\% credible interval band, estimated from ten such high-statistics random samples, is shown.
}
\label{fig_cent_selection}
\end{figure}

\begin{figure}
\includegraphics[scale=0.60]
{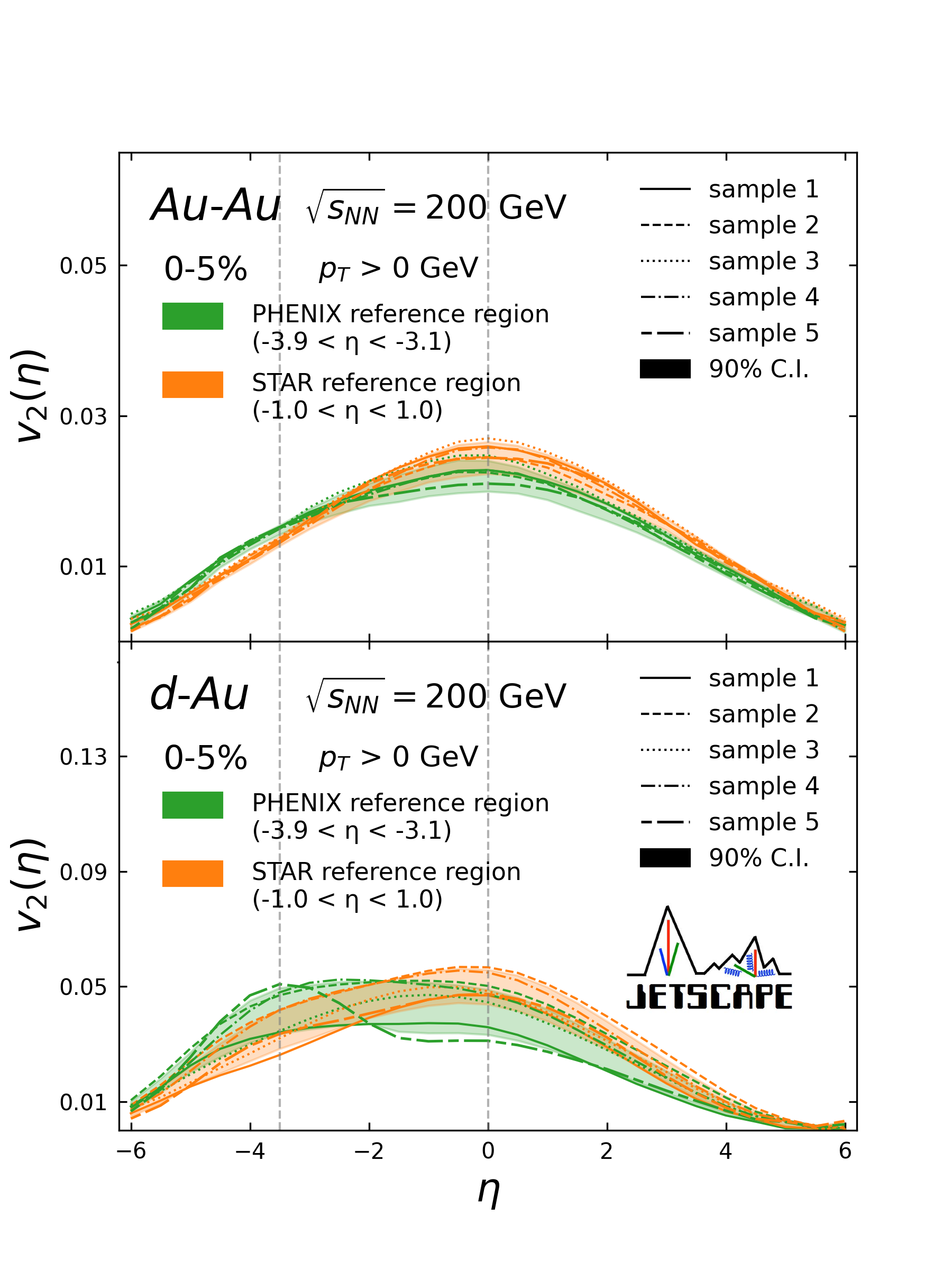}
\centering
\caption{The calibrated model calculations for $v_2$($\eta$) in Au-Au (top) and d-Au (bottom) for two generic choices of the reference rapidity region. The orange calculations use the mid-rapidity region (often used in STAR measurements) and the green calculations use a backward region (the Beam-Beam Counter acceptance used in many PHENIX measurements). The calculations are plotted for five individual parameter samples of the posterior simulated for ten thousand events each. In addition, a 90\% credible interval band, estimated from ten such high-statistics random samples, is shown.
}
\label{fig_v2_reference}
\end{figure}

\begin{figure}
\includegraphics[scale=0.60]
{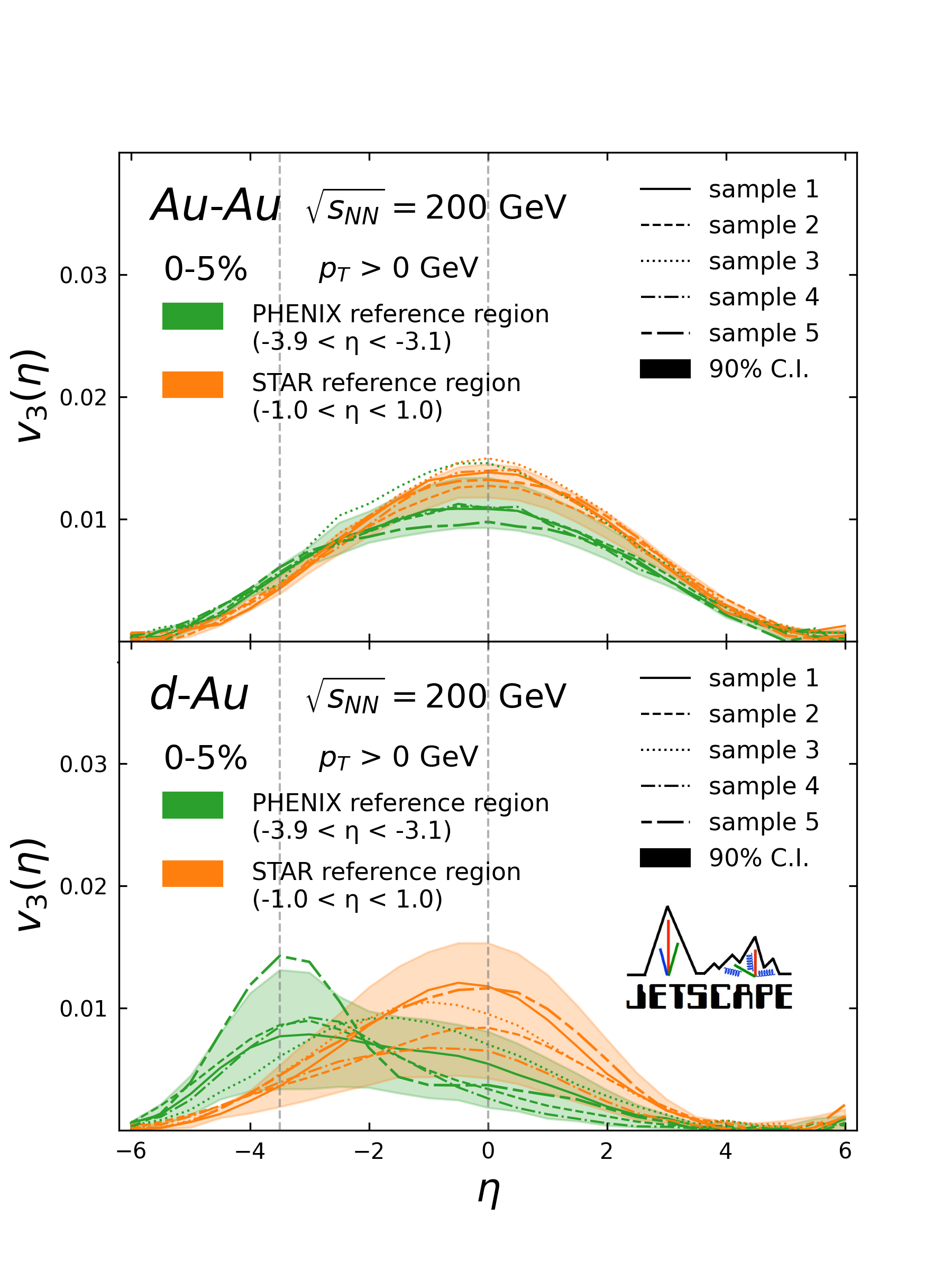}
\centering
\caption{The calibrated model calculations for $v_3$($\eta$) in Au-Au (top) and d-Au (bottom) for two generic choices of the reference rapidity region. The orange calculations use the mid-rapidity region (often used in STAR measurements) and the green calculations use a backward region (the Beam-Beam Counter acceptance used in many PHENIX measurements). The calculations are plotted for five individual parameter samples of the posterior simulated for ten thousand events each. In addition, a 90\% credible interval band, estimated from ten such high-statistics random samples, is shown.
}
\label{fig_v3_reference}
\end{figure}

\begin{figure}
\includegraphics[scale=0.60]{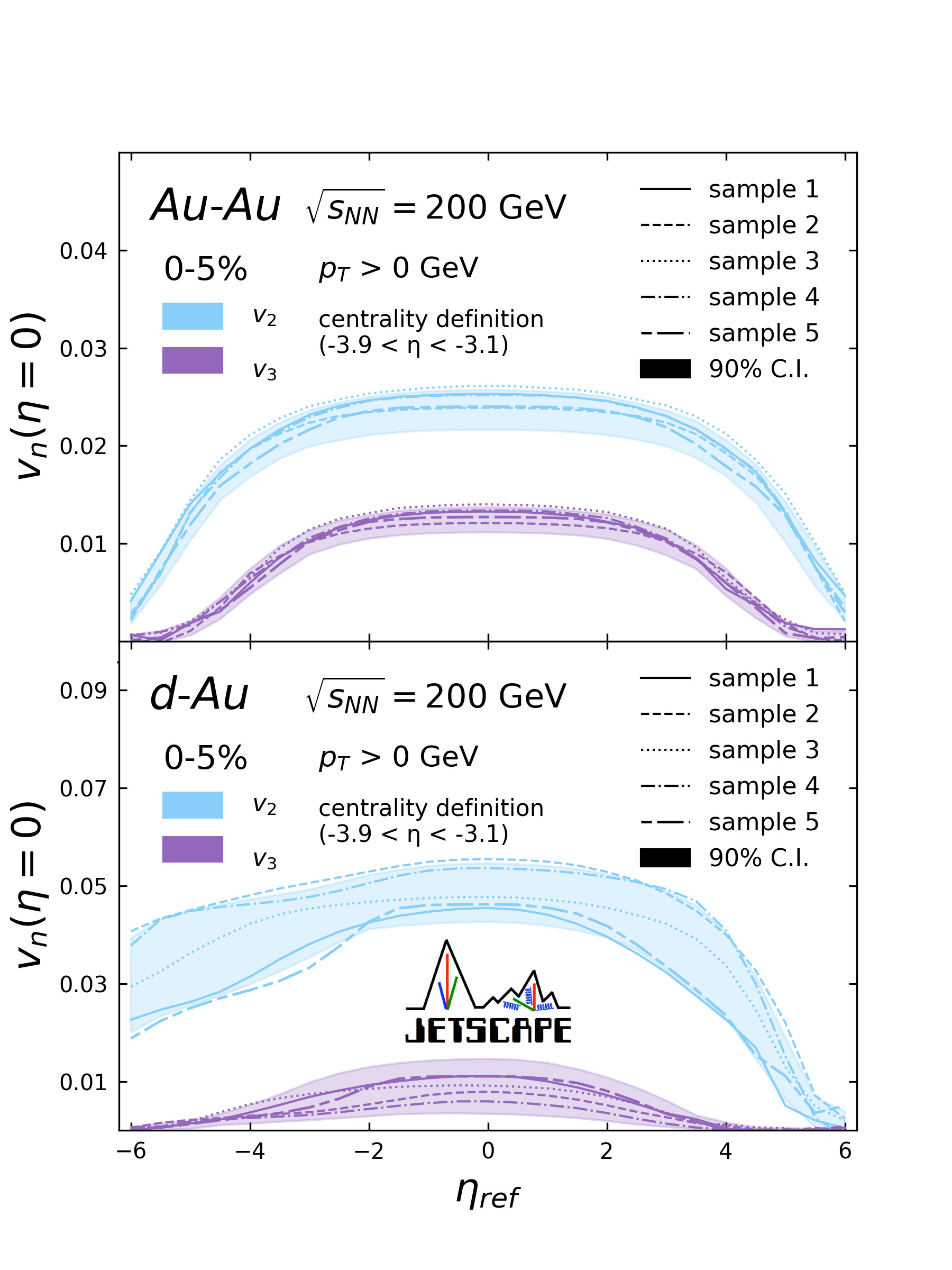}
\centering
\caption{The calibrated model calculations for flow coefficients in Au-Au and d-Au at mid-rapidity as a function of the reference rapidity region. The calculations are plotted for five individual parameter samples of the posterior simulated for ten thousand events each. In addition, a 90\% credible interval band, estimated from ten such high-statistics random samples, is shown.
}
\label{fig_vn_reference}
\end{figure}

\begin{figure*}
\includegraphics[scale=0.6]{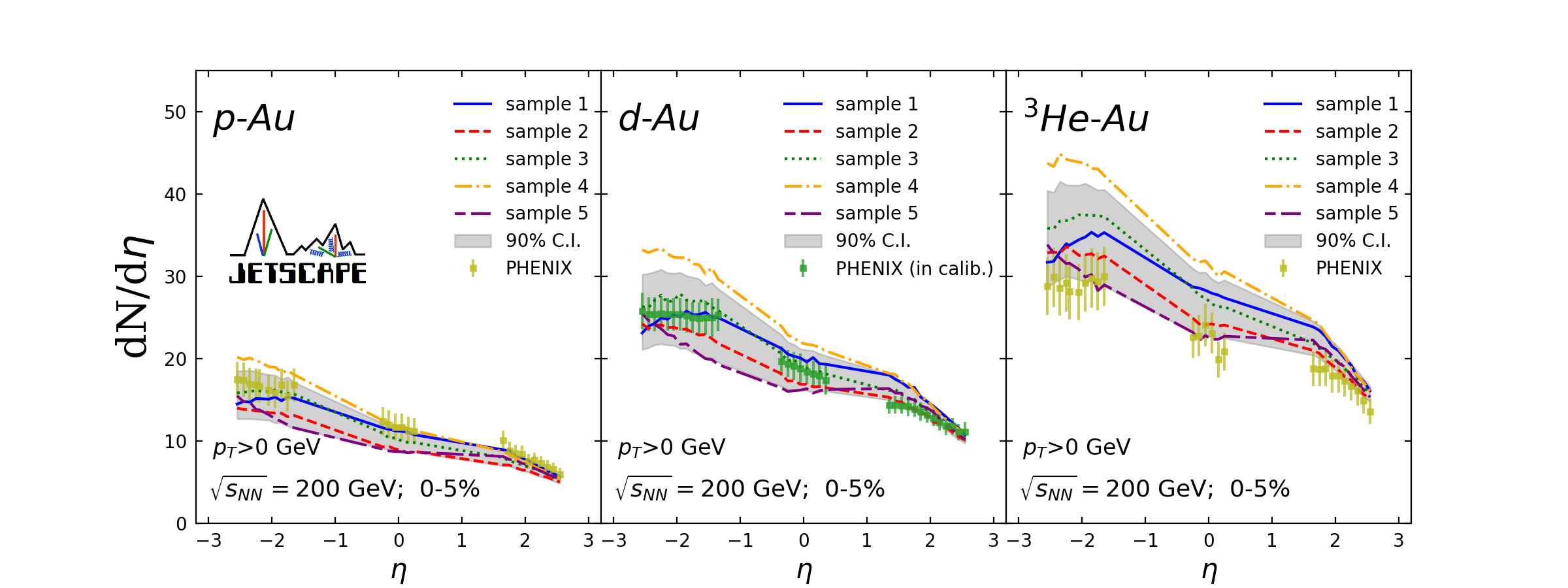}
\centering
\caption{The PHENIX measurements for the dN/d$\eta$ in p-Au, d-Au, and $^3$He-Au compared to a sampling of the posterior calibrated on the default dataset. The calibration dataset includes the dark green data points (d-Au) and does not include the olive green data points (p-Au and $^3$He-Au). The calculations are plotted for five individual parameter samples of the posterior simulated for ten thousand events each. In addition, a 90\% credible interval band, estimated from ten such high-statistics random samples, is shown.
}
\label{PHENIX_dNdeta}
\end{figure*}

\begin{figure*}
\includegraphics[scale=0.6]{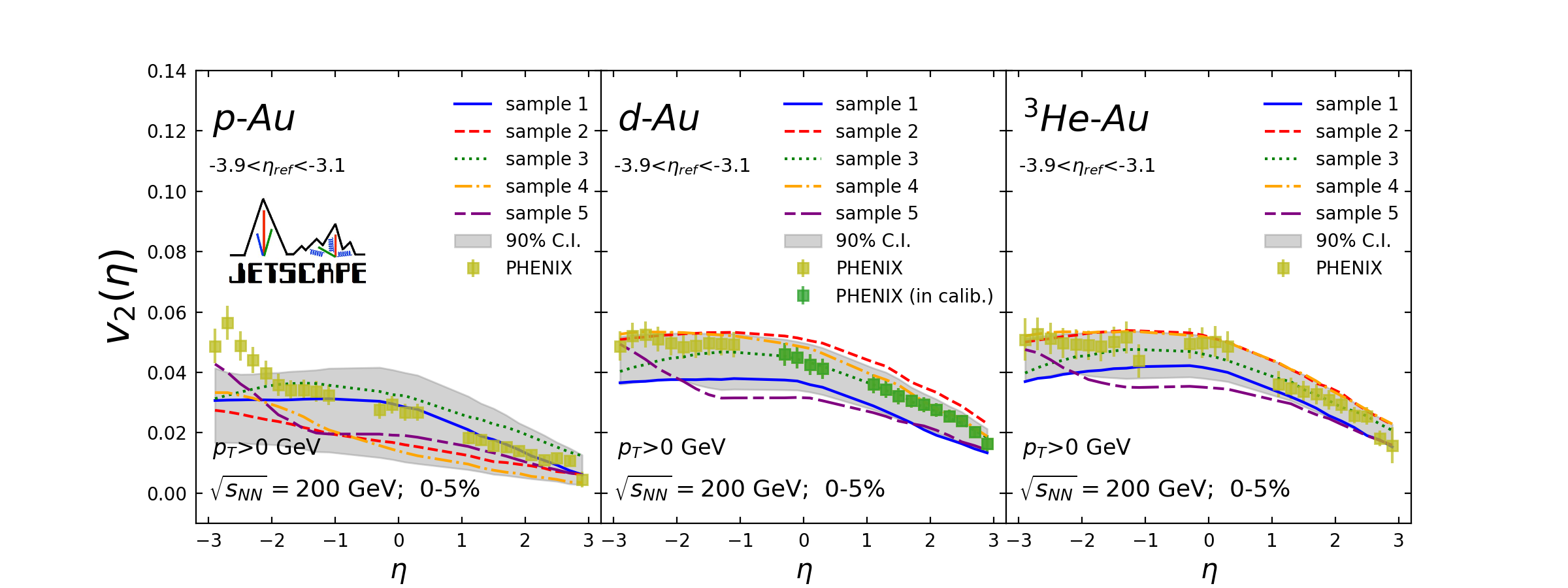}
\centering
\caption{The PHENIX measurements for the $v_2(\eta)$ in p-Au, d-Au, and $^3$He-Au compared to a sampling of the posterior calibrated on the default dataset. The calibration dataset includes the dark green data points (d-Au) and does not include the olive green data points (p-Au and $^3$He-Au). The calculations are plotted for five individual parameter samples of the posterior simulated for ten thousand events each. In addition, a 90\% credible interval band, estimated from ten such high-statistics random samples, is shown.
\label{PHENIX_v2}
}
\end{figure*}

\begin{figure*}
\includegraphics[scale=0.805]
{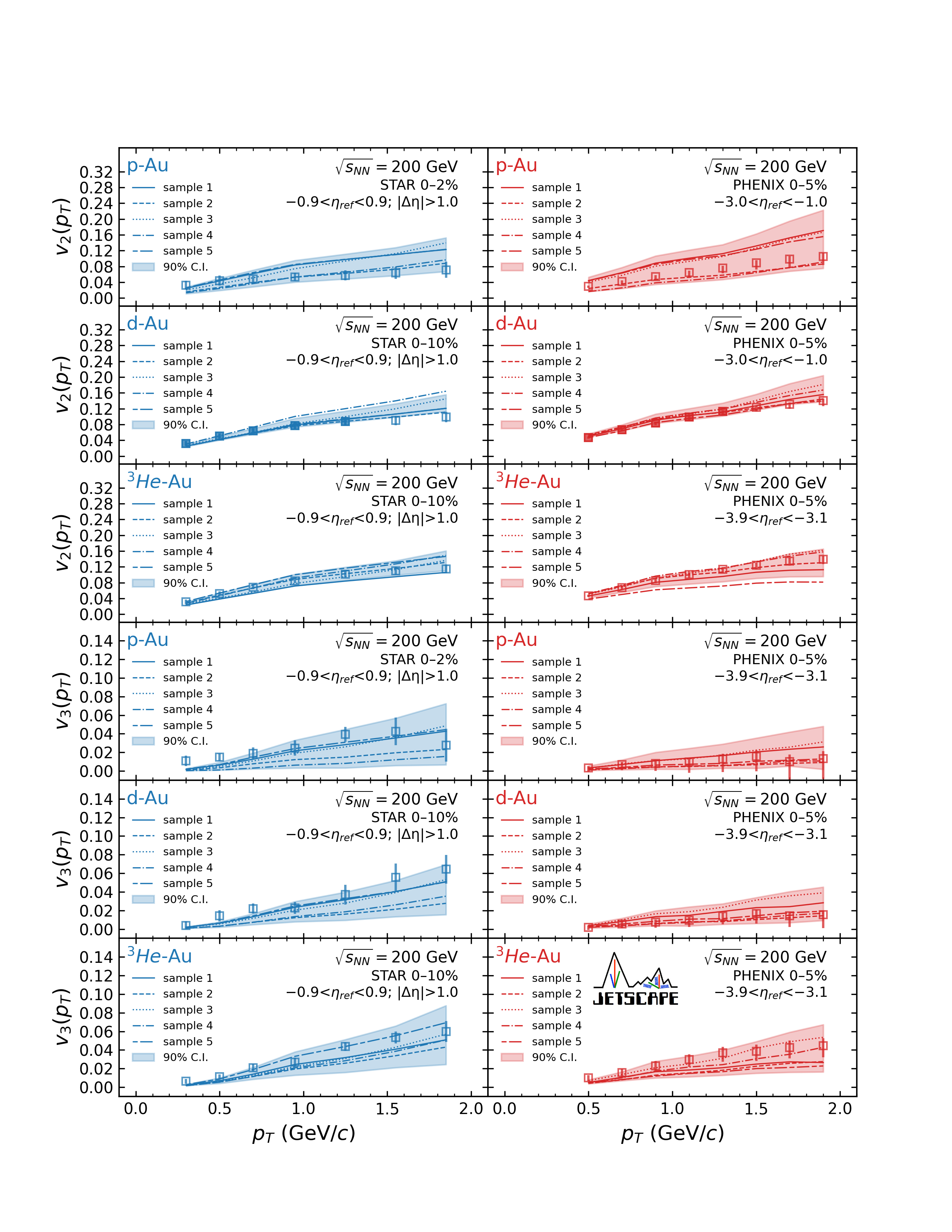}
\raggedright
\caption{The PHENIX (red) and STAR (blue) measurements for the $v_2$($p_T$) and $v_3$($p_T$) in p-Au, d-Au, and $^3$He-Au compared to a sampling of the posterior calibrated on the default dataset and calculated according to the analysis details of the two experiments. The calibration dataset includes the filled square data points (d-Au $v_2$ only) and does not include the open square data points. The calculations are plotted for five individual parameter samples of the posterior simulated for ten thousand events each. In addition, a 90\% credible interval band, estimated from ten such high-statistics random samples, is shown.
}
\label{combined_small_systems}
\end{figure*}

In the previous section we have seen the fits to experimental data and constraints on parameters they provide. In Fig.~\ref{fig-observable-posterior}, for example, we showed the fits to Au-Au and d-Au measurements, both explicitly used in the calibration and those not included. However, we need not restrict our model calculations to measurements in the calibration systems, or even to observables that have been measured at all. The calibrated model is a powerful tool that may be used to provide a wide variety of predictions, and we show in this section some of the results we can obtain with it. For all of the results in this Section we rely on the ten high-statistics (ten-thousand-event) posterior samples introduced in Sec.~\ref{sec:bayes:predictions} to represent the spread of the posterior for any given observable. The predictions from the first five of these samples\footnote{The parameter values for each of the five individually-plotted samples are provided in Appendix \ref{posterior_samples}.} are individually plotted in Figs.~\ref{fig_cent_selection},~\ref{fig_v2_reference},~\ref{fig_v3_reference},~\ref{fig_vn_reference},~\ref{PHENIX_dNdeta},~\ref{PHENIX_v2},~\ref{combined_small_systems},~\ref{fig_scaled_vn_reference}. In addition, the ten-sample ensemble is used to estimate--via linear interpolation--a 90\% credible interval band representative of the posterior, again shown in all the same Figures. This procedure introduces a relatively small uncertainty in the estimation of the 90\% band. However, it strikes a reliable compromise between showing the behavior of individual samples and showing a representation of the posterior without explicit emulation uncertainty--all clearly on the same plot--given limited visual and computational resources.

We choose to show predictions this way, rather than with a single MAP parameter set, which, though the best fit to calibration data, necessarily fails to capture the uncertainties quantified in the Bayesian analysis. We do not plot the statistical uncertainty on any of the individual sample predictions as it is always negligible compared to the width of the posterior itself. We underscore that using high-statistics model simulations means that the predictions are not explicitly affected by the emulator uncertainty, thus allowing us to explore observables that have larger emulator uncertainties. 

\subsection{\label{sec:longitudinal-structure}Observables in calibrated systems and the longitudinal structure of collectivity}

We first show the dependence of observables on the centrality selection method. When selecting centrality, experiments usually separate the kinematic region used to sort the events based on their multiplicity (and thereby centrality) and the region used to make the measurement itself. This separation is enforced to avoid correlation biases that arise if the two are done in overlapping regions. However, selection on any particular acceptance region may still bias measurements, especially when the multiplicity in the selection region is low. We show in Fig.~\ref{fig_cent_selection}, by means of individually-plotted posterior samples, as well as an estimated 90\% credible interval band, that the effect on the measured multiplicity is small in Au-Au but significant in d-Au when comparing two generic methods of centrality selection: by sorting at mid-rapidity (as in some STAR measurements) and at backward rapidity (as typically in PHENIX measurements, using the Beam-Beam Counter acceptance rapidity).\footnote{While the choices of generic ranges for the centrality selection and (as in the following paragraphs) the reference region of flow measurements do not reflect a particular experimental Collaboration's choices for all measurements, a roughly common range allows for easier comparison of our own calculations across systems and observables.} The multiplicity is shown to be larger in a rapidity region if the centrality selection was performed in the same or a nearby region, an effect long understood and demonstrated in minor apparent discrepancies between experimental measurements over the years, particularly in small systems.

Furthermore, we show in a series of figures (Figs.~\ref{fig_v2_reference}, \ref{fig_v3_reference}, \ref{fig_vn_reference}) the calibrated model's description of the longitudinal structure of collectivity---that is, the dependence of collective flow signals on the pseudo-rapidity and associated gaps between particles. We again plot the five simulated posterior samples, along with the 90\% credible interval ten-sample estimate, to represent the spread of the full posterior. In some cases, such as for the d-Au $v_2$($\eta$), the posteriors are wide relative to the differences between the probed observables, and thus overlap. However, even in these cases, the differences on a sample-by-sample basis are often significant. 

In measurements of azimuthal anisotropy the methods typically involve particles from two (or more) regions in the kinematic acceptance, often separated in pseudo-rapidity: the region of interest, where the measurement is quoted to have been made, and the reference region, used to pin down the orientation of the event, which of course varies event-by-event. In experiment, where the background due to jets, decays, or other correlations risks contaminating the flow signal (particularly in small systems), rapidity gaps are effective in removing these non-flow contributions. However, the use of rapidity gaps in these measurements also influences the flow signal itself, due to the non-trivial longitudinal structure of collectivity, leading to effects such as the decorrelation of the event planes in rapidity~\cite{Bozek:2010vz, Jia:2014ysa, Jia:2024xvl, Jia:2017kdq, Zhao:2022ugy, Ryu:2023bmx, CMS:2015xmx, ATLAS:2017rij, PHENIX:2021ubk, PHENIX:2022nht, Nie:2020trj, STAR:2022pfn}. 

In Fig.~\ref{fig_v2_reference} we show the  $v_2$($\eta$) in Au-Au and in d-Au for two generic reference flow regions commonly used in the PHENIX or STAR experiments. In Fig.~\ref{fig_v3_reference} we show similarly the  $v_3$($\eta$) in Au-Au and in d-Au for, like in Fig.~\ref{fig_v2_reference}, two generic reference flow regions commonly used in the PHENIX or STAR experiments. The calibrated model shows a strong dependence of the flow, both $v_2$($\eta$) and $v_3$($\eta$), on the rapidity region used as a reference, such that the signal increases with the proximity of the reference region. The effect is discernible even when the posterior is relatively wide (d-Au). Indeed, in the small system the difference between the average values of our simulated samples is larger. The effect is also larger for $v_3$ than for $v_2$ in both systems.

We see a similar trend in Fig.~\ref{fig_vn_reference}, where we show the dependence of the mid-rapidity value of $v_2$ and $v_3$ as a function of the reference rapidity region, in the full range from -6 to 6 units of pseudo-rapidity, in bins of 0.5. The dependence is again stronger in d-Au than in Au-Au and in $v_3$ than in $v_2$. We include a slightly different version of the plot in Appendix \ref{scaled_reference_flow} to better highlight the effect, where we show the fractional value of the observable scaled by the mid-rapidity-reference benchmark value.

In this subsection we have examined a set of results showing the non-trivial dependence of flow measurements on the full 3D structure of the collisions. In particular, we reiterate that a 3D analysis allows us to avoid the potential ambiguity faced by 2D analyses when attempting to describe flow measurements that depend on the relative rapidity of reference particles and particles of interest. In much the same way, the 3D description is crucial in describing asymmetric small system collisions, for which we make some more wide-ranging predictions in the next subsection.

\subsection{\label{sec:small_systems}Extending to p-Au and $^3$He-Au}

In this subsection we show predictions with the calibrated model for a number of small system measurements by STAR and PHENIX and compare them with data. Having calibrated the model on both a large and a small system, we gauge whether the model has gained sufficient insight to describe differences among small systems, namely p-Au, d-Au, and $^3$He-Au. We compare with two categories of observables, first with the rapidity-dependent measurements from PHENIX of the dN/d$\eta$ and $v_2$($\eta$) in the three systems and then with the $p_T$-dependent measurements from both STAR and PHENIX of $v_2$ and $v_3$ in those same three systems. While the two experiments measured nominally the same observables, they used distinct kinematic cuts, centrality selection, and reference regions. We see how much of these differences the calibrated model can account for.

In Fig.~\ref{PHENIX_dNdeta} we see the comparison with PHENIX measurements of dN/d$\eta$. All of the d-Au data points were used in calibration, and they are plotted in green. The data for the remaining systems are plotted in olive. The posterior as a whole (represented again by the same five high-statistics samples and the ten-sample estimate of the 90\% band in use throughout our results) is consistent with data in all three systems. Individual samples, however, especially in p-Au and $^3$He-Au, can still be far from the data. Similarly in Fig.~\ref{PHENIX_v2} we see that the comparison between the calibrated model and the $v_2$($\eta$) in the three systems is good. Here, only the forward and mid-rapidity points from the d-Au data were used in calibration, due to experimental uncertainties from the non-flow contribution---larger in small systems---in the backward region (proximity and overlap to the reference and centrality selection region would enhance the non-flow).  Nevertheless, we include those points in this comparison figure, as well as the points in that region measured in the other systems. We expect non-flow contribution in the data to enhance the flow signal in that region, an effect that should decrease with system size. In accordance with that expectation, we find the model calculation underestimating the data (though still consistent within the large experimental uncertainties)  only in the backward region of the p-Au measurement---the place of maximum non-flow---and in agreement with the data everywhere else. 

That said, we are interested in comparing with small system data from multiple experiments, and both STAR and PHENIX have published $p_T$-dependent $v_n$ measurements in these three systems~\cite{PHENIX:2018lia, STAR:2022pfn}. We note that the large ``discrepancy'' between the STAR and PHENIX $v_3$ measurements in d-Au has been a particular source of debate~\cite{STAR:2022pfn}. An earlier prediction of these two measurements using a hand-tuned version of our model was shown in~\cite{Zhao:2022ugy}, highlighting that the discrepancy can largely be accounted for by the experimental details in each measurement. In our results we note that the posterior (in the fifth row of Fig.~\ref{combined_small_systems}) is largely consistent with both measurements, though it more closely describes PHENIX data. 

As we can see in Fig.~\ref{combined_small_systems}, the posterior is indeed overall consistent with the data from STAR and PHENIX across all measurements. It is relatively wide compared to experimental uncertainties, especially for the $v_2$ in p-Au from both experiments, the $v_2$ in $^3$He-Au from PHENIX, and the $v_3$ in p-Au from STAR. 
Even though none of these $p_T$-dependent datasets in the additional small systems is clearly at odds with the predictions, the width of the posteriors holds promise for obtaining additional 
constraints by including the datasets in future calibrations.

\section{\label{sec:summary} Summary   }

In this work we presented a comprehensive (3+1)D Bayesian calibration of high-energy nuclear collision dynamics at the RHIC top energy, jointly analyzing large (Au–Au) and small (d–Au, with extensions to p–Au and $^3$He–Au) systems. Leveraging a selection of over two decades of measurements from PHENIX, STAR, PHOBOS, and BRAHMS, the study constrains both the longitudinal structure of the initial state and the transport properties of the QGP. The model combines a 3D Glauber initial condition with rapidity-dependent energy deposition and global energy conservation, relativistic viscous hydrodynamics using MUSIC, and a hadronic scattering phase using UrQMD.

The Bayesian analysis features emulation-facilitated model-to-data comparison across the prior range of a 20-dimensional parameter space. A broad data set is used to calibrate the model, as displayed in Table \ref{tab:table2}, including $\eta$-differential, $p_T$-differential, and integrated observables from the full range in rapidity and across multiple collision systems. 
The calibration dataset includes all observables that the model can reasonably describe and which can be reliably emulated. We use an additional set of measurements to compare with our calibrated model. The exclusion of the measurements in this additional dataset from calibration is due, in most cases, to the difficulty in emulating them. Recent developments in machine learning methods present great potential for enhancing emulator performance with precise uncertainty quantification. Recent work~\cite{li2025additive, ji2024conglomerate, ji2024graphical} has demonstrated that such methods, when appropriately implemented, can provide significant improvements for emulating nuclear collision observables. Integrating these methods in our Bayesian workflow in future calibrations could allow for extending the calibration dataset to many more observables, and to more fully exploiting those already included. 

We find that compared to mid-rapidity data alone, rapidity-dependent measurements provide additional constraints on the longitudinal profile of the initial state energy deposition, the transport coefficients, and the resulting longitudinal structure of correlations. In particular, their inclusion shows a preference for a larger specific shear and bulk viscosity, as shown in Fig.~\ref{fig_visc_full_midrap}.

The individual system calibrations show largely consistent constraints but do indicate some degree of model tension between Au-Au and d-Au. Remarkably, as shown in Fig.~\ref{fig_dAu_obs_post_mini}, the posterior from a calibration using only d-Au observables gives a reasonably good description of the Au-Au data. On the other hand, the Au-Au calibration, which includes the majority of all data points and a greater variety of observables, struggles to describe the $dN/d\eta$ in d-Au. These and other sources of tension are addressed in the analysis by performing additional calibrations using targeted subsets of the default dataset as well as by examining the local sensitivity of observables to changes in the parameters.

The decorrelation of event planes and differences in the strength of correlation observables as a function of rapidity (and relative rapidity) have been a topic of recent theoretical and experimental exploration ~\cite{Jia:2024xvl, Ryu:2023bmx, Jahan:2025cbp, PHENIX:2021ubk, STAR:2022pfn}. Using high-statistics non-emulated posterior samples, we demonstrate throughout Section \ref{sec:longitudinal-structure} the strong dependence of flow coefficients on the rapidity of both the region of interest and the reference, as well as the impact of the centrality selection region, essential considerations for consistent comparisons with experiments. The 3D analysis allows us to access this information embedded in rapidity-dependent data. It also enables direct comparisons of experimental measurements which differ non-trivially due to their incorporation of forward/backward information in the form of the relative rapidity and the centrality selection. We show the rapidity-dependent effects to be generally stronger in d-Au than in Au-Au, as perhaps best illustrated in Fig.~\ref{fig_scaled_vn_reference}. 

We make further predictions with the calibrated model (again without the need for explicit emulation) for additional smaller systems (p–Au, $^3$He–Au). We find a relatively good match across systems, demonstrating the capability of the model to discern the differences among small systems. In particular, we find the posterior to be consistent with rapidity- and $p_T$-differential PHENIX and STAR data in those systems, as shown in Fig. \ref{combined_small_systems}. We see also a reasonable agreement in, notably, d-Au, where the differences between the STAR and PHENIX measurements have long sparked fruitful discussion about the origin of collectivity in small systems.

While the results in this work are derived from a comprehensive modeling of heavy-ion collisions and systematic comparison with experimental data, they may be further extended in several directions. In terms of theoretical modeling, uncertainties related to the choice of initial state and pre-equilibrium models, of viscous corrections at particlization, or the parametrization of viscosity coefficients may be explored. As previously mentioned, improvements in model emulation can allow for more extensive comparisons and more precise and accurate constraints. Further, the modeling can be extended to include hard processes such that higher transverse momentum data can be compared with in conjunction to the low momentum datasets. Regarding experimental data, the model shows promise in describing measurements at LHC energies and the calibration could additionally benefit from comparison with the abundant and precise data collected at the LHC. Furthermore, the correlations present in experimental uncertainties could be treated more realistically, either through data-driven estimation techniques or by direct experimental input.

The results underscore that (3+1)D calibration with rapidity-dependent data is sensitive to initial-state longitudinal structure and QGP transport properties. Our calibrated model provides a tuned baseline for studying the longitudinal structure of correlations, and guiding future measurements, both at RHIC and LHC, that can further improve constraints on high-energy nuclear collision models. It also serves as a calibrated 3D background for jet energy loss studies, paving the way for joint hard-soft modeling of heavy-ion collisions. 

The posterior chains used to obtain results from the four main calibration datasets (default, mid-rapidity, Au-Au, and d-Au) discussed in this analysis are made publicly available for community use~\cite{mankolli_2026_18602345}.

\addtocontents{toc}{\protect\setcounter{tocdepth}{-1}}
\begin{acknowledgments}
This work was supported in part by the National Science Foundation (NSF) within the framework of the JETSCAPE collaboration, under grant number OAC-2004571 (CSSI:X-SCAPE) and OAC-2514008 (CSSI:C-SCAPE). It was also supported under NSF awards PHY-2111568 and PHY-2413003 (R.J.F., M.K., C.P.\ and A.S.), and award DMS-2316012 (S.M.); it was supported in part by the US Department of Energy, Office of Science, Office of Nuclear Physics under grant/contract numbers \rm{DE-AC02-05CH11231} (X.-N.W. and W.Z.), \rm{DE-AC52-07NA27344} (D.A.H. and R.A.S.), \rm{DE-SC0013460} (A.M., I.S., R.Da. and R.Do.), \rm{DE-SC0021969} (C.Sh.\ and W.Z.), \rm{DE-SC0024232} (C.Sh.\ and H.R.), \rm{DE-SC0012704} (B.S.), \rm{DE-FG02-92ER40713} (J.H.P.), \rm{DE-FG02-05ER41367} (S.A.B.), \rm{DE-SC0024660} (R.K.E), \rm{DE-SC0024347} (J.-F.P., G.S.R. and M.S.) and \rm{DE-FG05-92ER40712} (J.V., S.T. and A.Ma.). The work was also supported in part by the National Science Foundation of China (NSFC) under grant numbers 11935007, 11861131009 and 11890714 (X.-N.W.), by the Natural Sciences and Engineering Research Council of Canada under reference numbers \rm{SAPIN-2020-00048}, \rm{SAPIN-2024-00026}, and \rm{SAPIN-2023-00029} (C.G., S.J. and G.V.), by the University of Regina President's Tri-Agency Grant Support Program (G.V.), by the Canada Research Chair program (G.V. and A.K.) reference number CRC-2022-00146, by JSPS KAKENHI Grant Numbers~22K14041 and 25K07303 (Y.T.), by the Research Council of Finland, the Centre of Excellence in Quark Matter, by the European Research Council (ERC, grant agreements No. ERC-2023-101123801 GlueSatLight and No. ERC-2018-ADG-835105 YoctoLHC (H.R.) and No. ERC-2018-ADG-835105 YoctoLHC (I.S.)), and by the S\~{a}o Paulo Research Foundation (FAPESP) under projects 2017/05685-2, 2018/24720-6, and 2023/13749-1 (M.L.). C.Sh., J.-F.P., and R.K.E. acknowledge a DOE Office of Science Early Career Award. The content of this article does not reflect the official opinion of the European Union and responsibility for the information and views expressed therein lies entirely with the authors. 

This work used the Anvil HPC facility at Purdue University and Stampede2 at Texas Advanced Computing Center, through allocations PHY210068 and PHY230071 from the Advanced Cyberinfrastructure Coordination Ecosystem: Services \& Support (ACCESS) program~\cite{access_ref}, which is supported by U.S. National Science Foundation grants \#2138259, \#2138286, \#2138307, \#2137603, and \#2138296. This research also used services provided by the OSG Consortium~\cite{osg07,osg09,osg_url,osg_url2}, which is supported by the National Science Foundation awards \#2030508 and \#2323298.

\end{acknowledgments}
\addtocontents{toc}{\protect\setcounter{tocdepth}{3}}

\appendix

\section{\label{PCA}Principal Components}
As described in the main text, our emulators are trained to reproduce the values of a number of principal components of the linear decomposition of the model calculations. Each time the emulator is invoked to make a prediction, the principal components  are transformed back to observable space. 

A subset of principal components collectively describe a well-defined finite portion of the variance in the training model calculations. In principle, the inclusion of additional principal components will always cover a greater portion of the variance. However, we find that the returns diminish quickly, and only a few principal components are necessary to describe an overwhelming portion of the variance. Furthermore, the computational benefits of using PCA at all are reduced if too many principal components  are used. We therefore find a balance between these opposing effects by considering the portion of variance covered, the effects on emulation performance in plots like the ones in Appendix \ref{emulator_validation}, and closure tests in order to select the appropriate number of principal components. Ultimately we use 10 PCs for each of the Au-Au and d-Au emulators, accounting for more than 99\% of the variance of the training calculations in both systems. 

We ``truncate'' the rest of the principal components  and this leads to an uncertainty in the predictions due to this truncation that is not already accounted for by the emulator. We estimate the truncation error using a separate white noise kernel and add it to the emulator covariance to explicitly account for the uncertainty~\cite{Bernhard:2018hnz}.

\begin{figure}
\includegraphics[scale=0.565]{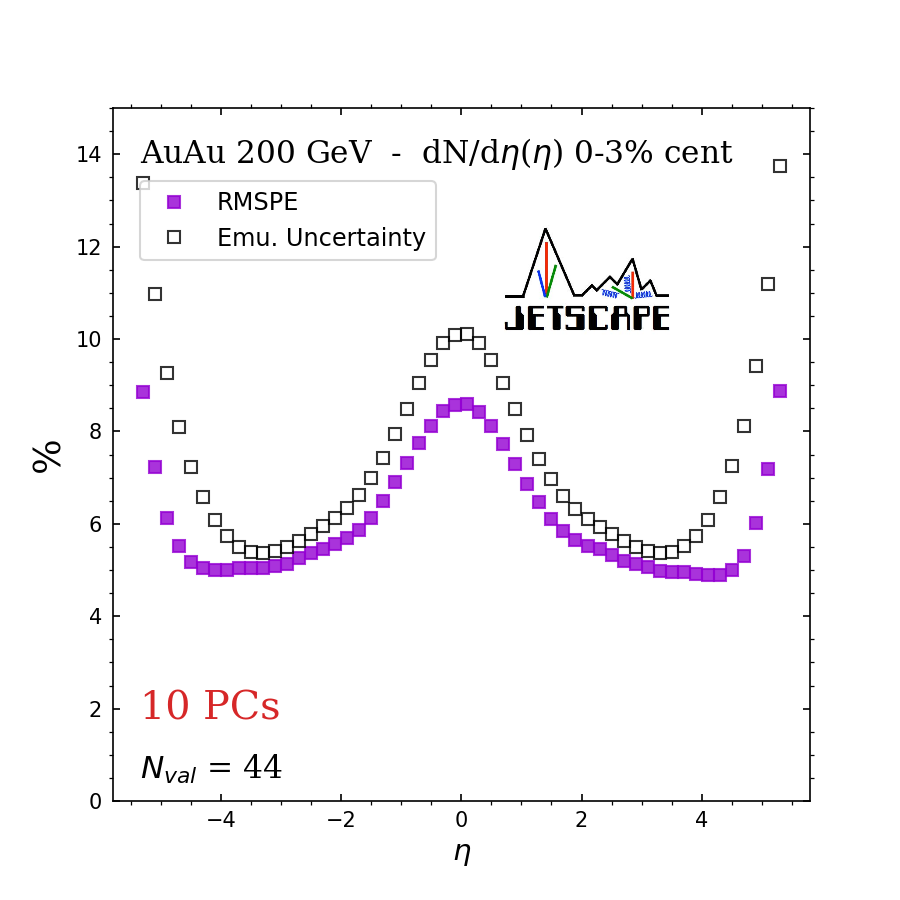}
\centering
\caption{The root mean square percent error (RMSPE, filled squares) in percent between model calculations and emulator predictions averaged across 44 validation points for each $\eta$ bin of the Au-Au dN/d$\eta$ in 0-3\% central collisions as in the measurement by PHOBOS. The empty squares show the emulator's estimate of the uncertainty in percent also averaged over all design points for each bin.
}
\label{emu_val_dNdeta}
\end{figure}

\begin{figure}
\includegraphics[scale=0.565]{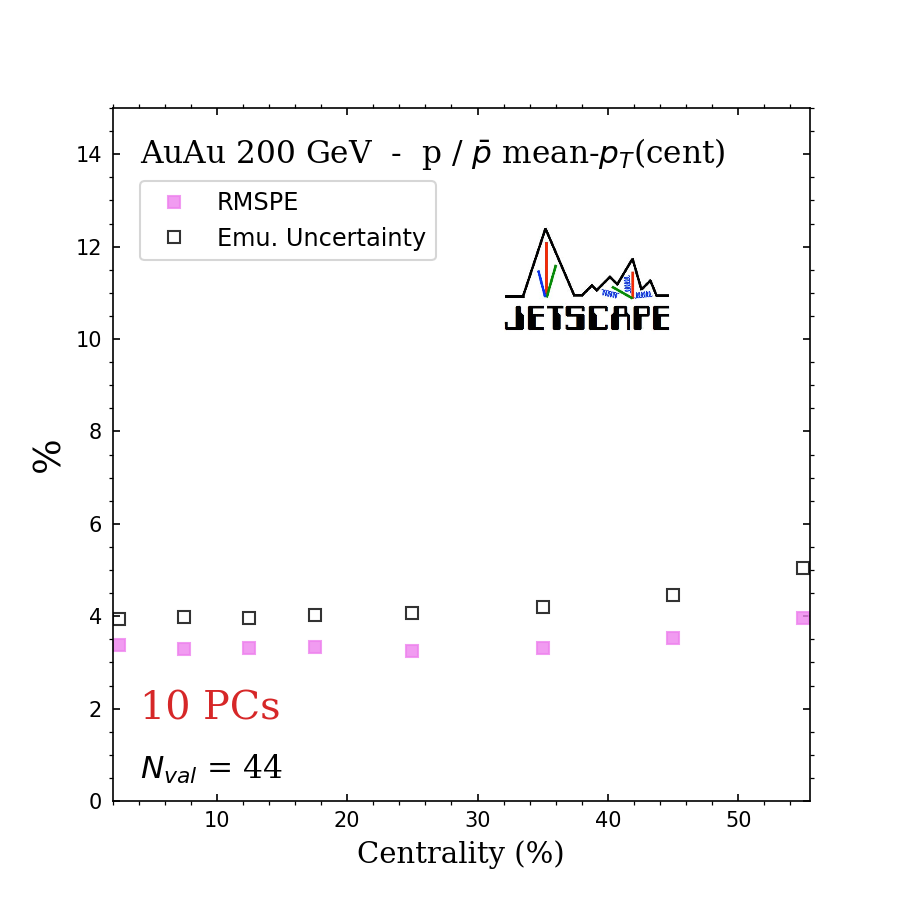}
\centering
\caption{The root mean square error (RMSPE, filled squares) in percent between model calculations and emulator predictions averaged across 44 validation points for each centrality bin of the Au-Au proton $\langle p_T \rangle$ as in the measurement by PHENIX. The empty squares show the emulator's estimate of the uncertainty in percent also averaged over all design points for each bin.
}
\label{emu_val_meanpT}
\end{figure}

Lastly, before and during PCA, we engage in ``post-processing'' of the model calculations. This includes transforming some of the observables by taking the logarithm of the multiplicity and transverse energy, as described in the main text (Section~\ref{sec:bayes:emulator}), and transforming all observables to have zero mean and unit variance across the training dataset. In addition, we examine the PCA distribution to identify and remove outliers in 2D plots of any pairs of principal components in the training dataset, since PCA yields linearly uncorrelated principal components. This step was shown to reduce emulation uncertainty as well as lead to improved closure. In this process of design curation we remove about 2\% of the training design points from each system's emulator training dataset.

\section{\label{emulator_validation}Emulator validation}

\begin{figure*}[!htbp]
\includegraphics[scale=0.6]{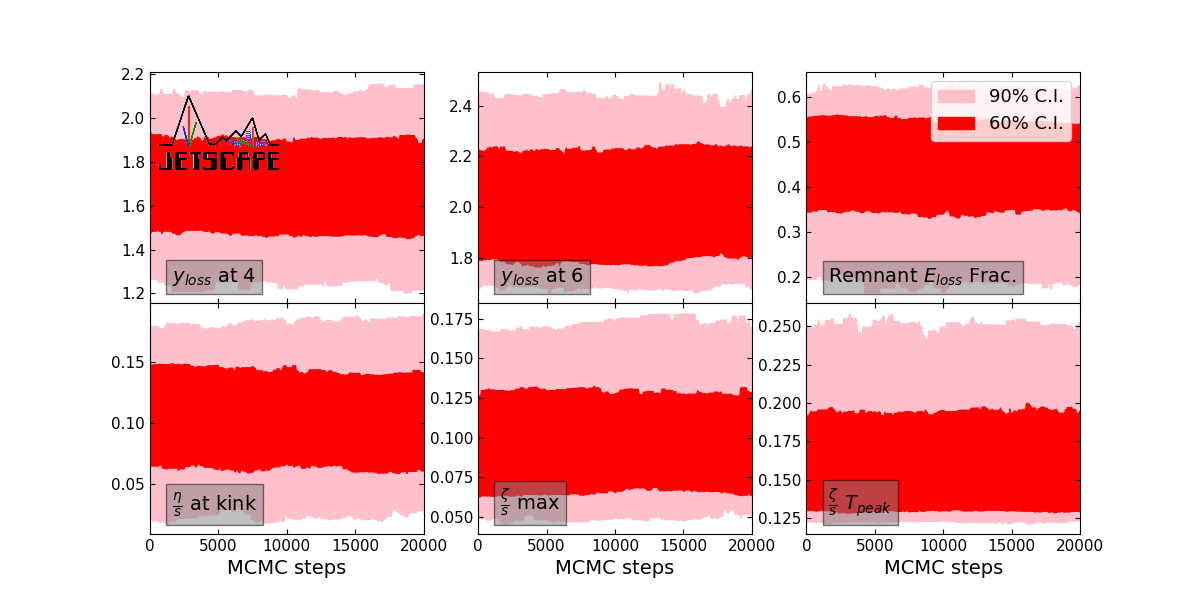}
\centering
\caption{The distributions of six sample parameters among walkers at particular steps of the MCMC chain following the burn-in phase. All the distributions remain stable after---and in some cases converging well before---the conclusion of the burn-in.
}
\label{convergence}
\end{figure*}

As a prerequisite to this analysis, we perform a thorough validation of the model emulation. We examine a number of quantifiers such as the shape of the prediction distributions around the mean, the dependence of the residuals on the magnitude of the mean, and the root mean square error (RMSPE) between the model predictions and the model calculations in the validation set. We use 44 of the design points as our validation set, keeping them separate from the training designs. We also evaluate how the RMSPE compares with the emulator's estimate of its uncertainty.

We use these quantifiers, and in particular the RMSPE, to determine whether the emulator performance is sufficient for observables to be included in the analysis. We perform an observable-by-observable assessment (where we define an ``observable" as all the bins of a differential observable---in $\eta$ or $p_T$---in a given centrality class, or, for integrated observables, all the centrality bins). While we do make some choices to exclude from the analysis certain experimental centrality or $p_T$ bins "within" an observable due to uncertainties related to the model scope, as described in Section~\ref{sec:exp_data}, all decisions stemming from the emulation performance are done at the level of the observable as a whole, and we do not in general discriminate between individual bins of a given observable.

We make these assessments after having taken several steps to optimize the emulator performance, such as the discretization of the viscosity parameters on a grid of temperature, the log-transformation of the multiplicity and transverse energy observables, and the choice of the number of principal components, among others. We pay particular attention to ensure that the emulator's estimate of its uncertainty does not underestimate significantly the uncertainty we estimate ``empirically'' as the RMSPE obtained from our validation points. We aim to keep the relative emulation uncertainty to no more than 15-20\% as a bin-average across each observable, although for most observables it is significantly smaller.

In subsection \ref{sec:bayes:emulator} we mention the discretized grid used to redefine the specific shear and bulk viscosity parameters for emulator training. We use ten uniformly distributed points across the prior range to serve as the new parameters replacing the original four parameters determining the functional form. The equivalent replacement is done for both the shear and bulk viscosity, totaling 20 new parameters used in emulator training due to viscosity discretization in place of the original eight. We find that an even finer grid, whether enhancing the density in a localized temperature region or uniform across the full range, does not significantly improve the emulation beyond this point.

In Figs.~\ref{emu_val_dNdeta} and \ref{emu_val_meanpT} we show the validation-set-average RMSPE for two sample observables (the Au-Au multiplicity for a central bin from PHOBOS, and the mean $p_T$ of protons as a function of centrality from PHENIX) as well as the emulator's estimate of the uncertainty across the range of pseudo-rapidity and centrality bins, respectively. We see that the dN/d$\eta$ RMSPE depends on pseudo-rapidity, being largest at mid-rapidity as well as at large values of $|\eta|$, and that it approximates decently well the emulator error across the range. Meanwhile the RMSPE of the integrated proton mean $p_T$ has little dependence on centrality, is in general low (below 5\%), and tracks well with the emulator uncertainty. The mean $p_T$ is an example of an observable that we are able to emulate well in the analysis, while some $\eta$ bins of the multiplicity, particularly for more peripheral centrality classes, or without optimization procedures like taking the logarithm, are more uncertain.

\renewcommand{\thesection}{\Alph{section}}

\section{\label{mcmc_convergence}MCMC convergence}

In order to ensure the convergence of the MCMC algorithm to the posterior distribution, we use a burn-in phase of 20,000 steps. We verify the convergence by showing in Fig.~\ref{convergence} the distribution of individual select parameters within the MCMC chain as a function of steps undertaken \textit{after the burn-in has concluded}. The bands at any given step indicate the distribution of the walkers for the respective parameters at that step. We can see that the distributions of the parameters have reached sufficient convergence by the end of the burn-in phase and remain stable as we sample the posterior.

\begin{figure}
\includegraphics[scale=0.58]{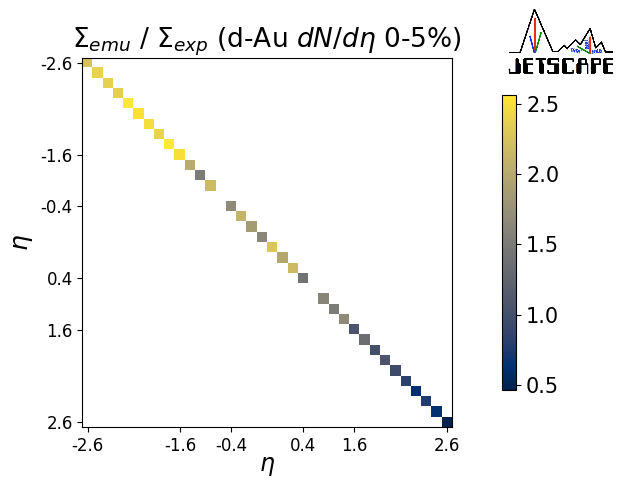}
\includegraphics[scale=0.58]{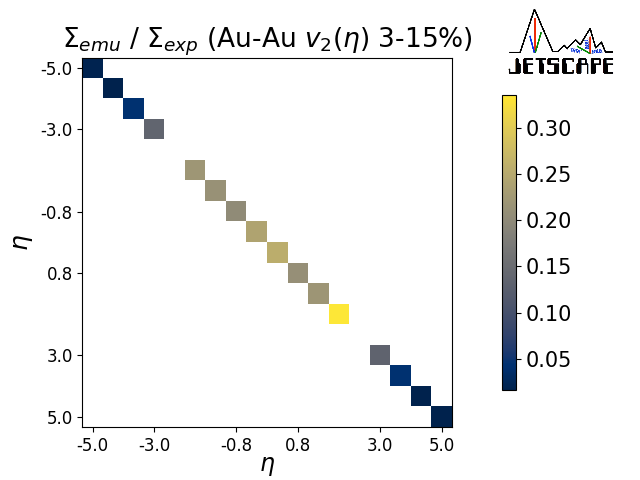}
\includegraphics[scale=0.58]{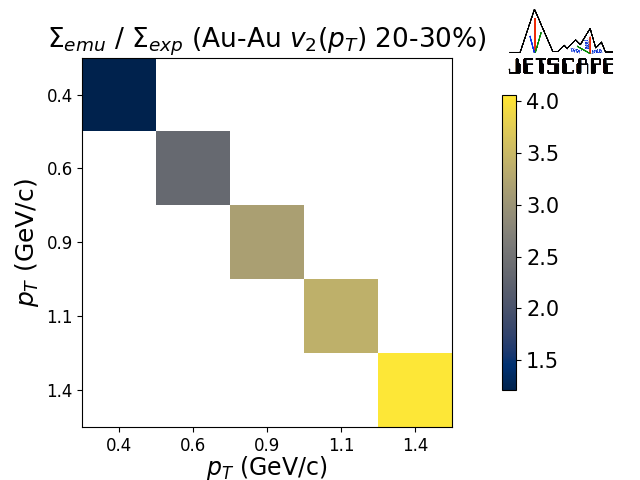}
\centering
\caption{The ratio between the diagonal of the emulation uncertainty matrix and the experimental (assumed diagonal) covariance for the 0-5\% centrality class of the d-Au multiplicity from PHENIX (top), the 3-15\% centrality class of the Au-Au $v_2(\eta)$ from PHOBOS (middle), and the 20-30\% centrality class of the Au-Au $v_2(p_T)$ from PHENIX (bottom), highlighting three different cases of the relative contributions from each source of uncertainty.
}
\label{covariance_ratio}
\end{figure}

\begin{figure}
\includegraphics[scale=0.58]{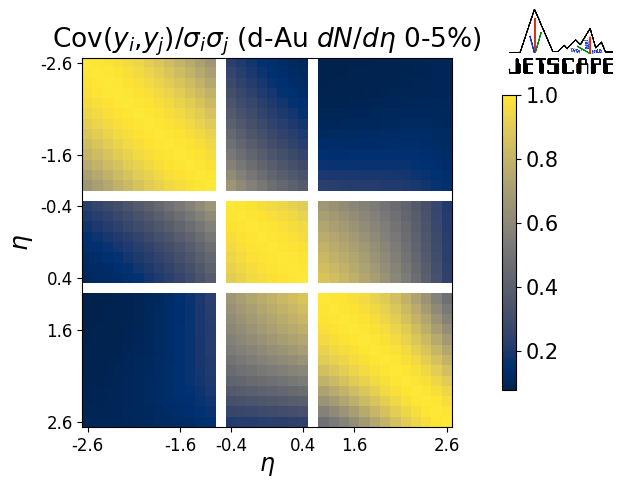}
\includegraphics[scale=0.58]{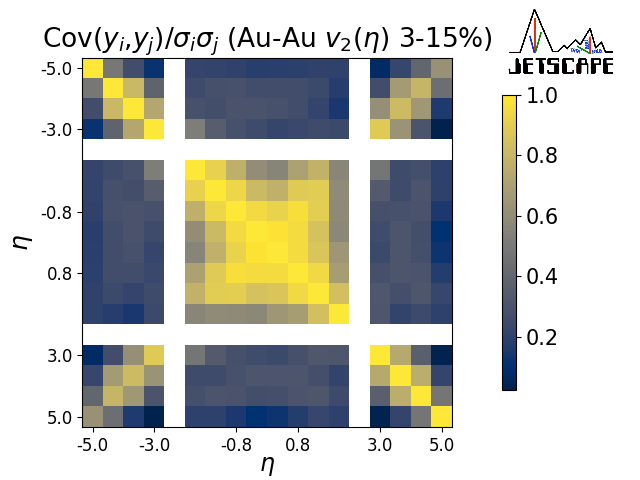}
\includegraphics[scale=0.58]{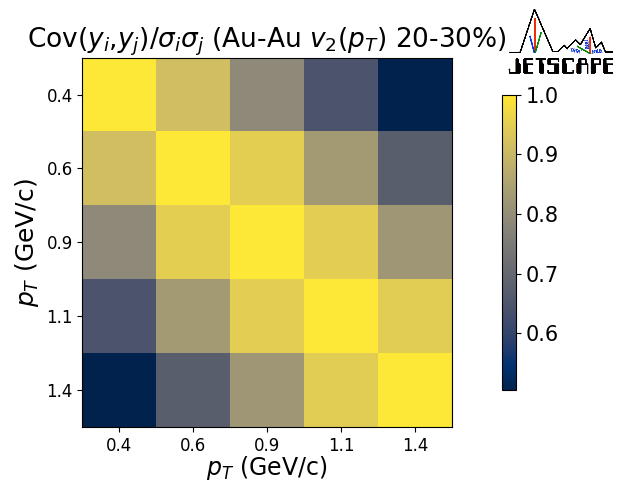}
\centering
\caption{The emulator uncertainty correlation matrix, including correlations between different $\eta$ and $p_T$ bins for the same observables shown in Fig. \ref{covariance_ratio}: the 0-5\% centrality class of the d-Au multiplicity from PHENIX (top), the 3-15\% centrality class of the Au-Au $v_2(\eta)$ from PHOBOS (middle), and the 20-30\% centrality class of the Au-Au $v_2(p_T)$ from PHENIX (bottom).
}
\label{covariance_emu}
\end{figure}

\begin{figure*}[htbp]
\includegraphics[scale=0.53]{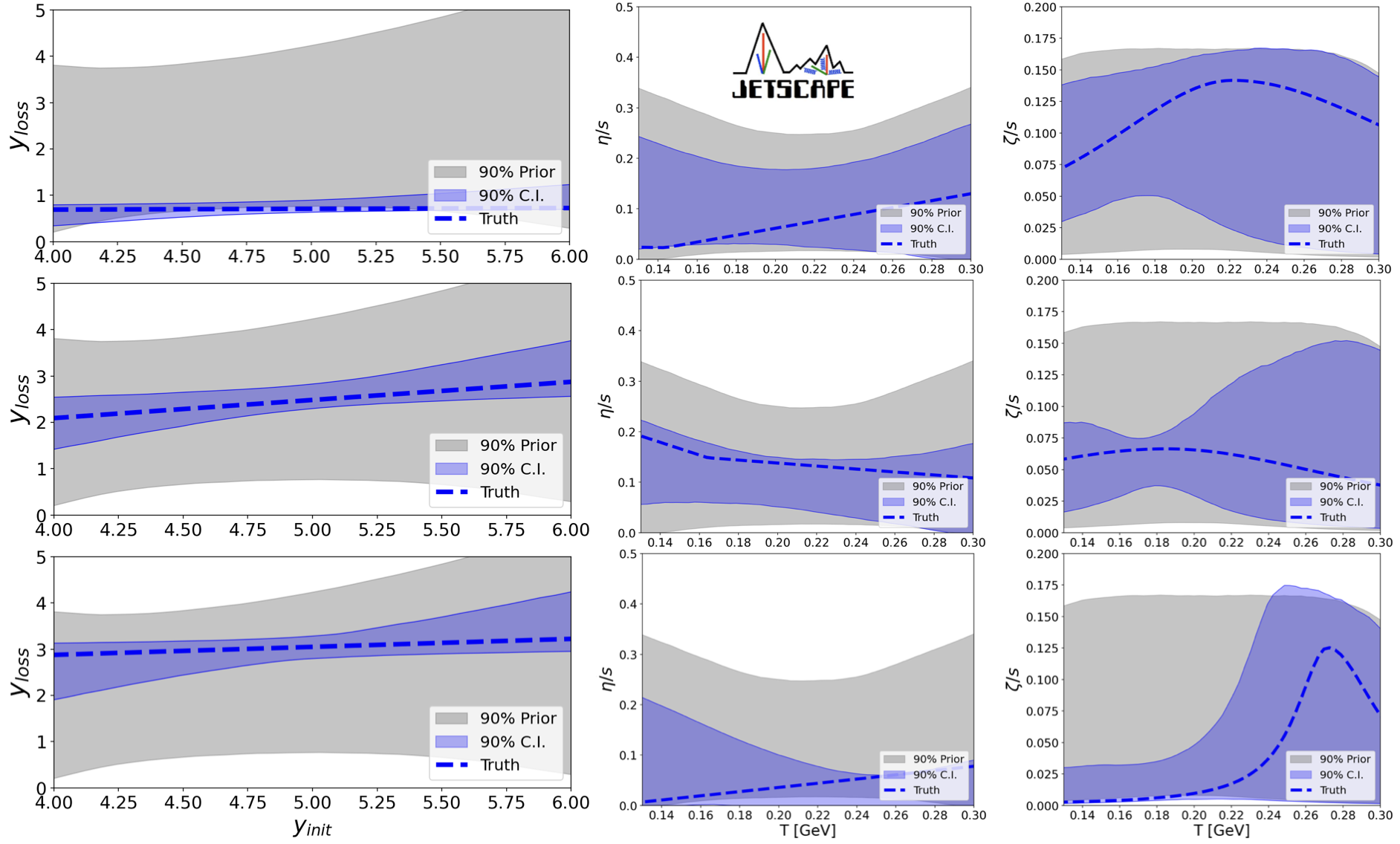}
\centering
\caption{The closure test posteriors for the rapidity loss as a function of incoming beam rapidity, and the shear and bulk viscosity as a function of temperature, from runs with the model calculations of 3 validation parameter sets used as pseudo-data.
}
\label{closure-tests}
\end{figure*}

\section{\label{covariance_matrix}Covariance matrix}
As described in Section \ref{sec:bayes} the covariance matrix contains, in principle, information about the uncertainties from both experimental data and the model. For an aggregation of the modeling uncertainty, we rely on the emulator's estimation, including for correlations between observables. The experimental uncertainties that enter into the diagonal of the covariance matrix come directly from experimental publications (for select few observables we adjust these values as described in Section~\ref{sec:exp_data} to account for methodological discrepancies). 
Previous works have found that prescriptions for the non-diagonal components of the covariance matrix can have counterintuitive effects~\cite{Soltz:2024gkm}.
We leave to future work a systematic examination of the benefits and pitfalls of various approximation schemes of experimental uncertainty correlations across rapidity, transverse momentum, centrality, and more.

As some examples of the relative contributions to the total covariance matrix from emulation and experimental uncertainties in our analysis, we show in Fig.~\ref{covariance_ratio} the relevant diagonal quantities for three observables. In the central bin of the d-Au dN/d$\eta$ from PHENIX we have an example of an observable for which the emulation or experimental uncertainties are relatively balanced. The 3-15\% centrality bin of the Au-Au $v_2(\eta)$ from PHOBOS is an example of an observable where the experimental uncertainties dominate, while the 20-30\% Au-Au $v_2(p_T)$ from PHENIX shows an observable dominated by the emulation uncertainty. We also show the strength of correlations in the emulation predictions for the same three observables in Fig.~\ref{covariance_emu}. Due to significant breaks in the continuity of the pseudo-rapidity binning, we plot the $\eta$-differential observables with two symmetric white gaps to indicate these breaks. The gaps are not drawn to scale and serve only as reminders. Within the sections that remain ``together", the bins are uniform in width. All three observables discussed in this section are included also in the aggregator plot in Fig.~\ref{covariance_all_points}, where the relative theoretical and experimental contributions to uncertainty are shown for all observables used in calibration.

\section{\label{closure_tests}Closure test posteriors}

In this appendix we show, as seen in Fig.~\ref{closure-tests}, the posteriors for select parameters from three closure-test parameter sets and compare them with the truth values.  We see in general good closure across parameters, as well as clear differences in the constraining power of pseudo-data between different parameters. We note that we see the best closure for the most sensitive parameters, in particular the rapidity loss parameters, while the viscosity parameters often show weaker constraints. Nevertheless, even the constraints on viscosity are significant, and we furthermore see additional indications of sensitivity in closure tests (not shown) where only a single parameter from a set (the width of the bulk viscosity function, for example) is varied. The posteriors in such emulator-assisted tests are responsive to these small changes in a single parameter.

Since we use the true experimental uncertainties for the pseudo-data at any given closure point, we do, in some cases, have a smaller pseudo-data uncertainty than the corresponding statistical uncertainty of the mean model calculations for that closure point. While the pseudo-data uncertainty may be arbitrarily small in general, if the statistical model uncertainty in generating the predictions for it is sufficiently large, closure tests with small uncertainties fail to constrain the model. Though we do not strictly enforce this requirement in our closure tests, we do restrict the analysis as a whole to observables we can compute with relatively low statistical uncertainty.

\section{\label{experimental_methods}A note on the experimental data}

In this appendix, we point out some additional technical details regarding accessing the data reported by select experimental papers and the decisions we made about how to include them in the analysis.

The BRAHMS dN/d$\eta$ measurement in Au-Au reported only statistical uncertainties in HEPData, but showed ``total'' uncertainties for three representative positive rapidity bins in a table in the paper~\cite{BRAHMS:2001llo}. Given that the total uncertainties were much larger than the statistical ones reported, we used the total uncertainties from the three bins to interpolate the values of the uncertainty for the fine-binned measurement. We reflected the uncertainty values to populate the negative-rapidity bins. This was the case for both the middle and large rapidity measurements reported in that paper and which we used to compare with our calibrated model. For this reason,
as well as some apparently inconsistent binning between the report on HEPData and the points in the paper, we did not include the measurement in the calibration set.

In the case of the STAR measurement of $v_2$($\eta$) in Au-Au~\cite{STAR:2004jwm}, the data, which was reported on the experiment website, did not show a breakdown of statistical and systematic uncertainties, and did not specify any details about the reported uncertainties. In this and any such cases, we used the reported uncertainties as one standard deviation total uncertainties.

In the case of the PHOBOS measurement of that same observable in Au-Au~\cite{PHOBOS:2004vcu}, the PHOBOS website reported ``90\% confidence intervals'' so those were accordingly scaled to obtain the 1-$\sigma$ values. For some of the bins, the uncertainties were asymmetric. In any such cases, we chose to use the maximum value between the upper and lower bounds as the $\sigma$ for a symmetric error, since we do not in general incorporate asymmetric uncertainties in the Bayesian analysis. We do, however, use asymmetric uncertainties as reported by experiment in plots of observables with which we only compare. 

The measurement of $v_2$($\eta$) by PHOBOS~\cite{PHOBOS:2004vcu} presented an additional challenge. For a few points there appeared to be a significant discrepancy between the plot in the paper and the reported observable value on the PHOBOS website. For all bins in that analysis, we decided to use the points from the plot in the paper by digitally extracting them. The reported uncertainties between plot and website appeared consistent, so we used them as reported. 

We digitized data from a plot for at least one additional measurement: the integrated STAR $v_3$ in Au-Au~\cite{STAR:2013qio}, though in this case it was the uncertainties that showed a possible discrepancy. In the HEPData entry only statistical uncertainties were reported; however, the plotted error bars in the paper indicated the systematic uncertainties were significant. We again gave precedence to the plot, and used the uncertainties in the figure as the total uncertainties.

We made efforts over the course of this study to reach out to experimental collaborators, including from all four of the long-serving experiments, as well as from sPHENIX, regarding questions of methodology and data reporting. We received fruitful feedback and guidance on many of the analyses we made use of.

\begin{figure*}
\includegraphics[scale=0.46]
{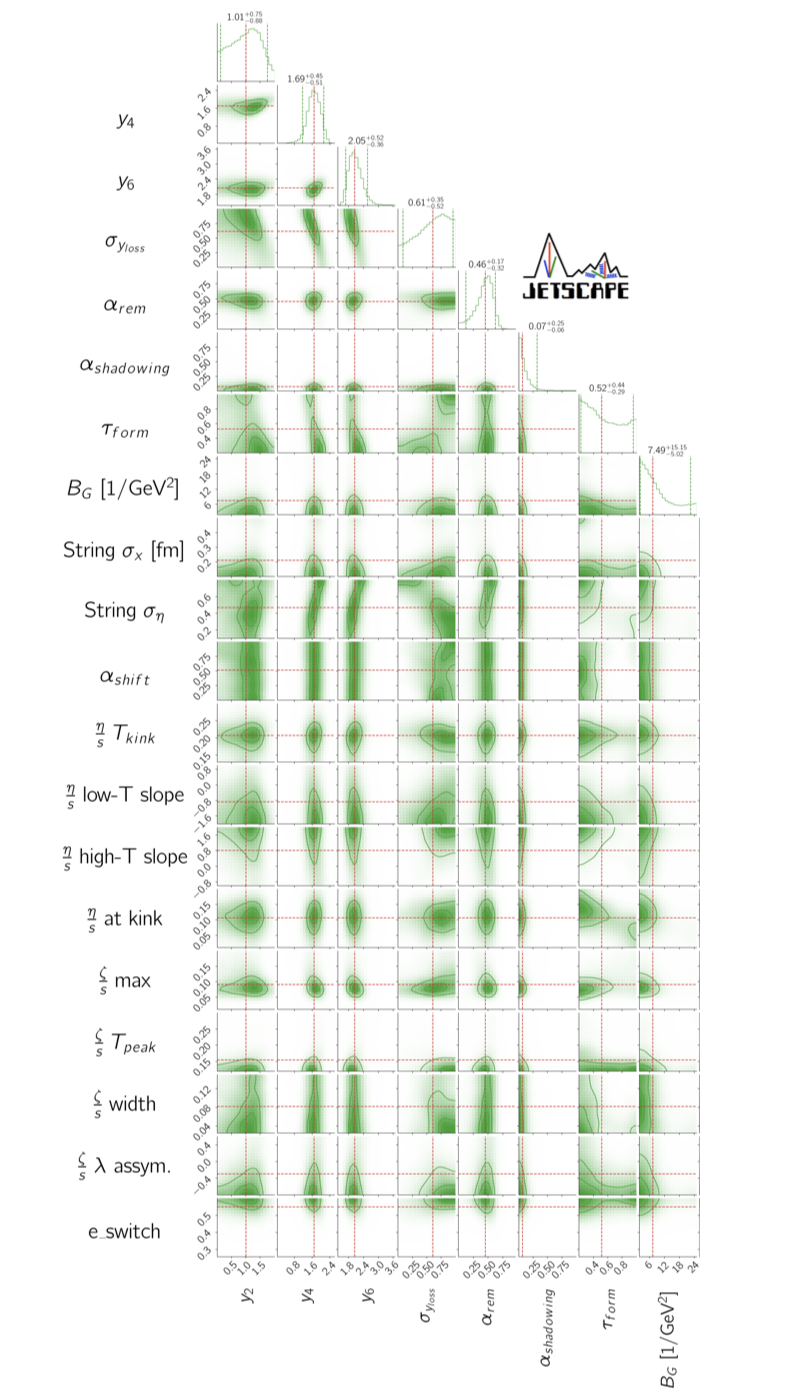}
\caption{A portion of the 1D and 2D marginalized parameter posteriors for the default calibration. Together with the plots in Fig.~\ref{fig_param_posterior_default_full_b} this represents the full, 20-dimensional posterior of parameters calibrated in the analysis. The plots across the main diagonal show the 1D marginalized posteriors for each of the parameters. The off-diagonal plots show the 2D correlations between pairs of parameters. The median and 90\% credible interval bounds are indicated by the red and green dashed lines, respectively.}
\label{fig_param_posterior_default_full_a}
\end{figure*}

\begin{figure*}[ht]
\includegraphics[scale=0.48]
{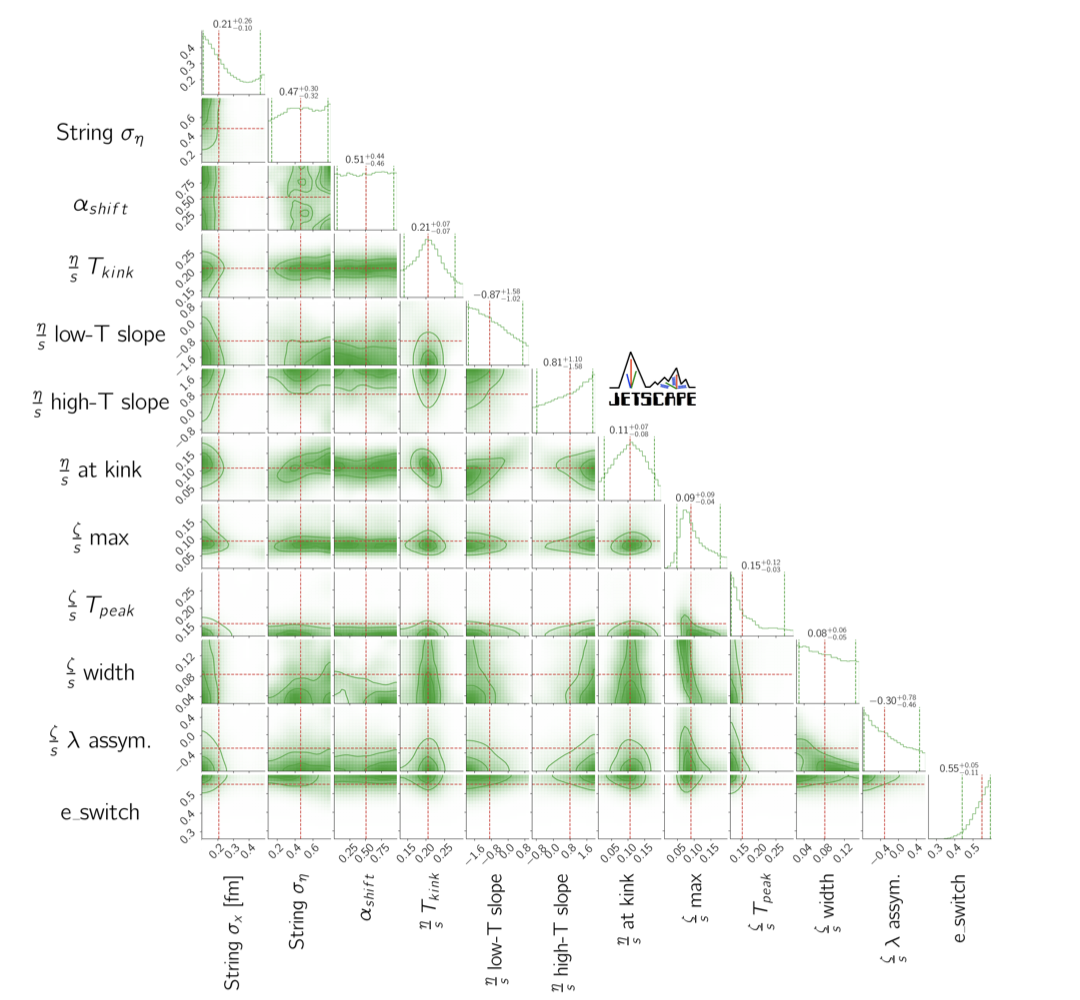}
\centering
\caption{The remaining portion of the 1D and 2D marginalized parameter posteriors for the default calibration. Together with the plots in Fig.~\ref{fig_param_posterior_default_full_a} this represents the full, 20-dimensional posterior of parameters calibrated in the analysis. The plots across the main diagonal show the 1D marginalized posteriors for each of the parameters. The off-diagonal plots show the 2D correlations between pairs of parameters. The median and 90\% credible interval bounds are indicated by the red and green dashed lines, respectively.
}
\label{fig_param_posterior_default_full_b}
\end{figure*}

\section{\label{corner_full}Full default posterior correlations}

In this appendix we show, in two parts, the full 2D correlations between all 20 of the model parameters in the default calibration in Figs.~\ref{fig_param_posterior_default_full_a} and ~\ref{fig_param_posterior_default_full_b}. This is a more complete picture of the correlations shown in Fig.~\ref{fig_param_posterior_default}, which showed only a selection of parameters for readability and focus.

\section{\label{observable_sensitivity}Observable sensitivity}

\begin{figure*}
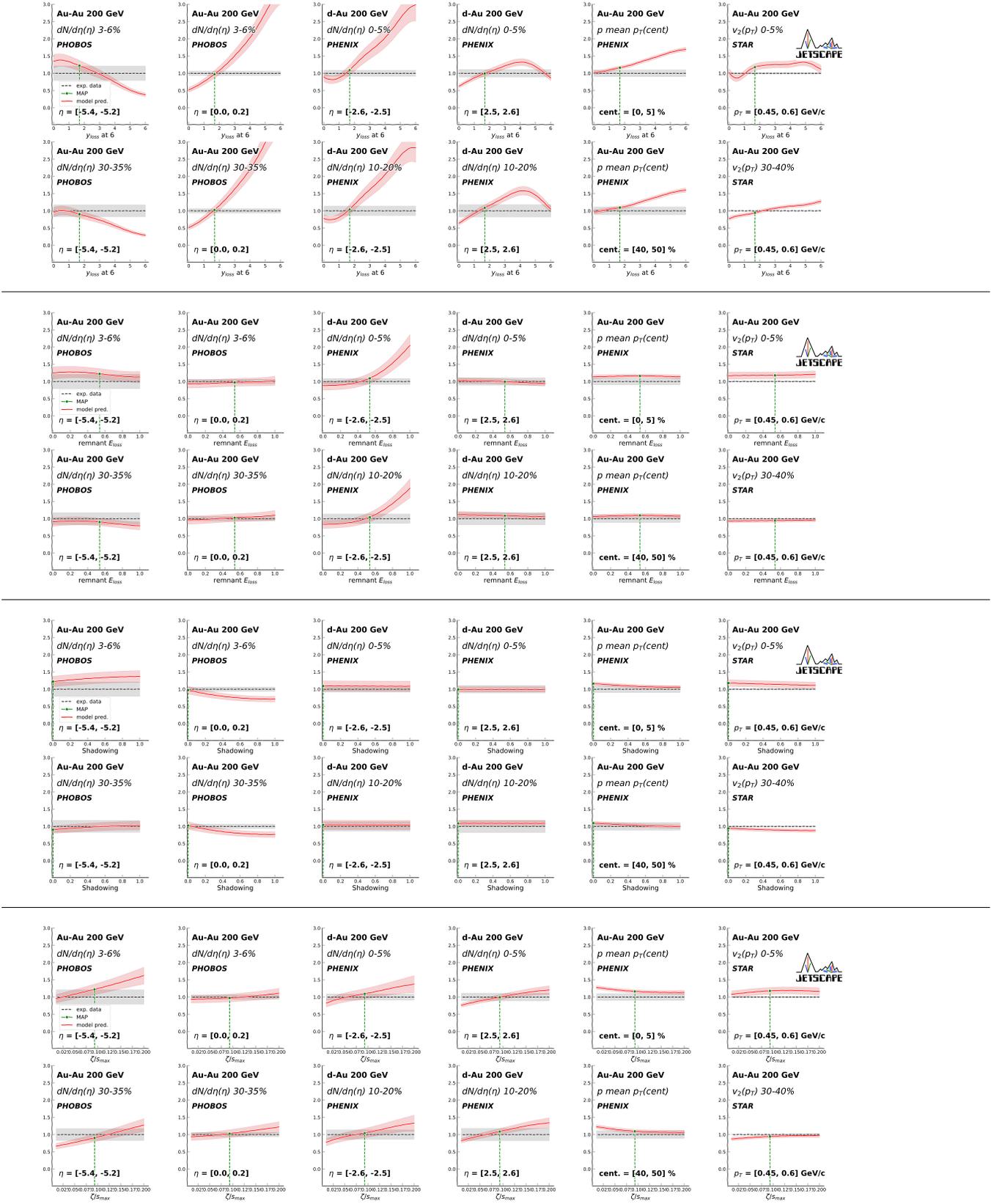

\includegraphics[scale=0.17]{obs_sensitivity_y6.png}
\rule[1ex]{\textwidth}{0.1pt}
\includegraphics[scale=0.17]{obs_sensitivity_remnant.png}
\rule[1ex]{\textwidth}{0.1pt}
\includegraphics[scale=0.17]{obs_sensitivity_shadowing.png}
\rule[1ex]{\textwidth}{0.1pt}
\includegraphics[scale=0.17]{obs_sensitivity_bulk.png}
\centering
\caption{The sensitivity of select observable bins to varying one of the rapidity loss parameters, the remnant energy loss factor, the shadowing factor, and the maximum of the bulk viscosity, calculated using the emulator. The red band indicates the emulated prediction and uncertainty, the gray band the experimental measurement and uncertainty, while the dashed line marks the MAP value of the parameter. Both the emulated predictions and the experimental data are scaled by the values of the respective experimental measurements to facilitate comparison across observables.
}
\label{obs_sensitivity}
\end{figure*}

We can use our trained emulator to efficiently probe the local sensitivity of observables to our model parameters. In Fig.~\ref{obs_sensitivity} in this appendix we show the impact on a sampling of observables due to changes in a single parameter at a time. This may in principle be done in any region of parameter space, though for these figures we choose the locus to be the MAP parameter set. We show the dependence on the maximum of the bulk viscosity, the shadowing factor, the rapidity loss, and the remnant energy loss factor. The plots show the emulator prediction at each parameter point and its corresponding uncertainty band in red, and the experimental measurement and its corresponding uncertainty in gray, each scaled by the value of the experimental data for that observable bin.

\section{\label{observable_posteriors}Full observable posteriors for independent system calibrations}

\begin{figure*}
\includegraphics[scale=0.365]{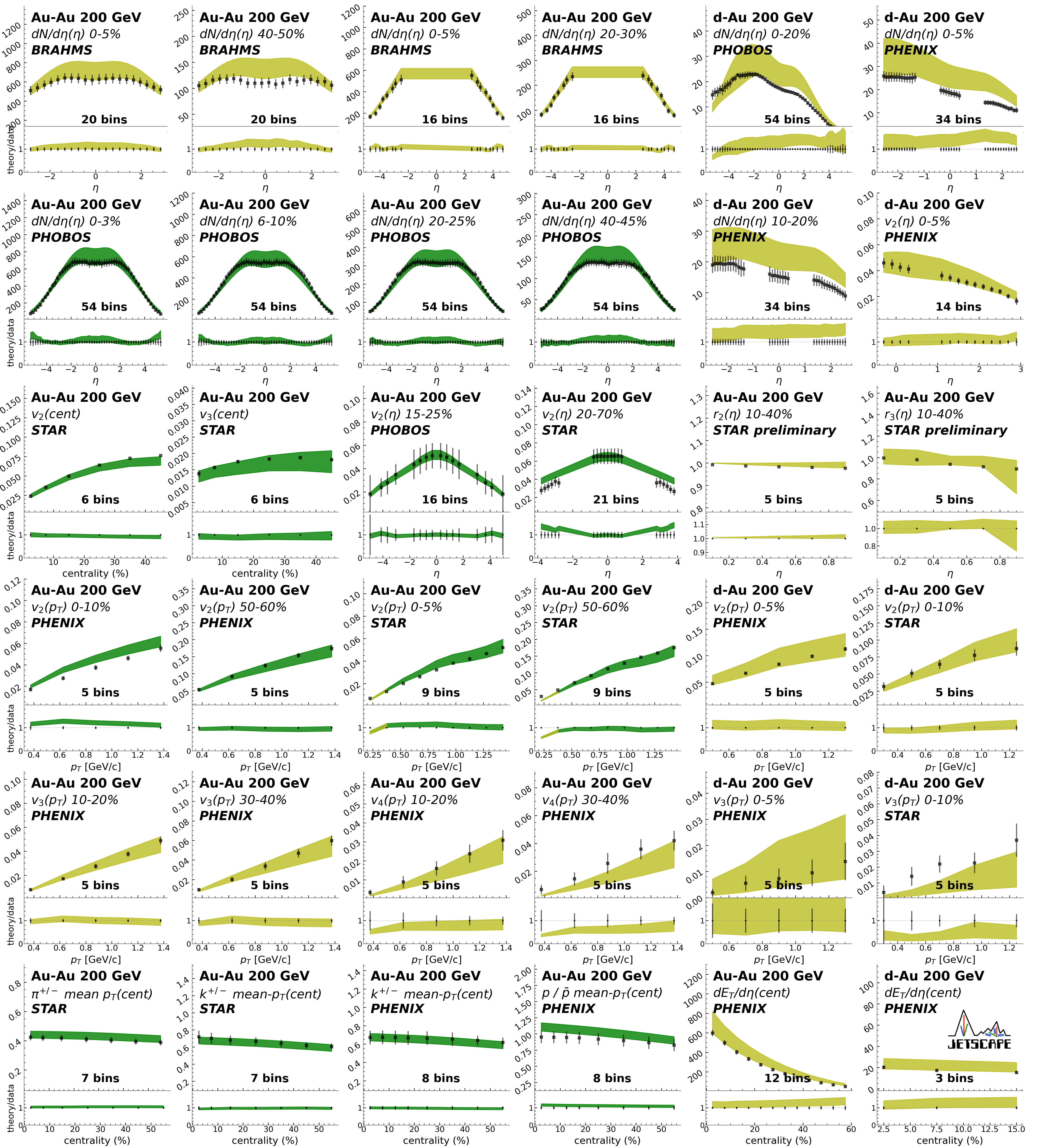}
\centering
\caption{The observable posterior for the data included in the Au-Au-only calibration, plotted in green, as well as predictions for some not included, plotted in olive. The solid markers are the experimental data, while the bands show a 90\% interval of the posterior calculated using the emulator. Each plot is labeled with the number of data bins corresponding to each observable in a given plot.}
\label{fig-observable-posterior-AuAu}
\end{figure*}

\begin{figure*}
\includegraphics[scale=0.365]{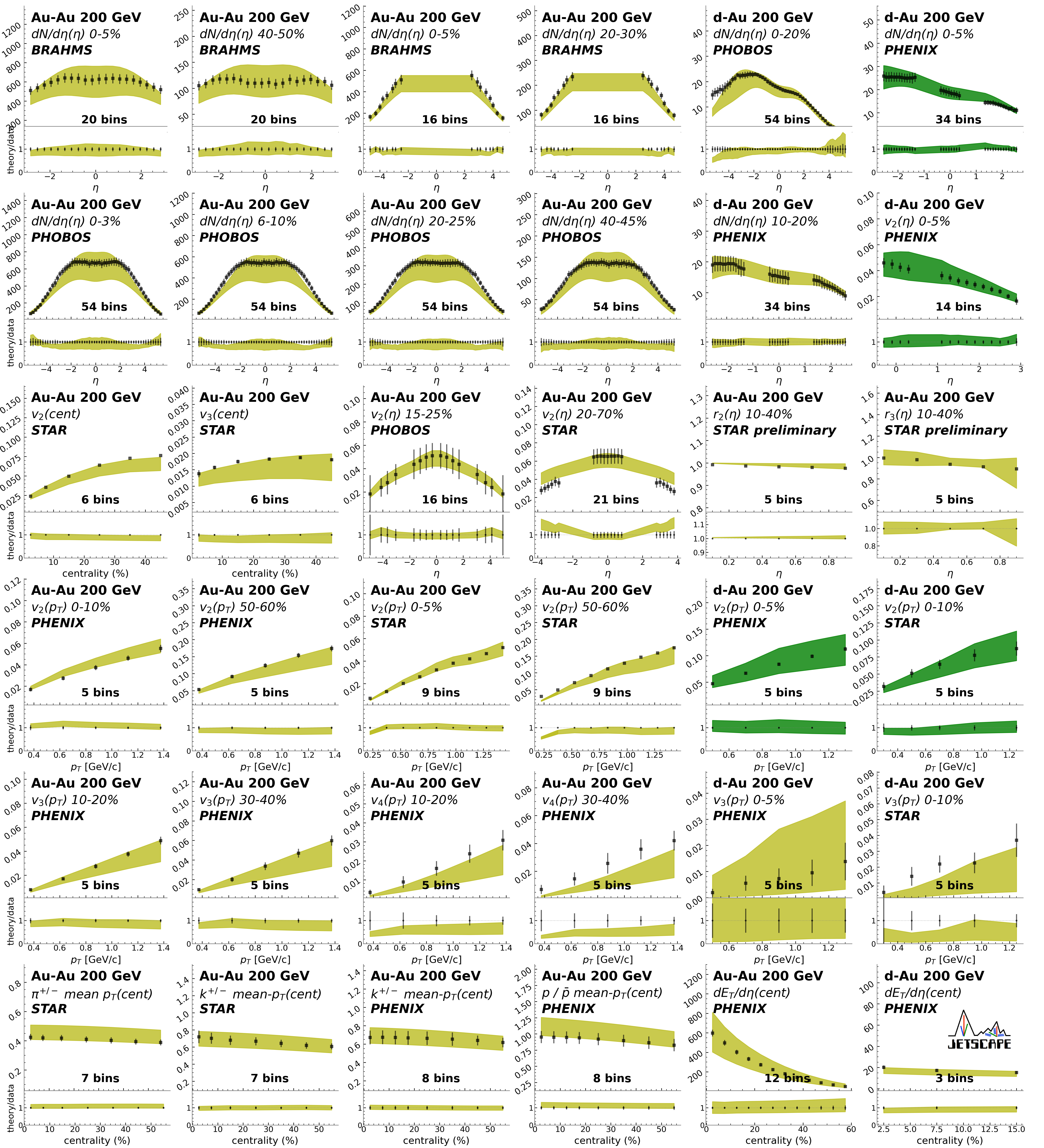}
\centering
\caption{The observable posterior for the data included in the d-Au-only calibration, plotted in green, as well as predictions for some not included, plotted in olive. The solid markers are the experimental data, while the bands show a 90\% interval of the posterior calculated using the emulator. Each plot is labeled with the number of data bins corresponding to each observable in a given plot.
}
\label{fig-observable-posterior-dAu}
\end{figure*}

In Section~\ref{sec:system_calibrations} we show and discuss the posterior for a selection of observables in the Au-Au and d-Au independent calibrations. In Figs.~\ref{fig-observable-posterior-AuAu} and \ref{fig-observable-posterior-dAu} in this appendix, we show for completeness a fuller set of observable posteriors for those calibrations as plotted in Fig.~\ref{fig-observable-posterior} for the default calibration, again using the emulator to estimate the model predictions for one hundred thousand samples of the respective posteriors.

\section{\label{systems_posteriors}Parameter posteriors for independent system calibrations}

\begin{figure*}
\includegraphics[scale=0.44]{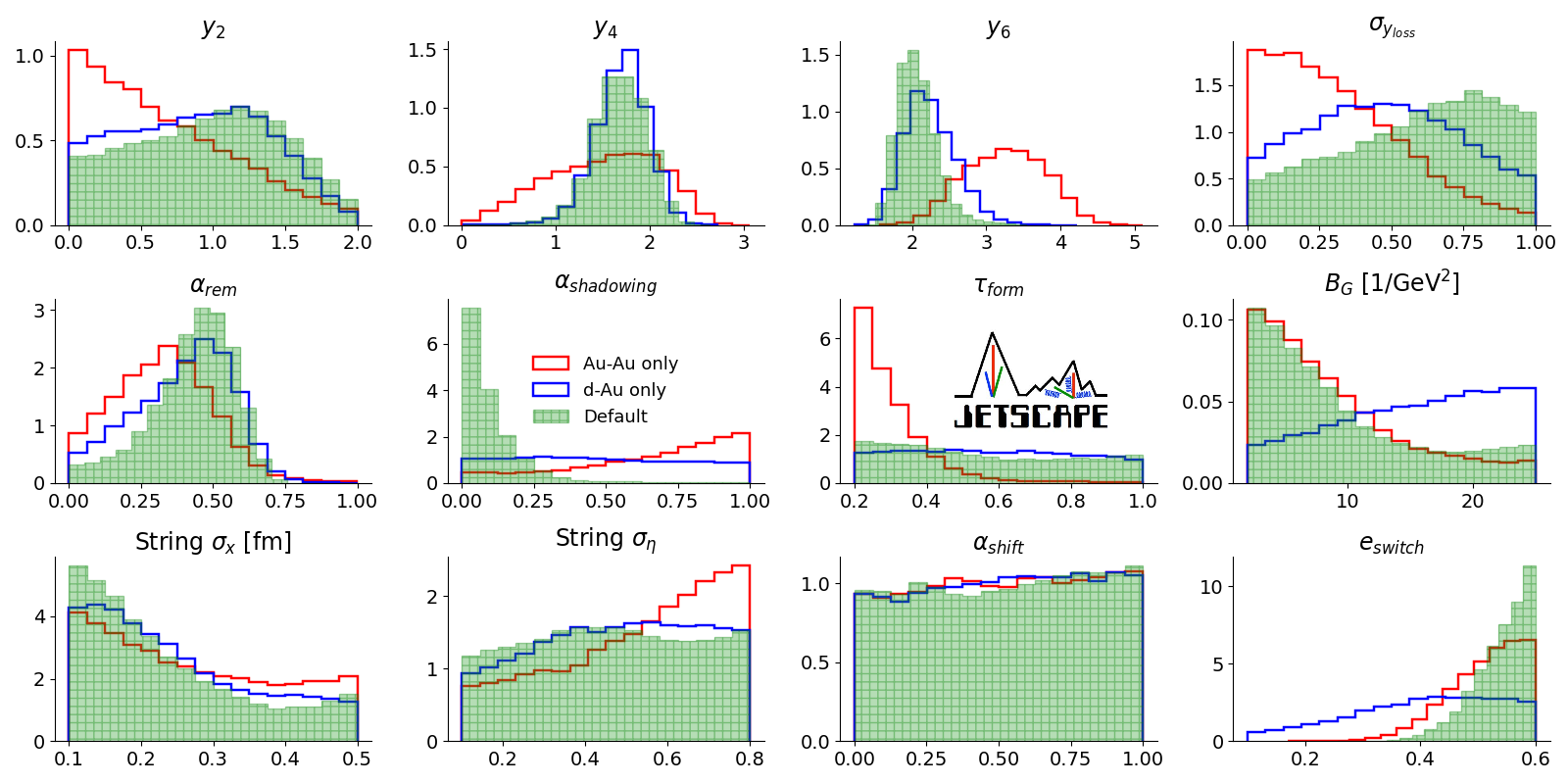}
\centering
\caption{The posteriors for the twelve model parameters (excluding the viscosity parameters), for three calibrations: the default, in hatched green, the Au-Au only calibration, in red, and the d-Au-only calibration in blue.
}
\label{fig_systems_12_params}
\end{figure*}

In Section~\ref{sec:system_calibrations} we show the posterior of the viscosity parameters for the independent Au-Au and d-Au calibrations in Fig.~\ref{fig_system_visc_post}. Here in Fig.~\ref{fig_systems_12_params} we show additionally the posteriors for the remaining 12 parameters in marginalized form.

\section{\label{posterior_samples}Parameter values of five posterior samples and MAP}

In Table \ref{tab:table3} we include the values of the 20 model parameters for each of the five samples of the posterior we have simulated and used to make predictions throughout Section~\ref{sec:predictions} as well as for the MAP parameter set. To reproduce the results included in this paper, these parameters should be used in conjunction with default parameters of the 3DGlauber \& iEBE-MUSIC framework, either through the native package or through X-SCAPE.

\begin{table*}[tbph]
\caption{\label{tab:table3}
All 20 model parameters for each of the five posterior samples used to make high-statistics posterior predictions throughout this analysis as well as the MAP parameter set and the prior range for each parameter.}
\begin{ruledtabular}
\begin{tabular}{ l |!{\vrule width -1pt}l !{\vrule width -1pt}|l !{\vrule width -1pt}|l !{\vrule width -1pt}|l !{\vrule width -1pt}|l !{\vrule width -1pt}|l !{\vrule width -1pt}|l } 

\rowcolor{gray!50}Parameters & Sample 1& Sample 2& Sample 3& Sample 4& Sample 5& MAP& Prior \\
\hline

$y_{2}$& 0.659 & 1.054 & 1.820 & 0.441 & 0.274 & 1.610 & [0,2]
\\
\rowcolor{gray!25}$y_{4}$& 1.357 & 1.684 & 1.936 & 1.812 & 1.032 & 1.685 & [$y_{2}$,4]
\\
$y_{6}$& 1.801 & 1.977 & 2.078 & 2.054 & 2.048 & 1.685 & [$y_{4}$,6]
\\
\rowcolor{gray!25}$\sigma_{y_{loss}}$& 0.990 & 0.420 & 0.354 & 0.774 & 0.912 & 0.682 & [0,1]
\\
$\alpha_{rem}$& 0.326 & 0.434 & 0.232 & 0.625 & 0.554 & 0.536 & [0,1]
\\
\rowcolor{gray!25}$\alpha_{shadowing}$& 0.104 & 0.034 & 0.032 & 0.122 & 0.095 & 0.001 & [0,1]
\\
$B_G$ [GeV$^{-2}$]& 24.020 & 6.020 & 21.860 & 4.440 & 21.220 & 3.960 & [2,25]
\\
\rowcolor{gray!25}$\sigma_x$ [fm]& 0.105 & 0.100 & 0.287 & 0.210 & 0.107 & 0.207 & [0.1,0.5]
\\
$\sigma_\eta$& 0.294 & 0.473 & 0.393 & 0.566 & 0.284 & 0.458 & [0.1,0.8]
\\
\rowcolor{gray!25}$\alpha_{shift}$& 0.927 & 0.779 & 0.075 & 0.510 & 0.011 & 0.477 & [0,1]
\\
$\tau_{form} [fm]$& 0.822 & 0.453 & 0.219 & 0.557 & 0.952 & 0.424 & [0.2,1]
\\
\rowcolor{gray!25}$(\eta/s)$ $T_{kink}$ [GeV]& 0.277 & 0.227 & 0.229 & 0.260 & 0.200 & 0.206 & [0.13,0.3]
\\
$m_{low}$ [GeV$^{-1}$]& 0.838 & -0.766 & -1.705 & -1.396 & -0.779 & -1.999 & [-2,1]
\\
\rowcolor{gray!25}$m_{high}$ [GeV$^{-1}$]& 0.500 & 1.351 & 1.880 & 1.218 & 1.211 & 1.999 & [-1,2]
\\
$(\eta/s)_{kink}$& 0.123 & 0.145 & 0.041 & 0.063 & 0.036 & 0.108 & [0.01,0.2]
\\
\rowcolor{gray!25}$(\zeta/s)_{max}$& 0.147 & 0.103 & 0.055 & 0.090 & 0.134 & 0.092 & [0.01,0.2]
\\
$(\zeta/s)$ $T_{max}$ [GeV]& 0.130 & 0.153 & 0.232 & 0.169 & 0.152 & 0.180 & [0.12,0.3]
\\
\rowcolor{gray!25}$w_{\zeta}$ [GeV]& 0.068 & 0.081 & 0.127 & 0.026 & 0.055 & 0.067 & [0.025,0.15]
\\
$\lambda_{\zeta}$& -0.358 & -0.644 & -0.334 & 0.004 & -0.673 & -0.799 & [-0.8,0.6]
\\
\rowcolor{gray!25}$e_{switch}$ [GeV/fm$^3$]& 0.591 & 0.527 & 0.476 & 0.423 & 0.588 & 0.520 & [0.1,0.6]
\end{tabular}
\end{ruledtabular}
\end{table*}

\section{\label{scaled_reference_flow}Dependence of flow coefficients on the flow reference}

\begin{figure}
\includegraphics[scale=0.60]{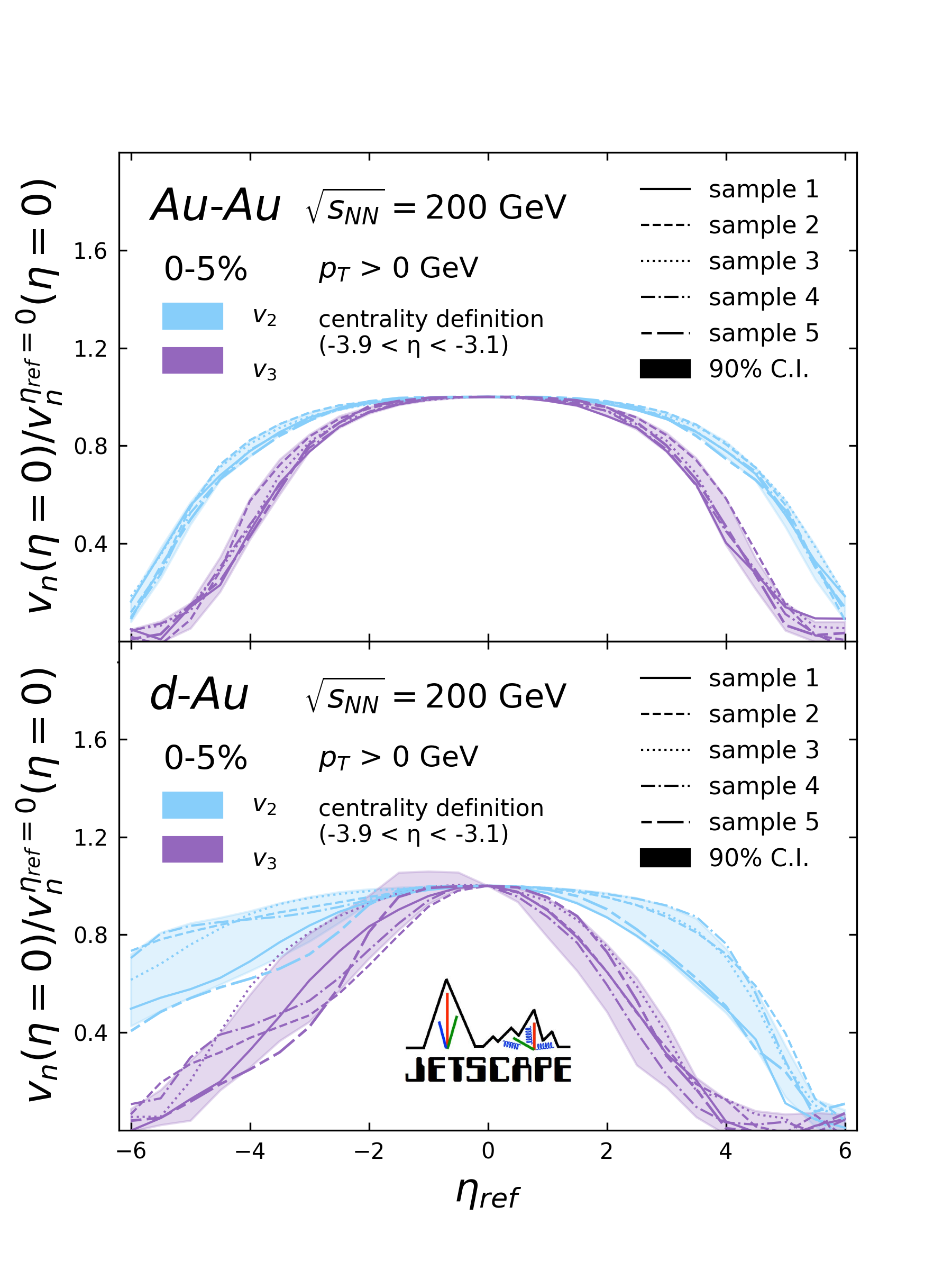}
\centering
\caption{The calibrated model calculations for flow coefficients in Au-Au and d-Au at mid-rapidity as a function of the reference rapidity region, scaled by the value of the coefficients using a mid-rapidity reference. The calculations are plotted for five individual parameter samples of the posterior simulated for ten thousand events each. In addition, a 90\% credible interval band, estimated from ten such high-statistics random samples, is shown.
}
\label{fig_scaled_vn_reference}
\end{figure}

The dependence of the flow coefficients $v_2$ and $v_3$ at mid-rapidity as a function of the reference rapidity was shown in Fig.~\ref{fig_vn_reference}. Here, in Fig.~\ref{fig_scaled_vn_reference}, we show a version of that result scaled by the value of the coefficients using, as a benchmark, a mid-rapidity reference. The scaled version allows for an easier comparison of the impact of the rapidity region of the reference across the full rapidity range and between systems. We note that the apparent decrease in $v_n$ at mid-rapidity with the distance of the reference from mid-rapidity is a convolution of multiple effects, including the decorrelation of the event planes in rapidity as well as the rapidity dependence of the magnitude of flow itself.

\begin{table}[htbp]
\caption{All 21 model parameters for the re-tuned MAP parameter set to fit 5.02 TeV Pb-Pb LHC data.}
\label{tab:table4}
\centering
\begin{minipage}{0.45\linewidth}
\centering
\begin{tabular}{ll}
\rowcolor{gray!50} Parameters & MAP \\ 
\hline
\rowcolor{gray!25} $y_{2}$ & 1.610 \\ 
$y_{4}$ & 1.685 \\ 
\rowcolor{gray!25} $y_{6}$ & 1.685 \\ 
$y_{10}$ & 2.800 \\ 
\rowcolor{gray!25} $\sigma_{y_{loss}}$ & 0.682 \\ 
$\alpha_{rem}$ & 0.536 \\ 
\rowcolor{gray!25} $\alpha_{shadowing}$ & 0.001 \\ 
$B_G$ [GeV$^{-2}$]& 3.960 \\ 
\rowcolor{gray!25} $\sigma_x$ [fm] & 0.207 \\ 
$\sigma_\eta$ & 0.458 \\ 
\end{tabular}
\end{minipage}%
\hfill 
\begin{minipage}{0.45\linewidth}
\centering
\begin{tabular}{ll}
\rowcolor{gray!50} Parameters & MAP \\ 
\hline
\rowcolor{gray!25} $\alpha_{shift}$ & 0.477 \\ 
$\tau_{form}$ & 0.424 \\ 
\rowcolor{gray!25} $(\eta/s)$ $T_{kink}$ [GeV] & 0.206 \\ 
$m_{low}$ [GeV$^{-1}$]& -1.999 \\ 
\rowcolor{gray!25} $m_{high}$ [GeV$^{-1}$]& 1.999 \\ 
$(\eta/s)_{kink}$ & 0.108 \\ 
\rowcolor{gray!25} $(\zeta/s)_{max}$ & 0.092 \\ 
$(\zeta/s)$ $T_{max}$ [GeV] & 0.180 \\ 
\rowcolor{gray!25} $w_{\zeta}$ [GeV]& 0.067 \\ 
$\lambda_{\zeta}$ & -0.799 \\ 
\rowcolor{gray!25} $e_{switch}$ [GeV/fm$^3$] & 0.120 \\ 
\end{tabular}
\end{minipage}
\end{table}

\begin{figure}[htbp]
\includegraphics[scale=0.60]{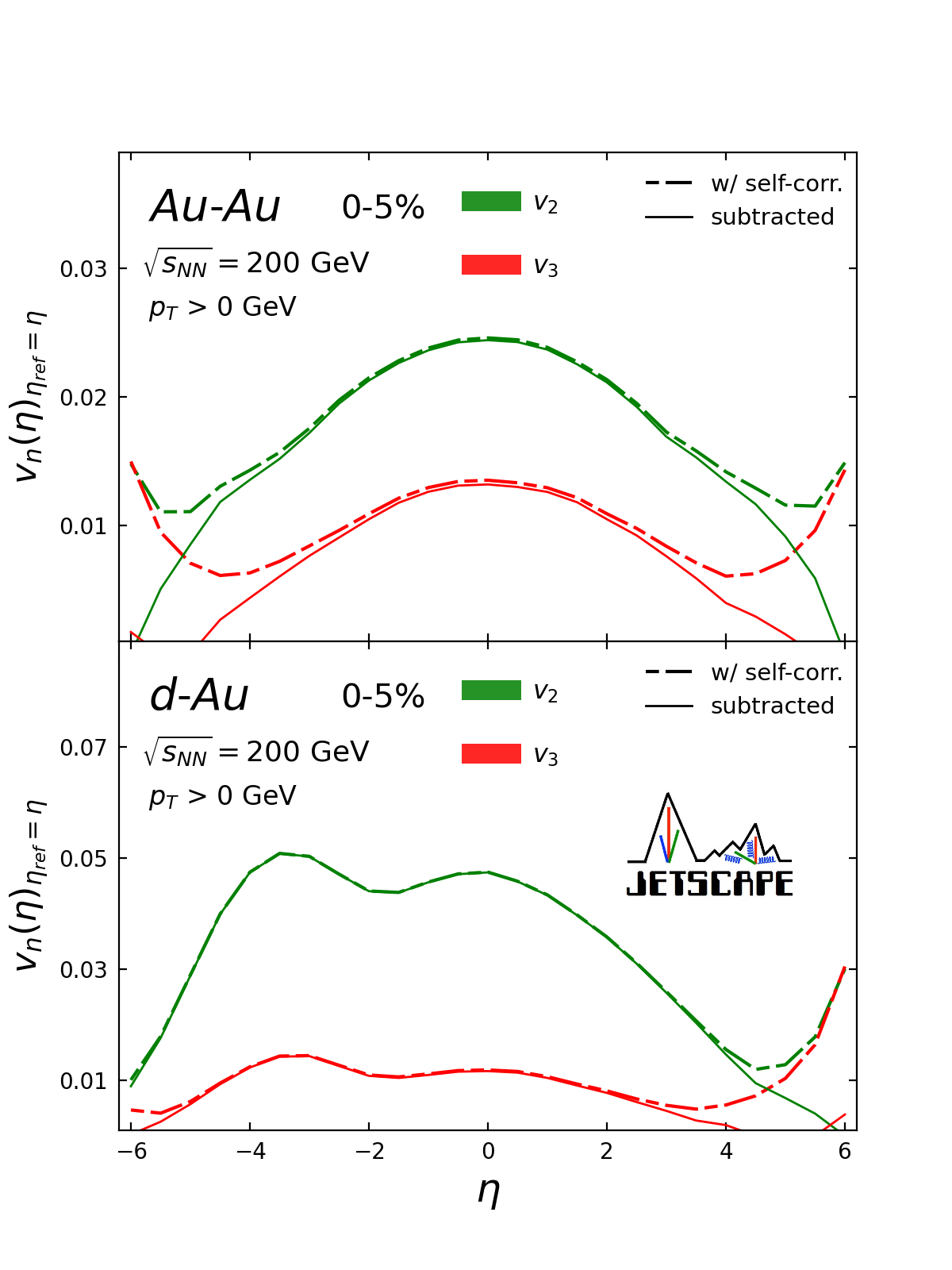}
\centering
\caption{The calibrated model calculations for $v_2$($\eta$) and $v_3$($\eta$) in Au-Au (top) and d-Au (bottom) for the case where the reference rapidity region coincides with the rapidity bin of interest, for all $\eta$, in bins of 0.5 units. The calculations are shown for sample 3 of the default posterior distribution for 0-5\% most central collisions in both systems. The dashed lines represent the calculation with the contribution due to auto-correlations among particles included (labeled ``w/ self-corr.''), while the solid lines represent the calculation with the contribution explicitly subtracted (labeled ``subtracted'').
}
\label{fig_v2_v3_selfcorr}
\end{figure}

\begin{figure}[htbp]
\includegraphics[scale=0.60]{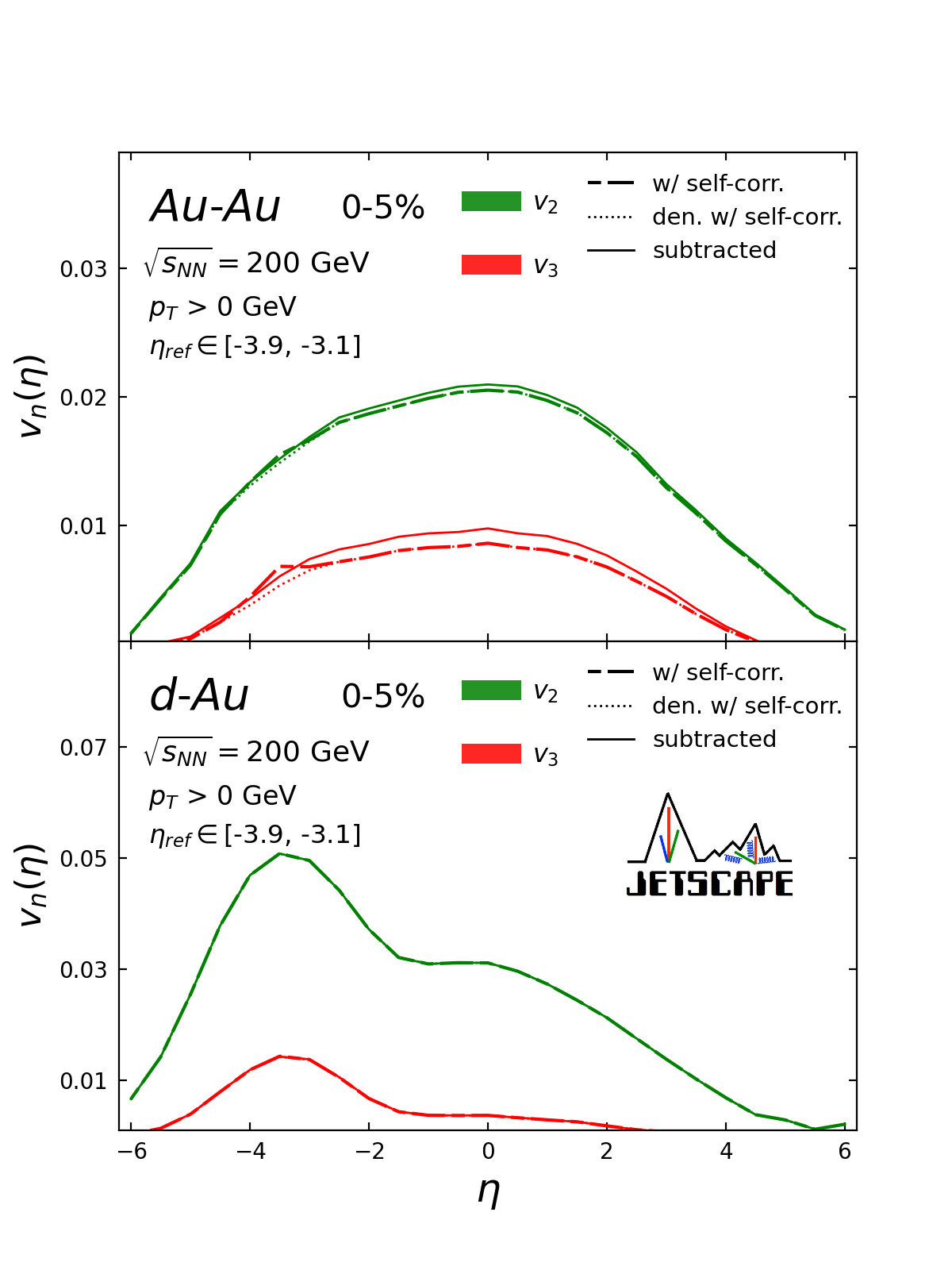}
\centering
\caption{The calibrated model calculations for $v_2$($\eta$) and $v_3$($\eta$) in Au-Au (top) and d-Au (bottom) using the generic PHENIX Beam-Beam Counter reference region ([-3.9, -3.1]), for all $\eta$ in bins of 0.5 units. The calculations are shown for sample 3 of the default posterior distribution for 0-5\% most central collisions in both systems. The dashed lines represent the calculation with the contribution due to auto-correlations included (labeled ``w/ self-corr.''), while the dotted lines show the observables having subtracted only the numerator auto-correlations (which are non-zero only for $\eta$ in the Beam-Beam Counter region) labeled as ``den. w/ self-corr.''. The solid lines (labeled ``subtracted'') meanwhile represent the fully-subtracted calculations, additionally removing auto-correlations originating from the denominator correlator. Note that the subtraction from the denominator naturally leads to a larger $v_n$. In the d-Au (bottom) panel all three curves essentially overlap. In the Au-Au (top) panel, where the effect is more significant, the dashed and dotted lines overlap everywhere except in the $\eta$-region that coincides with the reference, because the Q-vectors in the numerator correlator overlap in only that region.}
\label{fig_v2_v3_subtracted_denom}
\end{figure}

\section{\label{same_rap}Same-rapidity correlations}

In this appendix we show the dependence of $v_n$ on $\eta$ if the reference rapidity region coincides with the region of interest. This is sometimes the case in experimental measurements---particularly in some older measurements where only a single detector, confined to a limited rapidity region, was used to obtain $v_n$. In some measurements from STAR, for example, the observable was measured at mid-rapidity while enforcing an $\eta$-separation between individual pairs of particles within the mid-rapidity region covered by the central TPC. If such measurements are compared with---as they are in this paper---care must be taken to explicitly account for the potential auto-correlations that may arise from correlating particles from the same rapidity region. In Fig.~\ref{fig_v2_v3_selfcorr} we show our calculations of $v_n(\eta)$ when $\eta_{ref}=\eta$ (in bins of width 0.5 in $\eta$) for one sample of the default posterior. To highlight the potential impact of auto-correlations if not accounted for properly, we show the calculations including those contributions, as well as after subtraction. Overall, the $v_n$ values show similar $\eta$-dependence as when the regions of reference and interest are separated, with perhaps a slightly more peaked shape around mid-rapidity. We can see that the impact from auto-correlations is relatively small for large values of the $v_n$ and where the particle multiplicity is largest (notably at mid-rapidity). The impact grows as we move to large absolute rapidities for both of those reasons and becomes significant around $|\eta|$=4, and dominant quickly beyond that. We note also the smaller impact on the d-Au $v_n$ compared to Au-Au, chiefly attributable to the larger values of the observable for central events in that system.

\begin{figure}[htbp]
\includegraphics[scale=0.4]{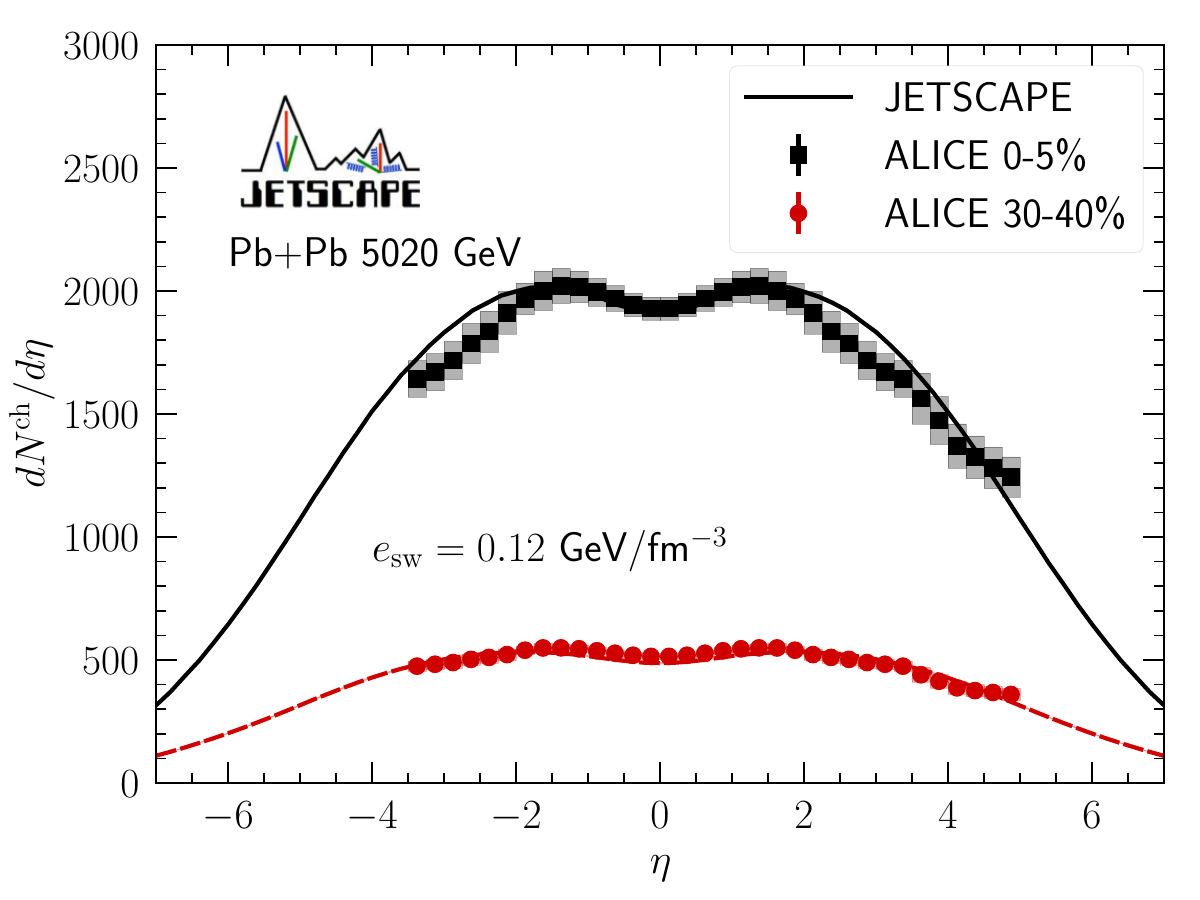}
\includegraphics[scale=0.4]{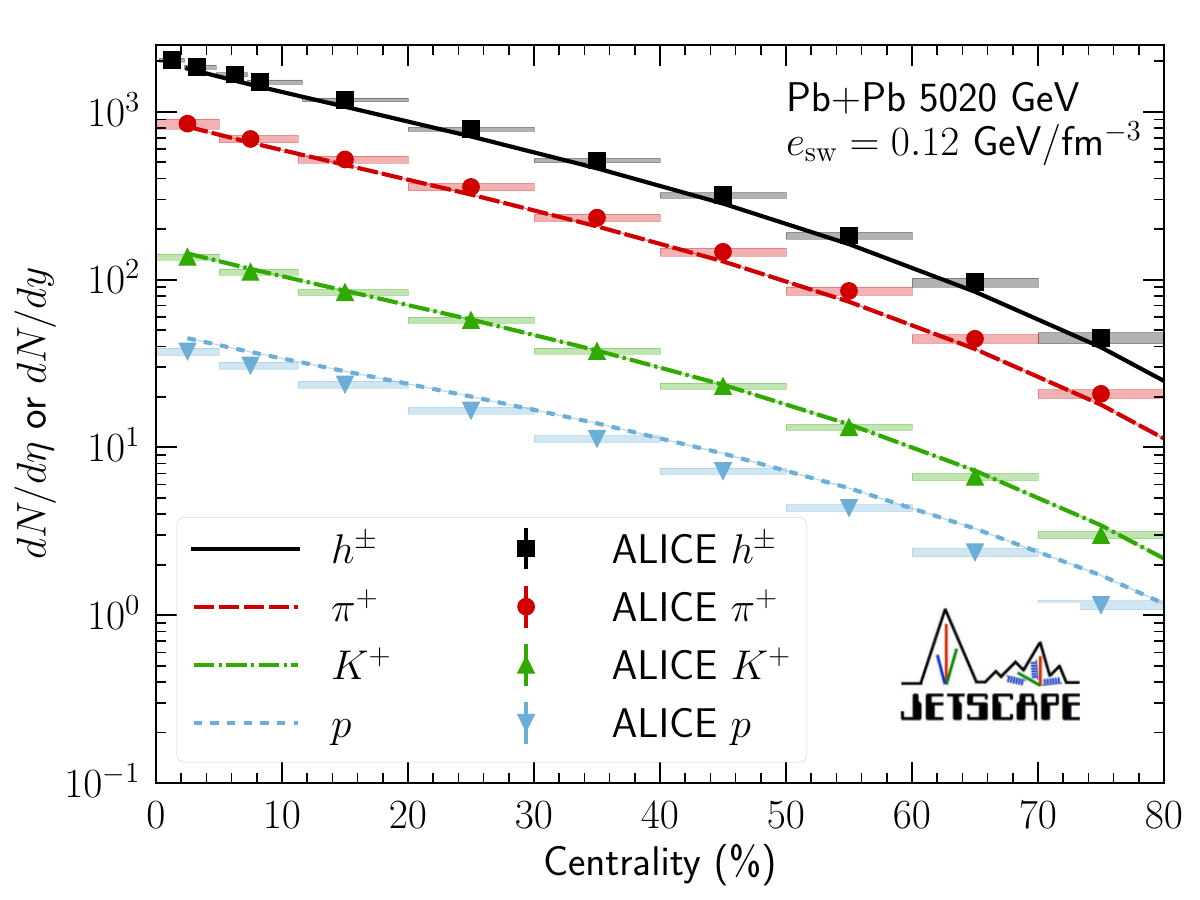}
\centering
\caption{The model calculations of the multiplicity as a function of $\eta$ and the integrated charged and identified hadrons for Pb-Pb at 5.02 TeV for the re-tuned parameter set obtained from our calibration's MAP compared to experimental measurements from ALICE. 
}
\label{fig:LHC_predictions}
\end{figure}

We also show in this appendix a cross-check of the impact of self-correlations due specifically to the reference correlator in the denominator of the scalar product definition. We define the scalar product as 
\begin{equation*}
\label{eq:scalarproduct}
v_n = \frac{\langle Q_A \cdot Q_{B}^* \rangle }
{\langle M_A \rangle \sqrt{\langle Q_{B} \cdot Q_{B}^* \rangle }}
\end{equation*}
where we only invoke two ``subevents" separated in rapidity, namely A and B, corresponding respectively to the regions of interest and reference, and where $M_A$ stands for the multiplicity in subevent A and the averages are taken over all events in a centrality bin. 

In defining the anisotropy coefficient this way, we make the simplification, used throughout the analysis, of correlating particles from the same subevent to compute the reference correlator $Q_B \cdot Q_B^*$. Thus, among the methodological differences with various flow measurements that we treat as negligible in this analysis, we also include the contribution of auto-correlations to this denominator. In Fig.~\ref{fig_v2_v3_subtracted_denom} we show that the contribution from such auto-correlations is indeed negligible for both Au-Au and (especially) d-Au, even for small values of the $v_n$. The largest impact, as shown in this figure, is on the $v_3(\eta)$, an observable which does not in fact feature in our analysis in $\eta$-differential form and for which we in any case have relatively large emulation uncertainties. Lastly, since the region of interest and reference do not in general overlap in the analysis, we do not have an auto-correlation contribution coming from the numerator correlator.

\section{\label{LHC_predictions}A note on making LHC predictions}

The current analysis constrains our 3D hydrodynamic model with data collected at the RHIC top energy. While the sensitivity of the rapidity loss parameters varies with beam energy (and additional constraints must thus be obtained at the LHC energy), one might reasonably assume that the remaining parameters have no explicit dependence on the beam energy. Under this assumption, then, we can make predictions for LHC observables starting from our current posterior. 

While, as we have done throughout this paper, it is best to make model predictions using a sampling of the posterior (in order to show a realistic quantification of the uncertainty on those predictions), this may not be practical if an additional fitting procedure is required. Thus, given this is the case for the rapidity loss when making LHC predictions, it may be best to use only the MAP parameter set as a starting point. We lay out here the parameter set resulting from such a fit, which includes not only a $y_{10}$ parameter for the rapidity loss at $y_{beam}$=10 (with $y_{LHC}$= 8.58 for 5.02 TeV center of mass energy), but additionally a re-fitted $e_{switch}$. The latter parameter is not of strong significance for predicting most LHC observables. However, given the exclusion of proton yield data in our current calibration, this observable can be well-described at the LHC (as well as at RHIC) by our model only if the switching energy density is refit to a lower value. Previous studies found that the switching point is indeed especially sensitive to identified particle observables \cite{Bernhard2019, JETSCAPE:2020mzn}.

The parameter set resulting from this LHC fit to Pb-Pb data is shown in Table \ref{tab:table4} and, again, features only two changes from the MAP quoted in Table \ref{tab:table3}.

While this LHC tune is not a direct prediction from our posterior, it is a fit directly informed by our calibration's best fit to RHIC top energy data. For small LHC systems such as p-Pb, O-O, or Ne-Ne, the same (or a similar) fitting procedure may be used to make predictions. The model may be run through the iEBE-MUSIC \cite{Shen:2023awv} or X-SCAPE frameworks \cite{JETSCAPE:2024dgu}.

A comparison of LHC predictions using this tune and the experimental data from ALICE for the Pb-Pb dN/d$\eta$ and integrated charged and identified particle yields may be seen in Fig.~\ref{fig:LHC_predictions}.

\clearpage

\bibliography{apssamp}

\end{document}